\documentclass[twocolumn]{aastex63}

\renewcommand{\thefootnote}{\fnsymbol{footnote}}

\graphicspath{{./}{figures/}}

\received{Mar 8, 2020}
\revised{May 29, 2020}
\accepted{May 30, 2020}
\submitjournal{ApJ}

\shorttitle{NIR Brightening of Non-Var. OH/IR Stars}
\shortauthors{Kamizuka et al.}

\begin{document}

\title{Long-Term Near-Infrared Brightening of Non-Variable OH/IR Stars}

\correspondingauthor{Takafumi Kamizuka}
\email{kamizuka@ioa.s.u-tokyo.ac.jp}

\author[0000-0001-6268-619X]{Takafumi Kamizuka}
\affiliation{Institute of Astronomy, the University of Tokyo, 2-21-1 Osawa, Mitaka, Tokyo 181-0015, Japan}

\author{Yoshikazu Nakada}
\affiliation{Institute of Astronomy, the University of Tokyo, 2-21-1 Osawa, Mitaka, Tokyo 181-0015, Japan}

\author[0000-0001-7592-9285]{Kenshi Yanagisawa}
\affiliation{Division of Optical and Infrared Astronomy, National Astronomical Observatory of Japan, 2-21-1 Osawa, Mitaka, Tokyo 181-8588, Japan}

\author[0000-0001-5797-6010]{Ryou Ohsawa}
\affiliation{Institute of Astronomy, the University of Tokyo, 2-21-1 Osawa, Mitaka, Tokyo 181-0015, Japan}

\author{Yoshifusa Ita}
\affiliation{Astronomical Institute, Tohoku University, 6-3 Aramaki, Aoba, Aoba-ku, Sendai, Miyagi 980-8578, Japan}

\author[0000-0002-8435-2569]{Hideyuki Izumiura}
\affiliation{Subaru Telescope Okayama Branch Office, National Astronomical Observatory of Japan, 3037-5, Honjou, Kamogata, Asakuchi, Okayama 719-0232, Japan
}

\author{Hiroyuki Mito}
\affiliation{UTokyo Organization for Planetary and Space Science, the University of Tokyo, 7-3-1, Hongo, Tokyo 113-0033, Japan}

\author[0000-0002-8015-0476]{Hiroki Onozato}
\affiliation{Nishi-Harima Astronomical Observatory, Center for Astronomy, Institute of Natural and Environmental Sciences, University of Hyogo, 407-2 Nishigaichi, Sayo-cho, Sayo-gun, Hyogo 679-5313, Japan}

\author[0000-0002-8107-3783]{Kentaro Asano}
\affiliation{Institute of Astronomy, the University of Tokyo, 2-21-1 Osawa, Mitaka, Tokyo 181-0015, Japan}

\author[0000-0003-0735-578X]{Toshiya Ueta}
\affiliation{Department of Physics and Astronomy, University of Denver, 2112 E Wesley Ave., Denver 80208, Colorado, USA}

\author{Takashi Miyata}
\affiliation{Institute of Astronomy, the University of Tokyo, 2-21-1 Osawa, Mitaka, Tokyo 181-0015, Japan}


\begin{abstract}
Non-variable OH/IR stars are thought to have just left the asymptotic giant branch (AGB) phase. In this conventional picture, they must still show strong circumstellar extinction caused by the dust ejected during the AGB phase, and the extinction is expected to decrease over time because of the dispersal of the circumstellar dust after the cessation of the stellar mass loss. The reduction of the extinction makes the stars become apparently brighter and bluer with time especially in the near-infrared (NIR) range. We look for such long-term brightening of non-variable OH/IR stars by using 2MASS, UKIDSS, and OAOWFC survey data. As a result, we get multi-epoch NIR data taken over a 20-year period (1997--2017) for 6 of 16 non-variable OH/IR stars, and all six objects are found to be brightening. The {\it K}-band brightening rate of five objects ranges from 0.010 to 0.130 mag yr$^{-1}$, which is reasonably explained with the conventional picture. However, one OH/IR star, OH31.0$-$0.2, shows a rapid brightening, which cannot be explained only by the dispersal of the dust shell. Multi-color ({\it J}-, {\it H}-, and {\it K}-band) data are obtained for three objects, OH25.1$-$0.3, OH53.6$-$0.2, and OH77.9+0.2. Surprisingly, none of them appears to have become bluer, and OH53.6$-$0.2 is found to have been reddened with a rate of 0.013 mag yr$^{-1}$ in ({\it J}-{\it K}). Our findings suggest other mechanisms such as rapid changes in stellar properties (temperature or luminosity) or a generation of a new batch of dust grains.
\end{abstract}



\section{Introduction} \label{sec:intro}
The asymptotic giant branch (AGB) phase is a late evolutionary phase of low- and intermediate-mass stars with main-sequence masses of $\sim$1--10 ${\rm M}_\sun$ \citep{1996A&ARv...7...97H, 2005ARA&A..43..435H, 2014PASA...31...30K}. The stars in this phase show significant variability caused by stellar pulsation and mass ejection. Dust grains are created in mass-loss winds, and consequently, the circumstellar dust shell is developed around the central AGB stars. Among AGB stars, more massive and more evolved AGB stars are thought to experience stronger mass loss, resulting in the optically thicker dust shell causing greater obscuration of the central star \citep{1993ApJ...413..641V, 1995A&A...297..727B,2006A&A...447..553F}.

OH/IR stars are believed to be such objects undergoing the heaviest mass loss. The name {\it OH/IR} indicates that this kind of objects shows OH maser emission at 1612 MHz and are bright in the infrared rather than the optical \citep{1996A&ARv...7...97H}. These characteristics arise from a thick circumstellar shell created by massive mass-loss winds. In addition, these objects show long-period variability with periods from several hundred to over a thousand days \citep{1983A&A...124..123E, HH85, 1996A&ARv...7...97H}. Such a long-period variability is also expected for the evolved stars which are larger but less massive \citep{1993ApJ...413..641V}. These pieces of evidence support the hypothesis that OH/IR stars are evolved AGB stars.

On the other hand, some OH/IR stars are known to have no or little variability, and they are called {\it non-variable} OH/IR stars. The variability of OH/IR stars is often investigated with \replaced{OH-maser}{OH maser} emission, because their variability is difficult to monitor in the optical. Such non-variable OH/IR stars were discovered by \citet{HH85}. They monitored \replaced{OH-maser}{OH maser} emissions of bright OH/IR stars selected from Baud's catalog \citep{1981A&A....95..156B} to evaluate their variability and found that 25\% of their targets showed no or little variability. 

Such a non-variability is expected to occur after the end of the AGB mass loss. In the transition from the AGB phase to the following post-AGB phase, the significant pulsational variability and heavy mass ejection of the central star are thought to cease, since these characteristics are not observed from post-AGB stars \citep[e.g.][]{1994ApJS...92..125V, 1995A&A...299..755B}. As a result, the circumstellar dust shell starts to dissipate, and the central star obscured in the AGB phase gradually brightens in the optical and near-infrared (NIR). This reappearance of the stellar emission in the NIR has been suggested from Spectral Energy Distribution (SED) analyses. \citet{2007apn4.confE..52E} found that some non-variable OH/IR stars showed NIR excess while normal OH/IR stars did not show such an NIR excess. This NIR excess was interpreted as the emission from the star that was reemerging from the optically thick dust shell. This result suggests that the non-variable OH/IR stars are in transition from the AGB phase to the post-AGB phase. \citet{1987A&A...186..136B} also supported this picture from mid- to far-infrared SED modeling. He reported that non-variable OH/IR stars dominated a region in the {\it IRAS} two-color diagram ($\log{\left[\lambda F_\lambda(25)/\lambda F_\lambda(12)\right]}\gtrsim0.2, \log{\left[\lambda F_\lambda(60)/\lambda F_\lambda(25)\right]}\sim$$-0.6$--0.1) and their colors could be reproduced by models with a dispersing dust shell after the cessation of the AGB mass loss.

These studies are based on one-epoch observations, but now, we can investigate the long-term brightness evolution of these objects by using infrared archival data. In this paper, we investigate the long-term NIR (the {\it J}, {\it H}, and {\it K} band) brightness evolution of non-variable OH/IR stars to reveal their real-time evolution. Data in wavelength regions longer than the {\it K} band are also useful especially to investigate the evolution of the thermal emission from the circumstellar dust shell. However, the treatment of the thermal emission from the dust shell requires the detailed modeling of the radiative transfer processes which are beyond our scope. Therefore, we focus on only the NIR evolution in this paper. Below, in \S\ref{sec:identify}, we describe the target identification by searching for infrared counterparts of non-variable OH/IR stars. The search method for the long-term brightness evolution using NIR archival data is described in \S\ref{sec:variability}. \S\ref{sec:res} shows the results, and we discuss the origin of the observed brightness evolution in \S\ref{sec:disc}. \S\ref{sec:summary} summarizes this study.

\section{Target Identification} \label{sec:identify}
\begin{deluxetable*}{lllccl}
\tablecaption{
Names and positions of the target non-variable OH/IR stars.
\label{tab:target}}
\tablewidth{0pt}
\tablehead{
\colhead{Object} & \colhead{2MASS} & \colhead{SSTGLMC} & \colhead{RA} & \colhead{Dec} & \colhead{Position}\\
\colhead{name$^{\rm a}$} & \colhead{name} & \colhead{name} & \colhead{(deg; J2000)} & \colhead{(deg; J2000)} & \colhead{source}
}
\startdata
OH0.3$-$0.2   &                           & G000.3335$-$00.1805 & 266.778946 & $-28.744976$ & GLIMPSE \\
OH1.5$-$0.0$^b$   &                           & G001.4837$-$00.0612 & 267.337090 & $-27.698449$ & GLIMPSE \\
OH11.5+0.1  & J18103867$-$1852583       & G011.5599$+$00.0873 & 272.661187 & $-18.882896$ & GLIMPSE \\
OH15.7+0.8  &                           & G015.7010$+$00.7705 & 274.107406 & $-14.920483$ & GLIMPSE \\
OH17.7$-$2.0  & J18303070$-$1428570 &                           & 277.627928 & $-14.482512$ & 2MASS\\
OH18.3+0.4  &                           & G018.2956$+$00.4291 & 275.679577 & $-12.794966$ & GLIMPSE \\
OH18.5+1.4  & J18193549$-$1208081       & G018.5182$+$01.4129 & 274.897885 & $-12.135602$ & GLIMPSE \\
OH18.8+0.3  &                           & G018.7687$+$00.3016 & 276.022247 & $-12.436756$ & GLIMPSE \\
OH25.1$-$0.3  & J18381546$-$0709543       & G025.0568$-$00.3505 & 279.564387 & $-07.165077$ & GLIMPSE \\
OH31.0+0.0  &                           & G030.9439$+$00.0351 & 281.921407 & $-01.753168$ & GLIMPSE \\
OH31.0$-$0.2  &                           & G031.0123$-$00.2193 & 282.179217 & $-01.808404$ & GLIMPSE \\
OH37.1$-$0.8  &                           & G037.1184$-$00.8473 & 285.526137 & $+03.337738$ & GLIMPSE \\
OH51.8$-$0.2  & J19274203$+$1637239       & G051.8038$-$00.2248 & 291.925188 & $+16.623349$ & GLIMPSE \\
OH53.6$-$0.2  & J19312537$+$1813103       & G053.6303$-$00.2392 & 292.855788 & $+18.219504$ & GLIMPSE \\
OH77.9+0.2  & J20283067$+$3907009 &                           & 307.127821 & $+39.116924$ & 2MASS \\
OH359.4$-$1.3$^c$ &                           & G359.3798$-$01.2008 & 267.216181 & $-30.088645$ & GLIMPSE \\
\enddata
\tablecomments{$^a$ Names used by \citet{HH85}.\\
$^b$ \replaced{OH1.484$-$0.061}{OH001.484$-$00.061} is regarded as OH1.5$-$0.0 (see text for details).\\
$^c$ \replaced{OH359.980$-$1.201 is regarded as OH359.4$-$1.3 (see text for details)}{More precisely, 0.7$\arcsec$ northwest from the GLIMPSE position (see text for details)}.
}
\end{deluxetable*}

The targets are selected from the list of OH/IR stars monitored by \citet{HH85}. They measured variability amplitude of \replaced{OH-maser}{OH maser} emission in radio magnitude, which is defined as the logarithm of the peak maser flux density. We select 16 OH/IR stars with the variability amplitudes less than 0.3 as non-variable OH/IR stars. They are listed in Table \ref{tab:target}.

At first, we determine their precise coordinates by finding bright infrared objects in images taken with the {\it Spitzer Space Telescope} \citep[{\it SST};][]{2004ApJS..154....1W}. OH/IR stars are sometimes invisible even in the NIR because of heavy obscuration by the circumstellar dust but are very bright in the thermal infrared (TIR) beyond the NIR. Therefore, high-resolution TIR images are useful to determine the precise coordinates of OH/IR stars \citep[e.g.][]{2007ApJ...664.1130D, 2007apn4.confE..52E}. We use TIR images of the target non-variable OH/IR stars taken with the Infrared Array Camera \citep[IRAC;][]{2004ApJS..154...10F} on-board the {\it SST}. The {\it SST} data are now easily retrievable from the Spitzer Enhanced Imaging Products (SEIP) database on the NASA/IPAC Infrared Science Archive (IRSA)\footnote{https://irsa.ipac.caltech.edu/frontpage/}. Since all of the targets are found near the galactic plane, the IRAC images and source list made available by the Galactic Legacy Infrared Mid-Plane Survey Extraordinaire (GLIMPSE) project are useful for the target identification. We look for infrared bright objects in the IRAC images in the SEIP Cryogenic Release v3.0 (CR3) and the GLIMPSE catalog \citep{2009yCat.2293....0S} using the IRSA viewer service\footnote{https://irsa.ipac.caltech.edu/irsaviewer/} on the IRSA website.

The target positions are initially obtained from the SIMBAD service \citep{2000A&AS..143....9W} \footnote{http://simbad.u-strasbg.fr/simbad/} by querying with the object names used by \citet{HH85}. Then, we look for a source around the SIMBAD position of the target non-variable OH/IR stars in IRAC images by using the IRSA viewer. At the same time, the 2MASS images \citep{2006AJ....131.1163S}, 2MASS point source catalog \citep[PSC;][]{2003yCat.2246....0C}, and GLIMPSE catalog \citep{2009yCat.2293....0S} are also examined in the IRSA viewer. Infrared counterparts are identified by finding red bright objects in these data. 

\begin{figure*}
\gridline{\rotatefig{0}{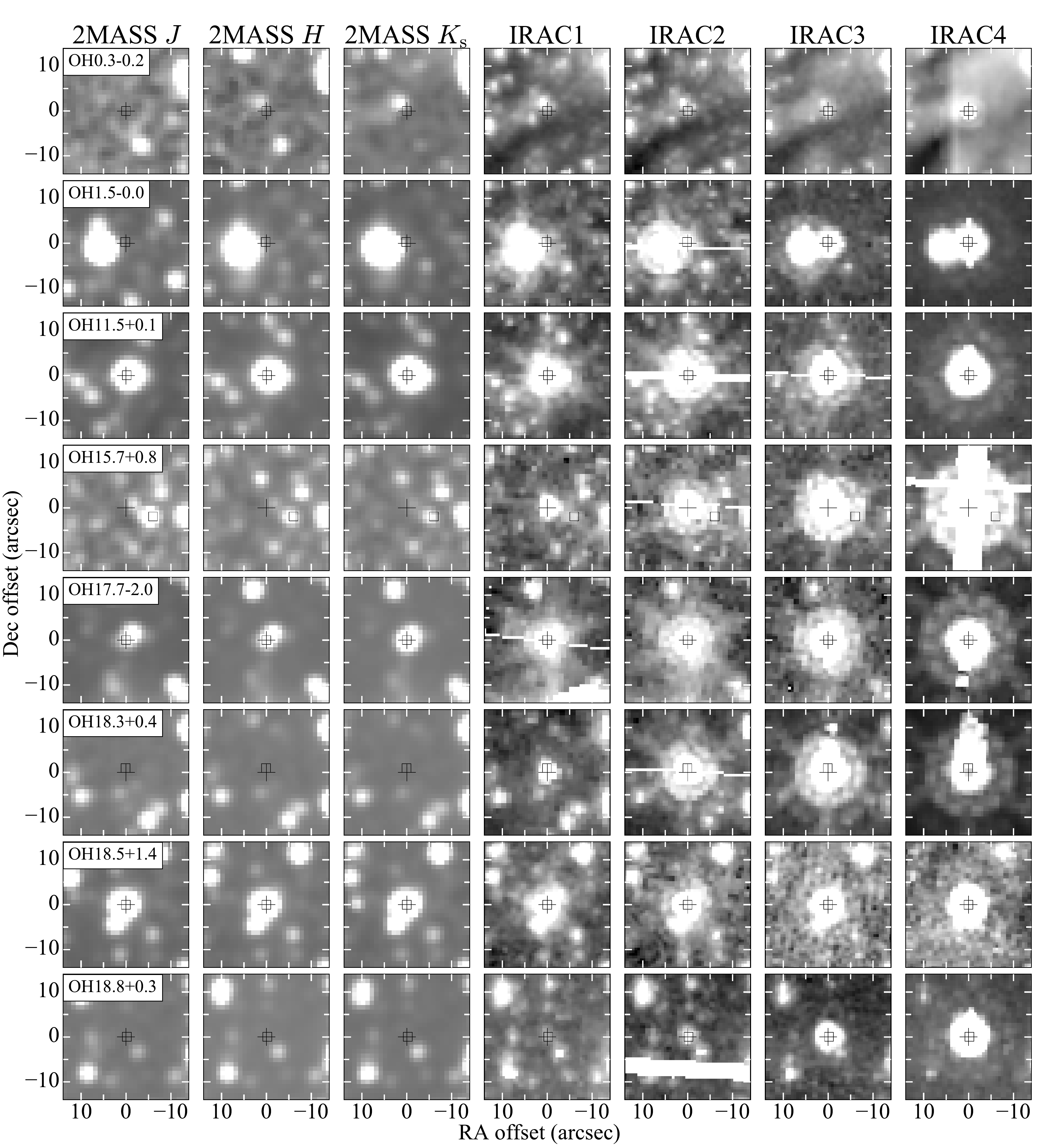}{1.0\textwidth}{}}
\vspace{-0.7cm}
\caption{2MASS and IRAC images of the identified objects. These images are retrieved from the IRSA viewer service. The images are shown in logarithmic scale. The grayscale range is adjusted for better visibility. The positions written in Table \ref{tab:target} are shown with cross marks in center. Horizontal and vertical axes show the RA and Dec offsets from those positions, respectively. The initially checked positions based on SIMBAD are shown with open squares. For OH1.5$-$0.0, the square mark shows the position of \replaced{OH1.484$-$0.061}{OH001.484$-$00.061} listed by \citet{1997A&AS..122...79S} (see text for details). \label{fig:id-a}}
\end{figure*}

\addtocounter{figure}{-1}
\begin{figure*}
\gridline{\rotatefig{0}{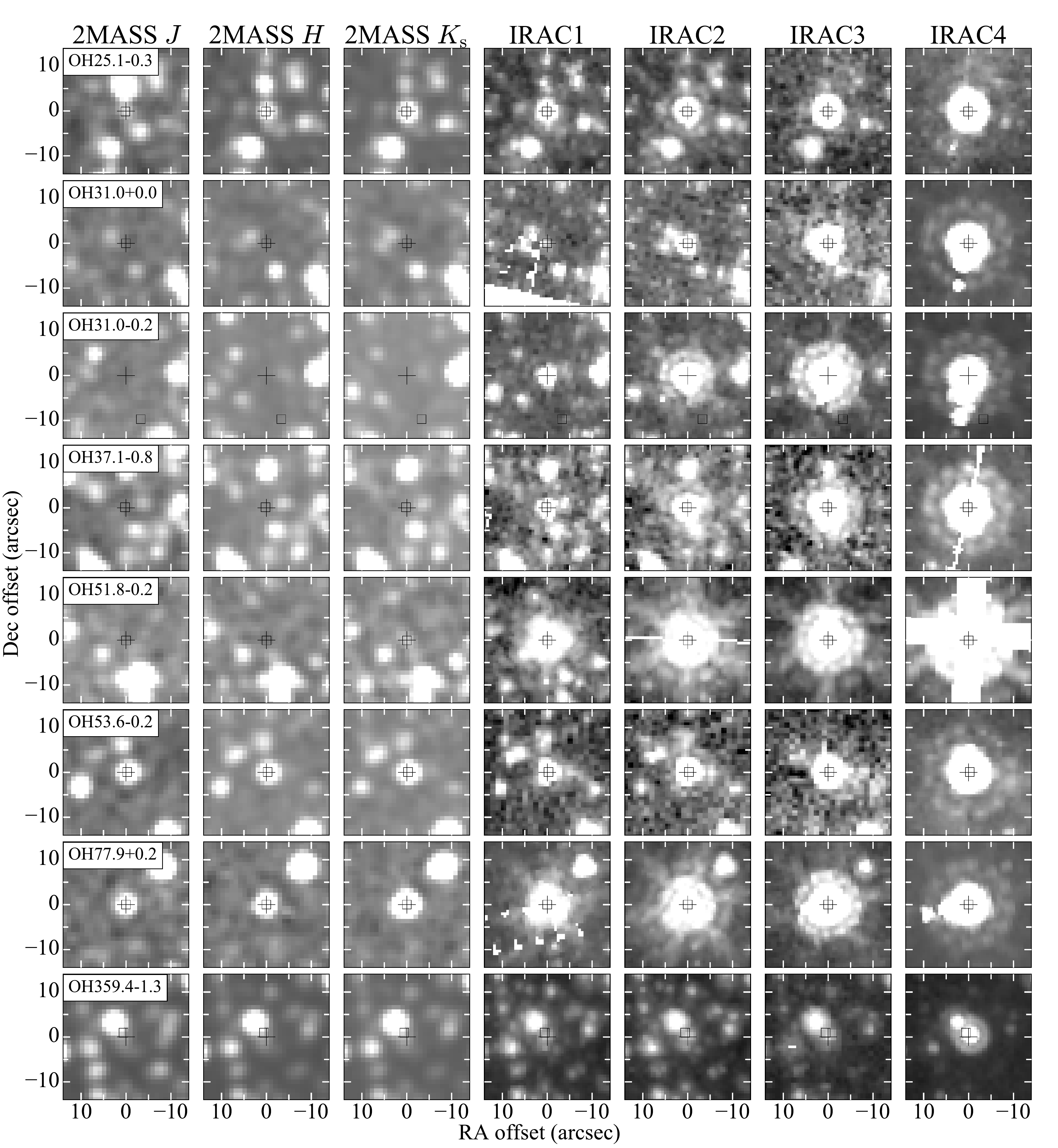}{1.0\textwidth}{}}
\vspace{-0.7cm}
\caption{\explain{Modified grayscale range of the images of OH359.4$-$1.3. The initial position marked with open square for OH359.4$-$1.3 was changed.}(Cont'd) 2MASS and IRAC images of the identified objects. \deleted{The square mark for OH359.4$-$1.3 shows the position of OH359.380$-$1.201 listed by \citet{1997A&AS..122...79S} (see text for details).} 
\label{fig:id-b}}
\end{figure*}

The results are shown in Table \ref{tab:target} and Figure~\ref{fig:id-a}. \replaced{Except for the following four objects, OH1.5$-$0.0, OH15.7+0.8, OH31.0$-$0.2, and OH359.4$-$1.3,}{Except for the following three objects, OH1.5$-$0.0, OH15.7+0.8, and OH31.0$-$0.2,} infrared counterparts are found close to the positions given by SIMBAD. The results are summarized as follows.

\paragraph{OH0.3$-$0.2}
A bright object is clearly found in the IRAC \replaced{band-4}{band 4} image (8.0 $\mu$m) at the SIMBAD position, while in the other images, another object is found in the northeast direction from the SIMBAD position. The former object becomes fainter in the shorter wavelengths, and the latter object gets brighter than the former object in the shorter wavelengths. We identify the former object, SSTGLMC G000.3335$-$00.1805 in the GLIMPSE catalog, as OH0.3$-$0.2 because of the redness and brightness in the TIR, and the positional agreement with SIMBAD. 

OH0.3$-$0.2 is also listed as OH0.335$-$0.180 in the SIMBAD database. The NIR spectrum of OH$0.335-0.180$ is reported by \citet{2006A&A...455..645V}. However, according to the observed position in their log \citep[Table A.3 in][]{2006A&A...455..645V}, the reported object seems to be a different object, 2MASS J17470734$-$2844282, brightly seen at the north edge of the 2MASS {\it H}-band image in Figure~\ref{fig:id-a}.

\paragraph{OH1.5$-$0.0 (OH001.484$-$00.061)}
\explain{paragraph title: replaced OH1.484$-$0.061 with OH001.484$-$00.061.}

Since this object is recorded without coordinates in the SIMBAD database, we searched for an OH/IR star around $(l, b) = (1.^{\circ}5, -0.^{\circ}0)$ in the \replaced{OH-maser}{OH maser} source catalog created by \citet{1997A&AS..122...79S} and found \replaced{OH1.484$-$0.061}{OH001.484$-$00.061} as the only object around the position. Therefore, we regard \replaced{OH1.484$-$0.061}{OH001.484$-$00.061} as OH1.5$-$0.0 and adopt the position of \replaced{OH1.484$-$0.061}{OH001.484$-$00.061} as the initial position for OH1.5$-$0.0. A bright infrared object, SSTGLMC G001.4837$-$00.0612, is found at this position, while the 2MASS images only show another object named 2MASS J17492131$-$2741555 in the east direction. This suggests that SSTGLMC G001.4837$-$00.0612 has very red color in the NIR. Based on the positional agreement, red color, and TIR brightness, we identify SSTGLMC G001.4837$-$00.0612 as OH1.5$-$0.0. 

\paragraph{OH11.5+0.1}
A very bright infrared object is found in both 2MASS and IRAC images at the SIMBAD position. It is 2MASS J18103867$-$1852583 and SSTGLMC G011.5599+00.0873 in the 2MASS PSC and the GLIMPSE catalog, respectively. We identify this object as OH11.5+0.1.

\paragraph{OH15.7+0.8}
There is an object at the SIMBAD position clearly seen in the 2MASS images. However, there are much brighter objects in the vicinity of the SIMBAD position in the IRAC images. This object at the SIMBAD position of OH15.7+0.8 is therefore rather blue, and hence, is unlikely to be an OH/IR star.

The reddest object in the vicinity of the SIMBAD position is thus the object at 6$\arcsec$ east of the SIMBAD position, which we consider to be OH15.7+0.8. This object is cataloged as SSTGLMC G015.7010+00.7705 in the GLIMPSE catalog.

IRAS18135$-$1456 is listed as an alternative name of OH15.7+0.8 in SIMBAD. Its NIR spectrum is reported by \citet{1995A&A...299...69O}. However, the target coordinate listed in their Table 1 corresponds to the position of a different object, 2MASS J18162412$-$1455138, seen at 18$\arcsec$ west from the SIMBAD position (out of the image in Figure~\ref{fig:id-a}). Therefore, the spectrum reported by \citet{1995A&A...299...69O} is probably not of OH15.7+0.8.

\paragraph{OH17.7$-$2.0}
A pair of bright objects is seen in the 2MASS images at the position of OH17.7$-$2.0 given by SIMBAD. SIMBAD suggests that OH17.7$-$2.0 is the southeast object of the pair, which corresponds to 2MASS J18303070$-$1428570. A bright object is found at the same position in the IRAC images retrieved from SEIP, and not listed in the GLIMPSE catalog. Therefore, we identify 2MASS J18303070$-$1428570 as OH17.7$-$2.0 and use the 2MASS position as the target position.

\paragraph{OH18.3+0.4}
At the SIMBAD position, a bright object is seen only in the IRAC images. This is SSTGLMC G018.2956+00.4291 in the GLIMPSE catalog. Based on the absence of the object in the 2MASS images, this object is very red, which is a usual characteristic of OH/IR stars. Therefore, we identify this object as OH18.3+0.4.

\paragraph{OH18.5+1.4}
One bright object is seen in both 2MASS and IRAC images. This is listed as 2MASS J18193549$-$1208081 and SSTGLMC G018.5182+01.4129 in the 2MASS PSC and the GLIMPSE catalog, respectively. We identify this object as OH18.5+1.4.

\paragraph{OH18.8+0.3}
In the IRAC images, a bright red object is found at the SIMBAD position, while no object is seen in the 2MASS images. We identify this red object, SSTGLMC G018.7687+00.3016 in the GLIMPSE catalog, as OH18.8+0.3.

\begin{figure*}[t!]
\gridline{\rotatefig{0}{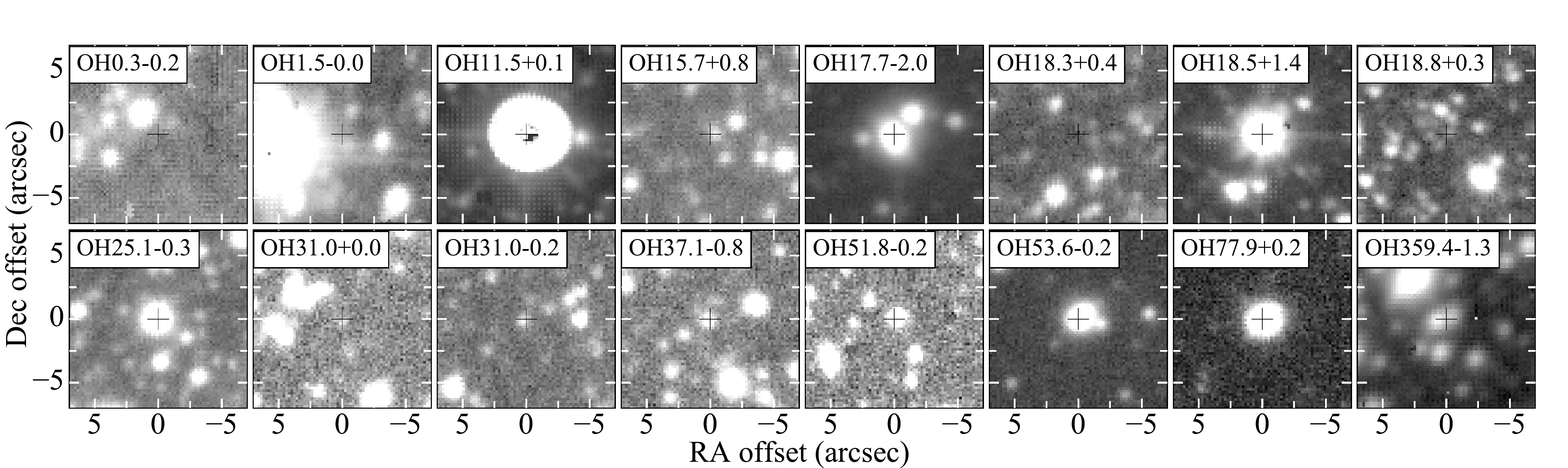}{1.0\textwidth}{}}
\vspace{-0.7cm}
\caption{\explain{Modified grayscale range. Changed center mark from circle to cross.}UKIDSS {\it K}-band images at the target positions. The images are shown in logarithmic scale. The grayscale range is adjusted for better visibility. \replaced{Center circles}{Cross marks} show the target positions as in Table \ref{tab:target}. No object is found for OH0.3$-$0.2, OH1.5$-$0.0, and OH18.3+0.4. OH18.8+0.3 and OH359.4$-$1.3 are \replaced{found}{detected} in the images, but they are not listed in the UKIDSS-GPS catalog. Other objects are found in these images and UKIDSS-GPS catalog. \label{fig:UKIDSSimgs}}
\end{figure*}

\paragraph{OH25.1$-$0.3}
A bright red object is found in both the 2MASS and IRAC images at the SIMBAD position. This object is 2MASS J18381546$-$0709543 and SSTGLMC G025.0568$-$00.3505 in the 2MASS PSC and the GLIMPSE catalog, respectively. This object is thought to be OH25.1$-$0.3.

\paragraph{OH31.0+0.0}
A bright red object is found at the SIMBAD position in the IRAC images, while no object is found in the 2MASS images. The object is cataloged as SSTGLMC G030.9439+00.0351 in the GLIMPSE catalog. We identify this object as OH31.0+0.0.

\paragraph{OH31.0$-$0.2}
No object is found at the SIMBAD position of OH31.0$-$0.2 in both 2MASS and IRAC images. However, we find a bright red object at 10$\arcsec$ northeast of the SIMBAD position in the IRAC images. We identify this object, listed as SSTGLMC G031.0123$-$00.2193 in the GLIMPSE catalog, as OH31.0$-$0.2.

\paragraph{OH37.1$-$0.8}
A bright red TIR object is seen in the IRAC images at the SIMBAD position. This object is listed as SSTGLMC G037.1184$-$00.8473 in the GLIMPSE catalog. We identify this object as OH37.1$-$0.8 based on the redness and brightness in the TIR. In the 2MASS images, this object is not seen, while a different object is seen in the west. This is 2MASS J19020603+0320170. We note that this 2MASS object is not OH37.1$-$0.8.

\paragraph{OH51.8$-$0.2}
A bright red TIR object is found in the IRAC images. This is SSTGLMC G051.8038$-$00.2248 in the GLIMPSE catalog. This object is faintly seen in the 2MASS {\it K}$_{\rm S}$-band image and listed as 2MASS J19274203+1637239 in the 2MASS PSC. We identify this object as OH51.8$-$0.2.

\paragraph{OH53.6$-$0.2}
One bright object is seen in all the 2MASS and IRAC images. This is 2MASS J$19312537+1813103$ and SSTGLMC G053.6303$-$00.2392 in the 2MASS PSC and the GLIMPSE catalog, respectively. We identify this object as OH53.6$-$0.2.

\paragraph{OH77.9+0.2}
One bright red object is seen in all the 2MASS and IRAC images. This object is listed in the 2MASS catalog as 2MASS J20283067+3907009 but not listed in the GLIMPSE catalog. Therefore, we use the position of 2MASS J20283067+3907009 as the target position. 

\paragraph{OH359.4$-$1.3}
\explain{Paragraph title was changed.}

\replaced{We do not find any object at the SIMBAD position in the IRAC images. However, according to the OH/IR star catalog given by \citet{1997A&AS..122...79S}, OH359.380$-$1.201 is the only object around $(l, b) = (359.^{\circ}4, -1.^{\circ}3)$, and we find a bright TIR object at that position in the IRAC images. This object is listed as SSTGLMC 359.3798$-$01.2008 in the GLIMPSE catalog. Therefore, we regard SSTGLMC 359.3798$-$01.2008 as OH359.4$-$1.3.}{A bright TIR source is found in the IRAC band 4 image at 0.7$\arcsec$ northwest of the GLIMPSE catalog position of SSTGLMC G359.3798$-$01.2008 (Table~\ref{tab:target}, Figure~\ref{fig:id-b}). This TIR source coincides well with OH359.380$-$01.201, the only OH maser source in the vicinity \citep{1997A&AS..122...79S}. The positional offset between this TIR source and the OH maser source is 0.3$\arcsec$, within the positional accuracy of the respective data (0.3$\arcsec$; \citet{SEIP}: 0.5$\arcsec$; \citet{1997A&AS..122...79S}). Hence, we regard this TIR object as the OH maser source. The UKIDSS {\it K}-band image at a higher spatial resolution shows two marginally resolved objects, the brighter one at 0.6$\arcsec$ east and the fainter one at $\sim$0.6$\arcsec$ northwest of the SSTGLMC G359.3798$-$01.2008 catalog position (Figure~\ref{fig:UKIDSSimgs}; also see the next section). Therefore, we consider the fainter of the two NIR objects in the UKIDSS {\it K}-band image as the TIR/OH maser source, i.e. OH359.4$-$1.3.}


\section{Archival Data Analysis} \label{sec:variability}
The long-term brightness evolution of the non-variable OH/IR stars is examined with the 2MASS PSC, UKIRT Infrared Deep Sky Survey (UKIDSS) data, and data taken with the Okayama Astrophysical Observatory Wide Field Camera \citep[OAOWFC;][]{2019PASJ..tmp..120Y}. Three stars in our sample, OH0.3$-$0.2, OH1.5$-$0.0, and OH359.4$-$1.3, are included in the observation area of the Vista Variables in the V\'{i}a L\'{a}ctea (VVV) survey \citep{2010NewA...15..433M}. However, we do not find their \replaced{counterparts}{photometric data} in the VVV Data Release 4 (DR4). Therefore, we use only 2MASS, UKIDSS, and OAOWFC data in this study.

As shown in Table \ref{tab:target}, seven objects are listed in the 2MASS catalog. Their NIR magnitudes from 1997 to 2001 can be investigated by using the 2MASS catalog. However, the {\it J}-, {\it H}-, and {\it K}$_{\rm S}$-band magnitudes of OH17.7$-$2.0 are not reliable because they have a quality flag of {\it U} or {\it E}. OH51.8$-$0.2 has a reliable {\it K}$_{\rm S}$-band magnitude with a quality flag of {\it B}, while its {\it J}- and {\it H}-band magnitudes have a quality flag of {\it U} and are unreliable. Other objects have reliable {\it J}-, {\it H}-, and {\it K}$_{\rm S}$-band magnitudes with a quality flag of {\it A} or {\it B}. The 2MASS catalog gives the total photometric error including not only measurement errors but also systematic errors such as calibration and flat errors as {\it [jhk]\_msigcom} \citep{2MASS.Suppl.}. We employ this {\it [jhk]\_msigcom} as the photometric error. We can also get precise observation dates from the catalog.

The UKIDSS project is described by \citet{2007MNRAS.379.1599L}. UKIDSS uses the UKIRT Wide Field Camera \citep[WFCAM;][]{2007A&A...467..777C} and a photometric system described by \citet{2006MNRAS.367..454H}. The pipeline processing and science archive are described by \citet{2004SPIE.5493..411I} and \citet{2008MNRAS.384..637H}. In the UKIDSS data, the target objects are observed in the Galactic Plane Survey \citep[GPS;][]{2008MNRAS.391..136L}. In the GPS program, the north galactic plane ($|b|<5^{\circ}$ in $l=$ $15^{\circ}$--$107^{\circ}$, $142^{\circ}$--$230^{\circ}$; $|b|<2^{\circ}$ in $l=-2^{\circ}$--$+15^{\circ}$) was surveyed by the UKIRT 3.8-m telescope once in the {\it J} and {\it H} bands and twice in the {\it K} band from 2005 to 2012. We use the source catalog and images opened in the 11th data release (DR11PLUS), which is available from the WFCAM Science Archive \citep[WSA;][]{2008MNRAS.384..637H} website\footnote{http://wsa.roe.ac.uk/index.html}. 

\begin{figure}
\epsscale{1.1}
\plotone{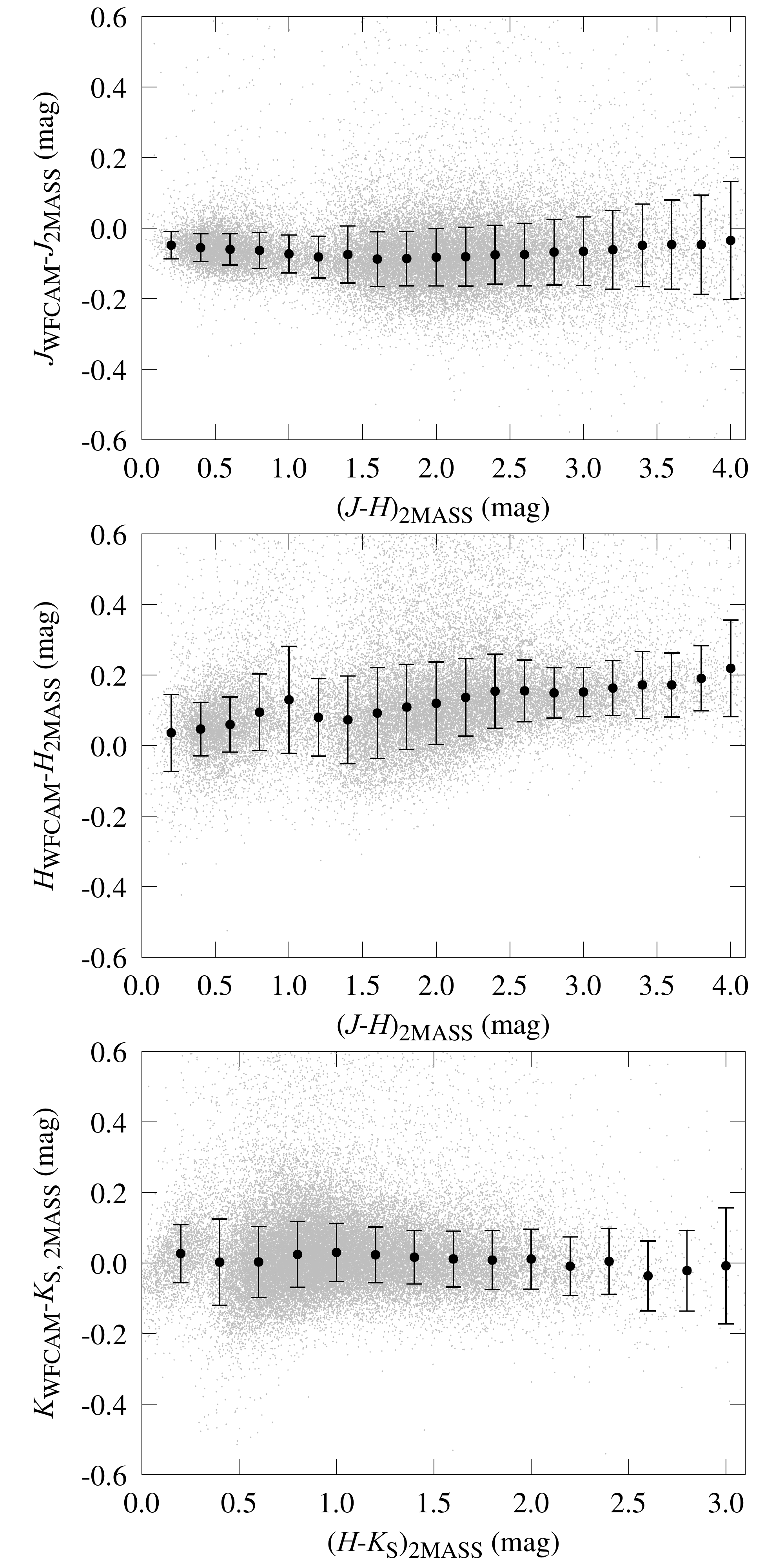}
\caption{Color dependence of the magnitude difference between WFCAM and 2MASS systems in the {\it J} (top), {\it H} (middle), and {\it K} (bottom) bands. Gray dots are the data retrieved from the UKIDSS-GPS and 2MASS catalogs. Black points and bars are the averages and standard deviations calculated in each 0.2-mag color bin with 3$\sigma$ clipping.}
\label{fig:sysdiff}
\end{figure}

Figure~\ref{fig:UKIDSSimgs} shows the UKIDSS images. Among the 16 targets, 13 objects except for OH0.3$-$0.2, OH1.5$-$0.0, and OH18.3+0.4 are found in the UKIDSS images. OH18.8+0.3 and OH359.4$-$1.3 are \replaced{found}{detected} in the UKIDSS images but are not listed in the GPS catalog. Therefore, only 11 objects have photometric data in the GPS catalog. Three of 11 objects, OH11.5+0.1, OH17.7$-$2.0, and OH18.5+1.4, have unreliable magnitudes because of saturation. The other eight objects have reliable magnitudes with error bits, {\it ppErrBits}, less than 256, which is a criteria to select good quality data same as used by \citet{2008MNRAS.384..637H}. Various photometric magnitudes measured with different sizes of aperture photometry are listed in the database. Among the magnitudes, {\it AperMag3}, the photometric magnitude measured with an aperture diameter of 2$\arcsec$, is the default magnitude in the catalog \citep{2008MNRAS.391..136L}. We adopt this {\it AperMag3} magnitude in this study. Uncertainties are also available in the database as {\it AperMag3Err}. However, according to \citet{2008MNRAS.391..136L}, this error is usually underestimated because it does not take into account systematic calibration errors. Such systematic errors are evaluated by \citet{2009MNRAS.394..675H}. They evaluated the repeatability of photometric measurements and concluded that an accuracy of 1--3\% could be achieved by applying appropriate corrections even under thin clouds. Therefore, in this study, we estimate the total photometric error by calculating the square root of the sum of squares (SRSS) of the measurement error ({\it AperMag3Err}) and 0.03-mag calibration error. The observation dates are obtained from the header of the FITS data for each target retrieved from the WSA website.

The OAOWFC is an NIR camera developed by modifying the 0.91-m telescope at the Okayama Astrophysical Observatory (OAO). The OAOWFC has been dedicated to a galactic plane monitoring survey in the {\it K}$_{\rm S}$ band since 2015. The survey region is $|b|<1^{\circ}$ in $l=$ $21^{\circ}$--$36^{\circ}$ and $75^{\circ}$--$85^{\circ}$. Two of 16 targets, OH25.1$-$0.3 and OH77.9+0.2, are found in the survey data. After basic reduction steps such as dark subtraction, flat correction, and World Coordinate System (WCS) matching, instrumental magnitudes are measured by the Source Extractor \citep[SExtractor;][]{1996A&AS..117..393B} and calibrated by comparing the 2MASS {\it K}$_{\rm S}$-band magnitudes.

In order to compare the magnitudes obtained by different instruments, system-conversion process must be applied. For the UKIDSS-GPS data, the system conversion was discussed by \citet{2009MNRAS.394..675H}. However, they investigated the system-conversion equation only in a blue region, where the 2MASS $(J-K_{\rm S})$ color, $(J-K_{\rm S})_{\rm 2MASS}$, is 0--1 mag. To deal with our red targets with $(J-K_{\rm S})_{\rm 2MASS}=$ 1.1--4.2 mag, we define our own system-conversion relation appropriate for red objects by using 2MASS and UKIDSS-GPS data. We cross-match objects in the Galactic center region ($|l|<1^{\circ}$ and $|b|<1^{\circ}$) listed in the 2MASS PSC (with quality flags of {\it A}, {\it B}, and {\it C}) and UKIDSS-GPS catalog (with magnitudes greater than 10 mag to avoid saturation) by using the CDS X-Match service\footnote{http://cdsxmatch.u-strasbg.fr/} with a search radius of 0.1$\arcsec$. 

The results of the cross-matching are shown in Figure~\ref{fig:sysdiff} as plots of the difference of the {\it JHK} magnitude between the two systems as a function of the corresponding 2MASS color. The gray dots are the data taken from the UKIDSS-GPS and 2MASS catalogs, and black points and bars show the averages and standard deviations calculated in each 0.2-mag color bin with 3$\sigma$ clipping. The results are converged within 10 iterations of the sigma clipping calculations. The averaged values are used to convert the 2MASS magnitudes to the WFCAM magnitudes. 

\begin{figure}
\epsscale{1.1}
\plotone{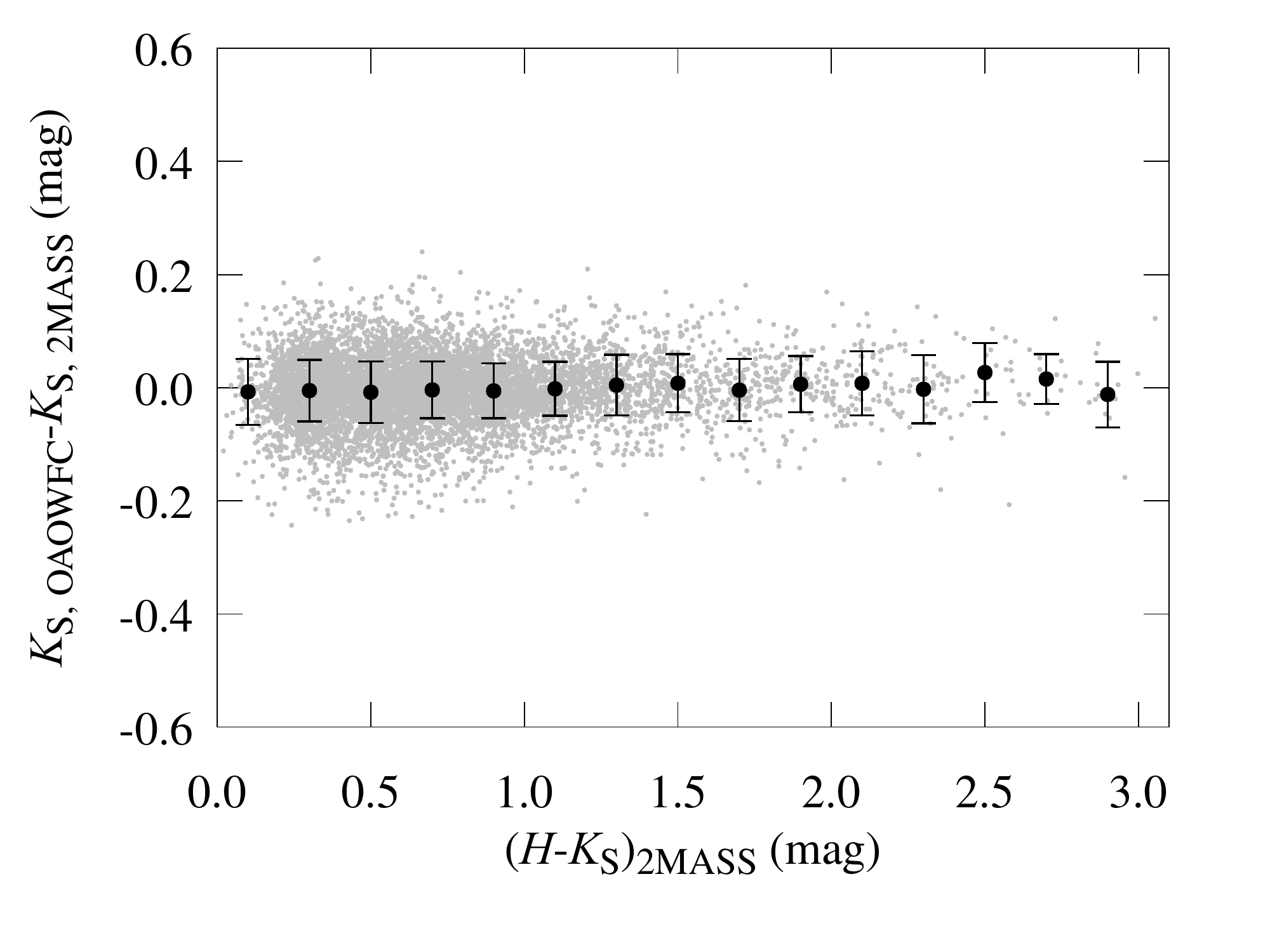}
\vspace{-0.6cm}
\caption{Same as Figure~\ref{fig:sysdiff} but for the {\it K}$_{\rm S}$-band magnitude difference between OAOWFC and 2MASS systems.}
\label{fig:sysdiffOAOWFC}
\end{figure}

As for the OAOWFC data, \citet{2019PASJ..tmp..120Y} give a system-conversion equation between the OAOWFC and 2MASS systems. However, the two objects observed by OAOWFC have redder colors, $(H-K_{\rm S})_{\rm 2MASS}\sim$1.7 mag, than the objects used for the derivation of their system-conversion equation, typically $(H-K_{\rm S})_{\rm 2MASS}\lesssim$1.2 mag. Therefore, we derive a new system-conversion relation which covers redder color regions. To cover redder regions, data taken in three fields in the galactic plane are used, because red objects obscured by interstellar dust exist in the galactic plane. After the basic calibrations, systematic {\it K}$_{\rm S}$-band magnitude difference against $(H-K_{\rm S})_{\rm 2MASS}$ is examined. In this process, data with large photometric error ($>$0.05 mag) are removed. Data which have significant magnitude differences between 2MASS and OAOWFC data (larger than five times of the photometric error) are also removed as unreliable data. The result is shown in Figure~\ref{fig:sysdiffOAOWFC}. The derived relation shows that the difference between 2MASS and OAOWFC systems is very small as shown by \citet{2019PASJ..tmp..120Y}. The OAOWFC magnitudes are corrected to WFCAM magnitudes by using both this relation and the relation between the WFCAM and 2MASS systems derived above.

\section{Results} \label{sec:res}
\begin{deluxetable*}{cccccc}
\tablecaption{
Multi-epoch data obtained from 2MASS, UKIDSS, and OAOWFC survey data.
\label{tab:res}}
\tablehead{
\colhead{Object} & 
\colhead{Band} & 
\colhead{Julian} & 
\colhead{Magnitude$^{\rm a}$} & 
\colhead{Data} &
\colhead{Rate of change$^{\rm b}$}\\
\colhead{name} &
\colhead{name} &
\colhead{date} &
\colhead{(mag)} &
\colhead{source} &
\colhead{(mag yr$^{-1}$)}
}
\startdata
OH15.7$+$0.8  
& \hspace{0.00cm}{\it K} & \hspace{0.00cm}2453887.0115 & \hspace{0.00cm}$15.644\pm0.104$ & \hspace{0.00cm}UKIDSS & --- \\
&                      & \hspace{0.00cm}2456033.0793 & \hspace{0.00cm}$14.880\pm0.034$ & \hspace{0.00cm}UKIDSS & $-0.130\pm0.019$\\
OH25.1$-$0.3  
& \hspace{0.00cm}{\it J} & \hspace{0.00cm}2451363.6290 & \hspace{0.00cm}$15.784\pm0.107$ & \hspace{0.00cm}2MASS & --- \\
&                      & \hspace{0.00cm}2453892.9832 & \hspace{0.00cm}$15.641\pm0.030$ & \hspace{0.00cm}UKIDSS& $-0.021\pm0.016$\\\
& \hspace{0.00cm}{\it H} & \hspace{0.00cm}2451363.6290 & \hspace{0.00cm}$13.270\pm0.044$ & \hspace{0.00cm}2MASS & --- \\
&                      & \hspace{0.00cm}2453892.9890 & \hspace{0.00cm}$13.202\pm0.030$ & \hspace{0.00cm}UKIDSS& $-0.010\pm0.008$\\
& \hspace{0.00cm}{\it K} & \hspace{0.00cm}2451363.6290 & \hspace{0.00cm}$11.539\pm0.035$ & \hspace{0.00cm}2MASS & --- \\
&                      & \hspace{0.00cm}2453892.9949 & \hspace{0.00cm}$11.408\pm0.030$ & \hspace{0.00cm}UKIDSS& $-0.019\pm0.007$\\
&                      & \hspace{0.00cm}2455828.7327 & \hspace{0.00cm}$11.353\pm0.030$ & \hspace{0.00cm}UKIDSS& $-0.010\pm0.008$\\
&                      & \hspace{0.00cm}2457135.1544--2457510.2033 & \hspace{0.00cm}11.241--11.503 & \hspace{0.00cm}OAOWFC& $-0.117\pm0.026$ \\ 
OH31.0$-$0.2  
& \hspace{0.00cm}{\it K} & \hspace{0.00cm}2453536.0609 & \hspace{0.00cm}$18.522\pm0.384$ & \hspace{0.00cm}UKIDSS& --- \\
&                      & \hspace{0.00cm}2455786.8618 & \hspace{0.00cm}$16.483\pm0.064$ & \hspace{0.00cm}UKIDSS& $-0.331\pm0.063$\\
OH37.1$-$0.8  
& \hspace{0.00cm}{\it K} & \hspace{0.00cm}2453619.7938 & \hspace{0.00cm}$15.882\pm0.044$ & \hspace{0.00cm}UKIDSS& --- \\
&                      & \hspace{0.00cm}2455784.8148 & \hspace{0.00cm}$15.528\pm0.038$ & \hspace{0.00cm}UKIDSS& $-0.060\pm0.010$\\
OH51.8$-$0.2  
& \hspace{0.00cm}{\it K} & \hspace{0.00cm}2451076.6704 & \hspace{0.00cm}$14.799\pm0.131^{\rm c}$ & \hspace{0.00cm}2MASS & --- \\
&                      & \hspace{0.00cm}2453891.0838 & \hspace{0.00cm}$14.938\pm0.032$ & \hspace{0.00cm}UKIDSS& --- \\
OH53.6$-$0.2  
& \hspace{0.00cm}{\it J} & \hspace{0.00cm}2450615.8619 & \hspace{0.00cm}$14.257\pm0.029$ & \hspace{0.00cm}2MASS & --- \\
&                      & \hspace{0.00cm}2453927.9395 & \hspace{0.00cm}$14.199\pm0.030$ & \hspace{0.00cm}UKIDSS& $-0.006\pm0.005$\\
& \hspace{0.00cm}{\it H} & \hspace{0.00cm}2450615.8619 & \hspace{0.00cm}$12.985\pm0.024$ & \hspace{0.00cm}2MASS & --- \\
&                      & \hspace{0.00cm}2453927.9451 & \hspace{0.00cm}$12.884\pm0.030$ & \hspace{0.00cm}UKIDSS& $-0.011\pm0.004$\\
& \hspace{0.00cm}{\it K} & \hspace{0.00cm}2450615.8619 & \hspace{0.00cm}$11.995\pm0.021$ & \hspace{0.00cm}2MASS & --- \\
&                      & \hspace{0.00cm}2453927.9503 & \hspace{0.00cm}$11.821\pm0.030$ & \hspace{0.00cm}UKIDSS& $-0.019\pm0.004$\\
&                      & \hspace{0.00cm}2455811.7852 & \hspace{0.00cm}$11.771\pm0.030$ & \hspace{0.00cm}UKIDSS& $-0.010\pm0.008$\\
OH77.9$+$0.2  
& \hspace{0.00cm}{\it J} & \hspace{0.00cm}2450985.9161 & \hspace{0.00cm}$15.110\pm0.042$ & \hspace{0.00cm}2MASS & --- \\
&                      & \hspace{0.00cm}2454357.8615 & \hspace{0.00cm}$15.006\pm0.030$ & \hspace{0.00cm}UKIDSS& $-0.011\pm0.006$\\
& \hspace{0.00cm}{\it H} & \hspace{0.00cm}2450985.9161 & \hspace{0.00cm}$13.572\pm0.027$ & \hspace{0.00cm}2MASS & --- \\
&                      & \hspace{0.00cm}2454357.8669 & \hspace{0.00cm}$13.389\pm0.030$ & \hspace{0.00cm}UKIDSS& $-0.020\pm0.004$\\
& \hspace{0.00cm}{\it K} & \hspace{0.00cm}2450985.9161 & \hspace{0.00cm}$11.888\pm0.017$ & \hspace{0.00cm}2MASS & --- \\
&                      & \hspace{0.00cm}2454357.8709 & \hspace{0.00cm}$11.719\pm0.030$ & \hspace{0.00cm}UKIDSS& $-0.018\pm0.004$\\
&                      & \hspace{0.00cm}2455826.8329 & \hspace{0.00cm}$11.662\pm0.030$ & \hspace{0.00cm}UKIDSS& $-0.014\pm0.011$\\
& & \hspace{0.00cm}2457603.2533-- 2457986.1875 & \hspace{0.00cm}11.601--11.756 & OAOWFC & $-0.031\pm0.026$ \\ 
\enddata
\tablecomments{$^a$ Magnitudes converted to WFCAM photometric system. The uncertainty of UKIDSS data is the SRSS of 0.03-mag systematic uncertainty and the photometric error quoted from UKIDSS-GPS catalog. \\
$^b$ For UKIDSS data, the rates are derived from the UKIDSS observation and one before. For OAOWFC data, the rates are derived by linear fitting of the OAOWFC data. \\
$^c$ This magnitude is not converted to the WFCAM photometric system because it does not have $(H-K_{\rm S})_{\rm 2MASS}$ required for system conversion.}
\end{deluxetable*}

\begin{figure*}
\gridline{\fig{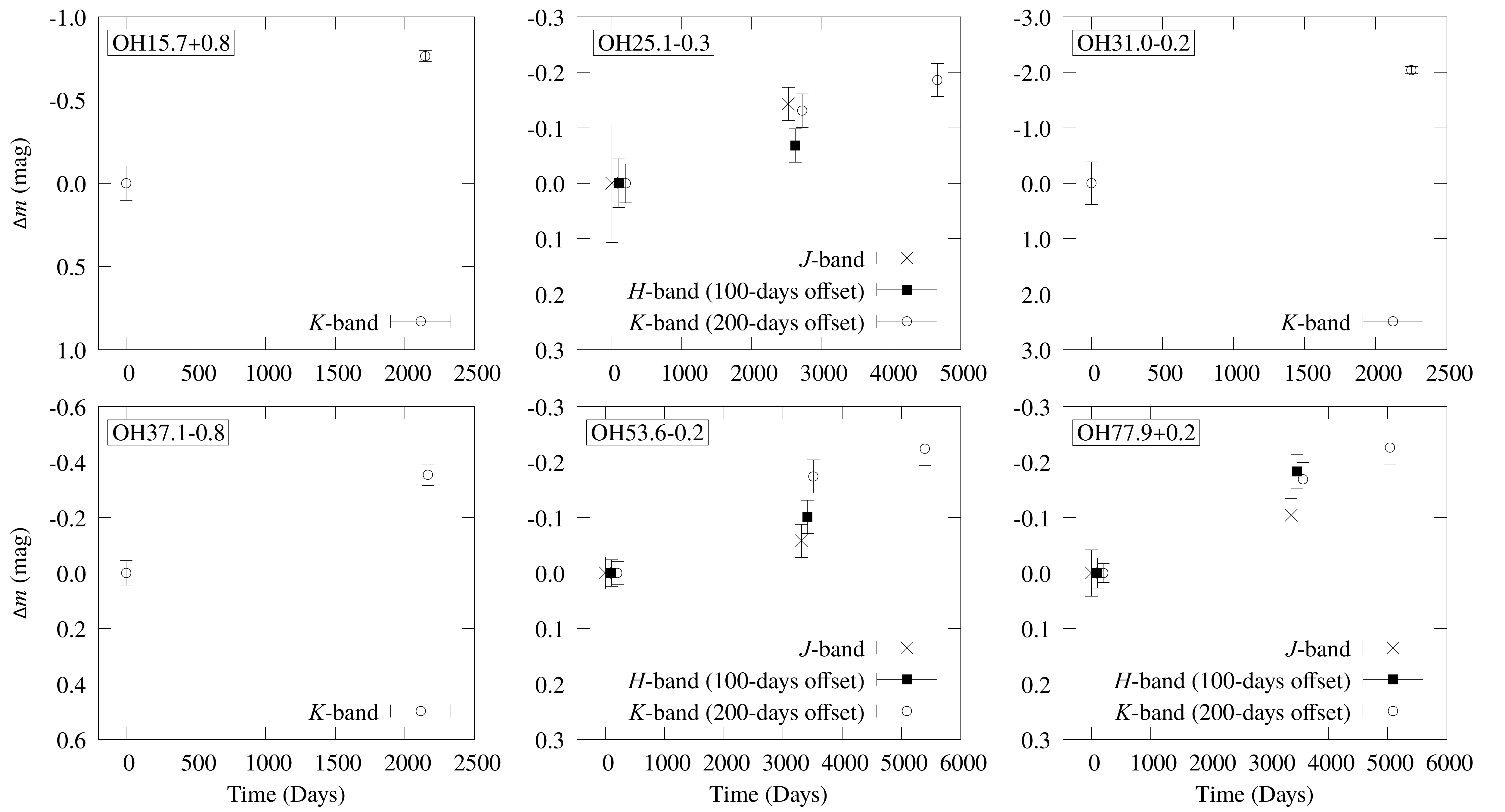}{0.92\textwidth}{}}
\vspace{-0.6cm}
\caption{NIR variability of the non-variable OH/IR stars obtained from 2MASS and UKIDSS data. Object names are shown in the top left corner in each panel. Horizontal axis shows the elapsed time since the first observation. Vertical axis shows the relative magnitude from the first observation. Cross marks, filled squares, and open circles show the {\it J}-, {\it H}-, and {\it K}-band data. Magnitudes are converted to the WFCAM system. For OH25.1$-$0.3, OH53.6$-$0.2, and OH77.9+0.2, the {\it H}- and {\it K}-band data are offsetted for clarity by 100 and 200 days, respectively.
\label{fig:variability}}
\end{figure*}

\begin{figure*}
\gridline{\fig{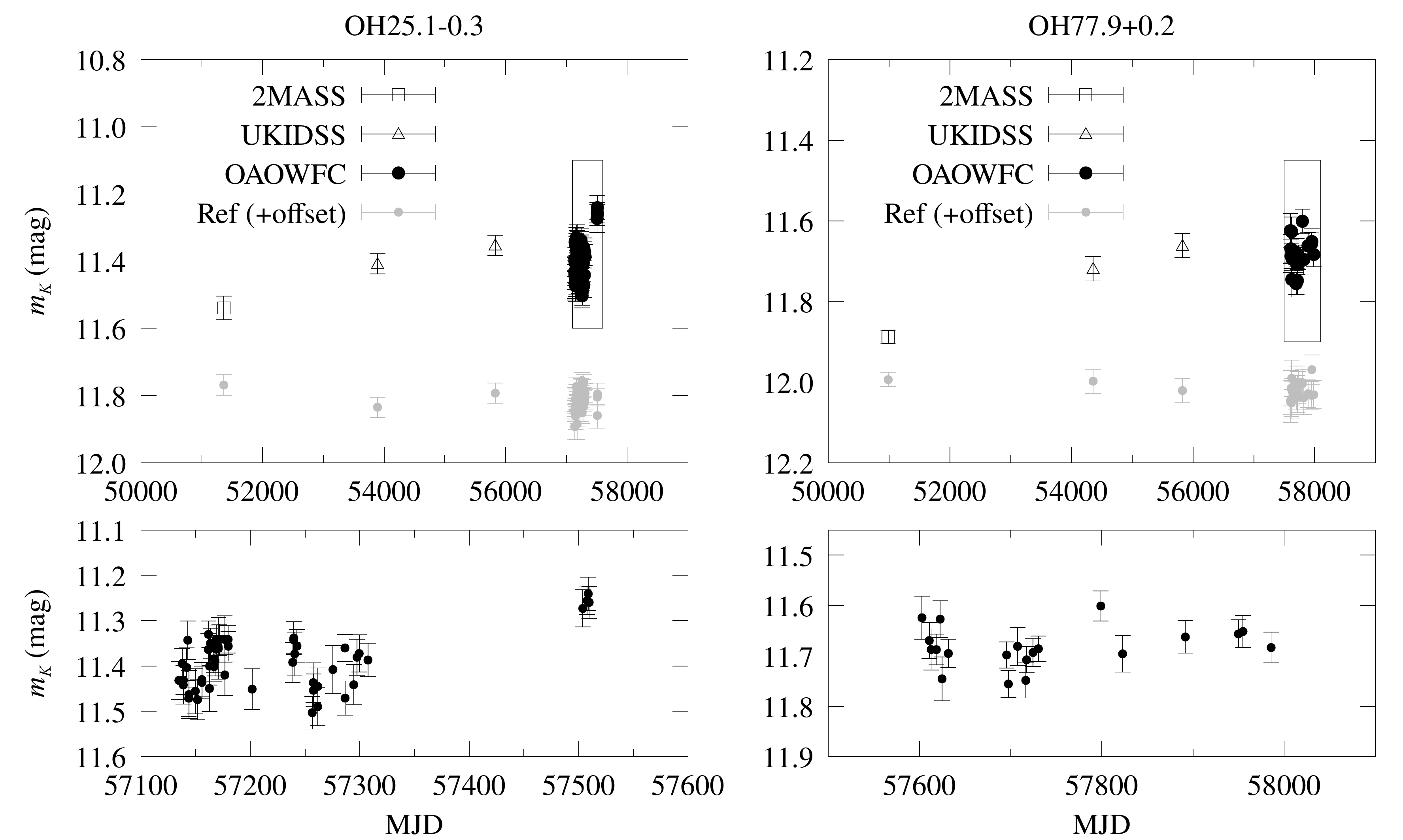}{0.8\textwidth}{}}
\vspace{-.7cm}
\caption{{\it K}-band light curves of OH$25.1-0.3$ (left) and OH$77.9+0.2$ (right) obtained by 2MASS, UKIDSS, and OAOWFC. Open square, open triangle, and filled circle show the 2MASS, UKIDSS, and OAOWFC data, respectively. Gray circles show offsetted magnitudes of nearby reference stars shown for comparison. Bottom panels show the magnified plots of the OAOWFC data. The magnified regions are shown with the squares in the top panels. All magnitudes are converted to the WFCAM system.
\label{fig:K-bandLC}}
\end{figure*}

\begin{figure*}
\gridline{\fig{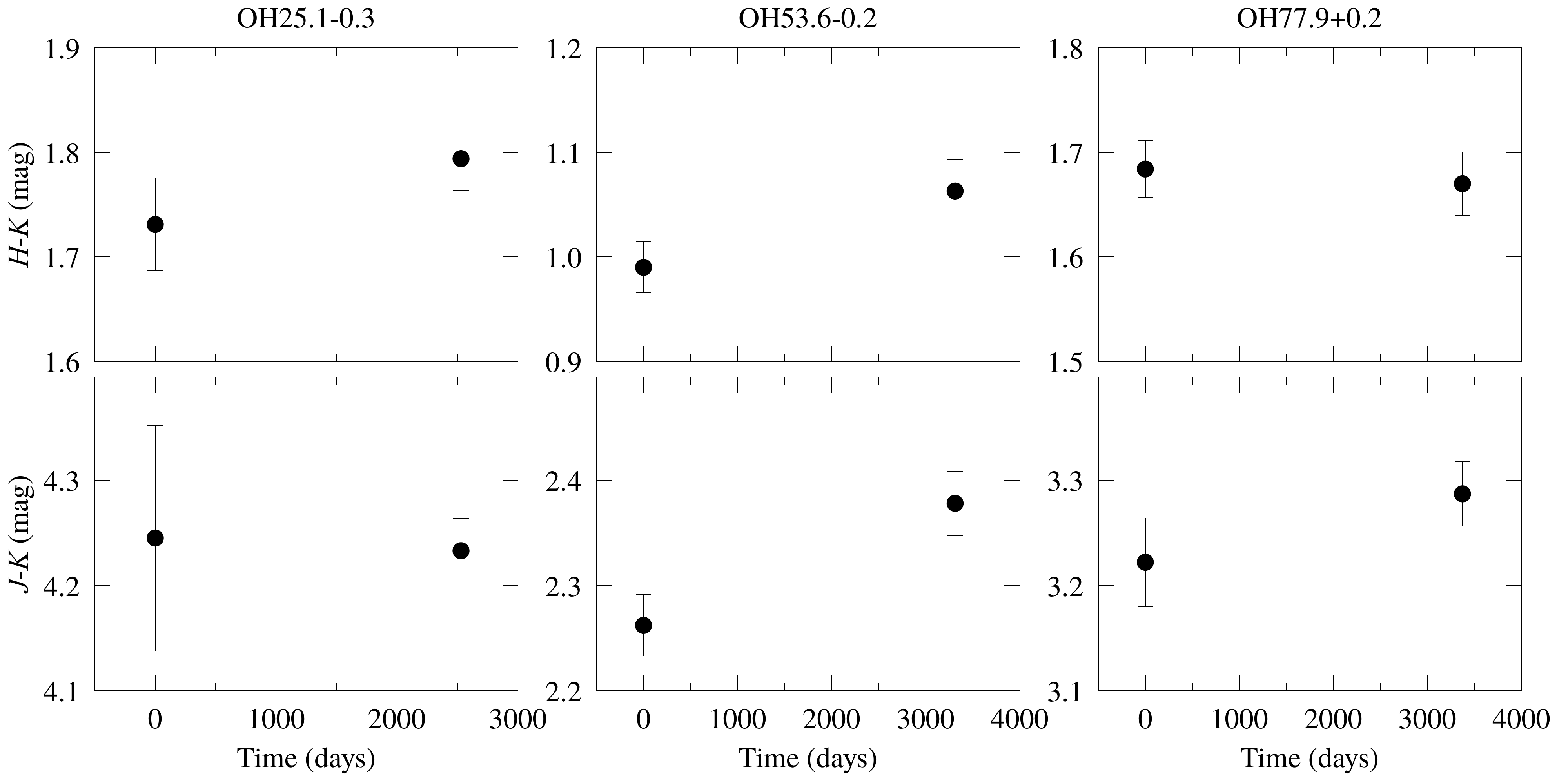}{0.95\textwidth}{}}
\vspace{-.7cm}
\caption{
Color change observed by 2MASS and UKIDSS. OH25.1$-$0.3, OH53.6$-$0.2, and OH77.9+0.2 are shown in the left, middle, and right panels, respectively. Top panels show $(H-K)$ variability, and bottom panels show $(J-K)$ variability. Horizontal axis shows the elapsed time from the first observation by 2MASS. Magnitudes are converted to the WFCAM system.
\label{fig:colorchange}}
\end{figure*}

NIR multi-epoch data are established for seven objects, OH15.7+0.8, OH25.1$-$0.3, OH31.0$-$0.2, OH37.1$-$0.8, OH51.8$-$0.2, OH53.6$-$0.2, and OH77.9+0.2. They are summarized in Table \ref{tab:res}. The table shows the object names, {\it J}-, {\it H}-, and {\it K}-band magnitudes (where appropriate), observation dates, and data sources. All the magnitudes except for the {\it K}-band magnitude of OH51.8$-$0.2 observed by 2MASS are converted to those of the WFCAM system. The {\it K}-band magnitude of OH51.8$-$0.2 cannot be converted to WFCAM magnitude, because it does not have $(H-K_{\rm S})_{\rm 2MASS}$ data required for the system conversion. Therefore, we can correctly examine NIR variability only for six objects except for OH51.8$-$0.2. 

Table \ref{tab:res} also shows the rate of change of the magnitude in the last column. The rate of change for the UKIDSS observations are derived by dividing the magnitude difference with the time interval between the first-epoch 2MASS or UKIDSS observation and the second-epoch UKIDSS observation. The first finding is that the six objects all show negative rates of change, i.e. brightening over a few thousand days. The brightening of these six objects is also clearly seen in Figure~\ref{fig:variability}, which shows the time evolution of the magnitude over multi-epoch observations. In the {\it J} band, the apparent brightening is found in OH25.1$-$0.3, OH53.6$-$0.2, and OH77.9+0.3 with significances of 1--2$\sigma$. In the {\it H} band, the apparent brightening is detected for OH25.1$-$0.3, OH53.6$-$0.2, and OH77.9+0.3 at the 1, 3, and 5$\sigma$ levels, respectively. The {\it K}-band brightening observed between 2MASS and UKIDSS observations is of the $>$$3\sigma$ significance. OH25.1$-$0.3, OH53.6$-$0.2, and OH77.9+0.2 are observed three times in 2MASS and UKIDSS observations, and the apparent brightening found in each of the two consecutive intervals is consistent within the uncertainties.

As for OH51.8$-$0.2, the difference of the {\it K}-band magnitude is +0.139 (i.e. darkening) between the first-epoch 2MASS observation and the second-epoch UKIDSS observation without the system conversion. As shown in Figure~\ref{fig:sysdiff}, the differences between the {\it K}-band magnitudes of 2MASS and UKIDSS observations are small, $-0.037$--$+0.030$ mag in the whole $(H-K_{\rm S})_{\rm 2MASS}$ range. If we evaluate the total uncertainty on the magnitude change as the SRSS of this system-conversion uncertainty and the photometric uncertainties of the 2MASS and UKIDSS observations, that is calculated to be 0.138--0.140 mag. Therefore, it is difficult to conclude if OH51.8$-$0.2 is darkening with a reasonable amount of certainty.

For OAOWFC data, the rate of change in Table \ref{tab:res} are derived by linear fitting of the OAOWFC monitoring data shown in Figure~\ref{fig:K-bandLC}. Figure~\ref{fig:K-bandLC} shows the long-term variations of the {\it K}-band magnitude for the two non-variable OH/IR stars, OH25.1$-$0.3 and OH77.9+0.2 (black plots), together with nearby reference stars (gray plots). Some OAOWFC data whose measured magnitudes sensitively change depending on the photometric aperture size are removed as unreliable data. The reference stars are selected with a criteria that the star is seen in the proximity of the target with a {\it K}-band magnitude close to that of the target. For OH25.1$-$0.3, 2MASS J18381571$-$0710026 is selected as the reference star. This reference star is at 9$\arcsec$ north from OH25.1$-$0.3 and has a {\it K}-band magnitude of $\sim$11.1 mag, which is close to the {\it K}-band magnitude of OH25.1$-$0.3 (11.2--11.6 mag). The reference star for OH77.9+0.2 is 2MASS J20282996+3907096, which exists at 12$\arcsec$ northwest from OH77.9+0.2. The {\it K}-band magnitude ($\sim$11.8 mag) is close to that of OH77.9+0.2 (11.6--11.9 mag). 

For OH25.1$-$0.3, a slight periodic variability with an amplitude of $\sim$0.1 mag and a period of $\sim$50 days may be seen in the OAOWFC light curve. Even if we take into account this possible variability by setting the magnitude errors to be 0.2 mag, a linear fitting of all 2MASS, UKIDSS, and OAOWFC data gives a significant negative rate of change of $-0.008\pm0.003$ mag yr$^{-1}$. Therefore, we conclude that OH25.1$-$0.3 brightened since the 2MASS era even with the possible periodic variability. This brightening is also supported by the comparison of light curves of the target and the reference star, which shows that such a brightening is only seen in the target light curve. In the target light curve, only the data around MJD = 57100--57350 are relatively faint compared to the trend seen in the 2MASS and UKIDSS data, while the data around MJD = 56000 seem to be on that trend. The discrepancy between the data around MJD = 57100--57350 and those around MJD = 56000 is the cause of the large rate of change shown for the OAOWFC data in Table \ref{tab:res}. If we assume that the faint magnitudes around MJD = 57100--57350 are caused by calibration problems and evaluate the rate of change since the second UKIDSS observation without these faint data, the rate of change is derived to be $-0.021\pm0.007$ mag yr$^{-1}$. This is consistent with the values derived from 2MASS and UKIDSS observations.

\begin{table}
\centering
\caption{Observed rates of color change.} \label{tab:ccr}
\begin{tabular}{ccc}
\tablewidth{0pt}
\hline
\hline
Object & Color & Rate of change \\
       &       & (mag yr$^{-1}$)\\
\hline
OH25.1$-$0.3 & $H-K$ & $+0.009\pm0.010$ \\
             & $J-K$ & $-0.002\pm0.017$ \\
OH53.6$-$0.2 & $H-K$ & $+0.008\pm0.006$ \\
             & $J-K$ & $+0.013\pm0.006$ \\
OH77.9+0.2   & $H-K$ & $-0.002\pm0.006$ \\
             & $J-K$ & $+0.007\pm0.007$ \\
\hline
\end{tabular}
\end{table}

As for OH77.9+0.2, any periodic variability like OH25.1$-$0.3 is not seen in the OAOWFC light curve. Although the overall magnitudes observed by OAOWFC is slightly fainter than those expected from the 2MASS-UKIDSS trend, the OAOWFC magnitudes are significantly brighter than the 2MASS magnitude. Since the brightening trend is seen only in the target light curve, it can be concluded that OH77.9+0.2 also brightened since the 2MASS era. The rate of change derived from the OAOWFC light curve is consistent with those derived from the 2MASS and UKIDSS observations as in Table \ref{tab:res}. Therefore, the magnitude offset between the 2MASS-UKIDSS trend and OAOWFC light curve may be caused by systematic error (see \S\ref{sec:oaowfcunc}).

As shown in Table \ref{tab:res}, we have multi-color multi-epoch data for three objects, OH25.1$-$0.3, OH53.6$-$0.2, and OH77.9+0.2, enabling us to study their color changes. Figure~\ref{fig:colorchange} shows the time variations of $(H-K)$ and $(J-K)$ colors between the 2MASS and UKIDSS observations. The rates of the color change are summarized in Table \ref{tab:ccr}. Conventionally, we have expected that a non-variable OH/IR star becomes brighter and bluer as its mass-loss rate drops at the end of the AGB evolution. However, contrary to this picture, we find that the three non-variable OH/IR stars kept their colors constant or became redder over a few thousand days. This is the second finding of this study. OH25.1$-$0.3 and OH77.9+0.2 does not show significant color change in both $(H-K)$ and $(J-K)$. On the other hand, the color of OH53.6$-$0.2 in the UKIDSS observation is redder than that in the 2MASS data. The significance levels of the color changes are 1.4 and 2.2$\sigma$ in $(H-K)$ and $(J-K)$, respectively. 

\section{Origin of the Brightness Evolution} \label{sec:disc}
To investigate the origin of the observed long-term brightness evolution, we make a simple model based on the conventional picture and compare it with the observed rates of brightening and color change.

\subsection{A simple model}
In the conventional picture, it is thought that non-variable OH/IR stars have ceased the dust supply to the circumstellar dust shell, and the dust shell is expected to expand its inner cavity. Therefore, we assume a spherical dust shell created by a stellar wind with a constant expansion velocity, $v_{\rm exp}$, and a constant dust mass-loss rate, $\dot{M}_{\rm d}$, and that its inner radius, $r_{\rm in}$, is increasing with time, $t$, after the cessation of dust production.

At a wavelength, $\lambda$, the apparent magnitude of a star with an intrinsic apparent magnitude, $m_{\lambda,{\rm i}}$ is written as,
\begin{equation}
m_\lambda = m_{\lambda,{\rm i}} + A_{\lambda,{\rm c}} + A_{\lambda,{\rm i}},
\label{eq:mag}
\end{equation}
where $A_{\lambda,{\rm c}}$ and $A_{\lambda,{\rm i}}$ are the extinction caused by the circumstellar dust and interstellar dust, respectively. Here, we omit the thermal emission from the circumstellar dust shell, because it is difficult to model without mid- to far-infrared data simultaneously taken with the NIR data. This assumption is thought to be reasonable for a situation where enough time elapsed since the cessation of the dust supply and hot dust grains ($\gtrsim$700 K) disappeared (e.g. $t\gtrsim$10 yr for a star with $L=10^4 {\rm L}_{\sun}$ and $v_{\rm exp}=10$ km s$^{-1}$). If such hot dust grains exist in the dust shell, their thermal emission can contribute to the NIR brightness especially at longer wavelengths. Such a possible contribution of the dust emission will be addressed in \S\ref{sec:beyond}.

$A_{\lambda,{\rm c}}$ is described as, 
\begin{equation}
A_{\lambda,{\rm c}} = -2.5 \log{[\exp{(-\sigma_{\lambda} N)}]}\approx1.09\sigma_{\lambda} N,
\label{eq:ext}
\end{equation}
where $\sigma_{\lambda}$ and $N$ are the absorption cross-section and the column density of dust grains in the dust shell. 

$\dot{M_{\rm d}}$ is related with the number density of dust grains, $n_{\rm d}$, as,
\begin{equation}
\dot{M_{\rm d}} = 4\pi r^2 v_{\rm exp} m_{\rm d} n_{\rm d},
\end{equation}
where $r$ is the distance from the star, and $m_{\rm d}$ is the mass of a single dust particle. 
Since we assume a constant $v_{\rm exp}$ and a constant $\dot{M}_{\rm d}$, $n_{\rm d} r^2$ becomes a constant, $C=\dot{M_{\rm d}}(4\pi v_{\rm exp}m_{\rm d})^{-1}$.

Then, $N$ is written as,
\begin{equation}
N = \int_{r_{\rm in}}^{r_{\rm out}} n_{\rm d}(r) dr = C\left(\frac{1}{r_{\rm in}}-\frac{1}{r_{\rm out}}\right),
\end{equation}
where $r_{\rm in}$ and $r_{\rm out}$ are the inner and outer radii of the dust shell.
If we assume that $r_{\rm out}$ is sufficiently larger than $r_{\rm in}$, $N$ is approximated to be
\begin{equation}
N \approx \frac{C}{r_{\rm in}} = \frac{C}{r_{\rm in,0}+v_{\rm exp}t},
\label{eq:approxN}
\end{equation}
where $r_{\rm in,0}$ is the inner radius of the dust shell at $t$=0.

By substituting Equation~(\ref{eq:approxN}) to Equation~(\ref{eq:ext}) and introducing a crossing-time parameter, $t_{\rm cross}$ = $r_{\rm in,0}/v_{\rm exp}$, we get an expression of $A_{\lambda,{\rm c}}$ as,
\begin{equation}
A_{\lambda,{\rm c}}(t) = \frac{A_{\lambda,{\rm c}}(0)}{1+t/t_{\rm cross}}. \label{eq:ext-last}
\end{equation}

If we assume that the properties of the central star and interstellar extinction (i.e. $m_{\lambda, {\rm i}}$ and $A_{\lambda,{\rm i}}$) are constant, the rate of change of the apparent magnitude is described as, 
\begin{equation}
\frac{dm_\lambda}{dt}\left(t\right)=\frac{dA_{\lambda,{\rm c}}}{dt} =  -\frac{A_{\lambda,{\rm c}}(0)}{t_{\rm cross}\left(1+t/t_{\rm cross}\right)^{2}}. \label{eq:CR}
\end{equation}

Another important observable is the rates of color change. In the model, the color in a wavelength range of $\lambda$--$\lambda'$ is,
\begin{equation}
C_{\lambda\lambda'} = C_{\lambda\lambda',{\rm i}} + E_{\lambda\lambda',{\rm c}} + E_{\lambda\lambda',{\rm i}},
\label{eq:CCEE}
\end{equation}
where $C_{\lambda\lambda',{\rm i}}$ is the intrinsic color of the central star. $E_{\lambda\lambda',{\rm c}}$ and $E_{\lambda\lambda',{\rm i}}$ are the color excess in $\lambda$--$\lambda'$ caused by the circumstellar and interstellar dust, respectively. If we assume that the stellar properties and interstellar extinction (i.e. $C_{\lambda\lambda',{\rm i}}$ and $E_{\lambda\lambda',{\rm i}}$) are constant, the rate of color change is described by using Equation~(\ref{eq:ext}) and (\ref{eq:CCEE}) as,
\begin{equation}
\frac{dC_{\lambda\lambda'}}{dt}\left(t\right) = \left(\frac{\sigma_{\lambda}}{\sigma_{\lambda'}}-1\right)\frac{dA_{\lambda',c}}{dt}.
\label{eq:dct}
\end{equation}

The wavelength dependence of the absorption cross-section is often expressed as a power-law function. If the power-law index is written as $p$ $(\sigma_\lambda\propto\lambda^{p})$, the rate of color change is written as,
\begin{equation}
\frac{dC_{\lambda\lambda'}}{dt}\left(t\right) = \left[\left(\frac{\lambda}{\lambda'}\right)^{p}-1\right]\frac{dA_{\lambda',c}}{dt}.
\label{eq:dct2}
\end{equation}

We compare the observed results with this model in the following sections.

\subsection{{\it K}-band brightening rate}\label{sec:disc-brighteningrate}
The {\it K}-band brightening rate can be calculated based on Equation~(\ref{eq:CR}) by substituting appropriate values of the crossing time and initial {\it K}-band circumstellar extinction, $A_{K,{\rm c}}(0)$, to $t_{\rm cross}$ and $A_{\lambda,{\rm c}}(0)$, respectively.

According to the SED analysis by \citet{1992ApJ...389..400J}, the inner radii of OH/IR stars are 1.3--15$\times10^{12}$ m, and the expansion velocities are in a range of 2.3--24.2 km s$^{-1}$. The combinations of these values lead to the crossing times of 2.8--24 yr. The actual expansion velocities of the six objects can be estimated from the velocity differences of the \replaced{OH-maser}{OH maser} emission peaks. Based on the observations by \citet{HH85}, their expansion velocities are estimated to be 11.0--14.7 km s$^{-1}$ and actually in the parameter range above. The optical depth at 9.7 $\mu$m, \added{$\tau_{9.7}$,} the peak of the silicate feature, is also given by \citet{1992ApJ...389..400J} and ranges from 0.03 to 19.6. The extinction in the {\it K} band can be estimated from \replaced{the}{this} 9.7-$\mu$m optical depth to be $\lesssim$10 mag by assuming the silicate opacity proposed for OH/IR stars by \citet{1999MNRAS.304..389S}. 

\begin{figure}
\epsscale{1.15}
\plotone{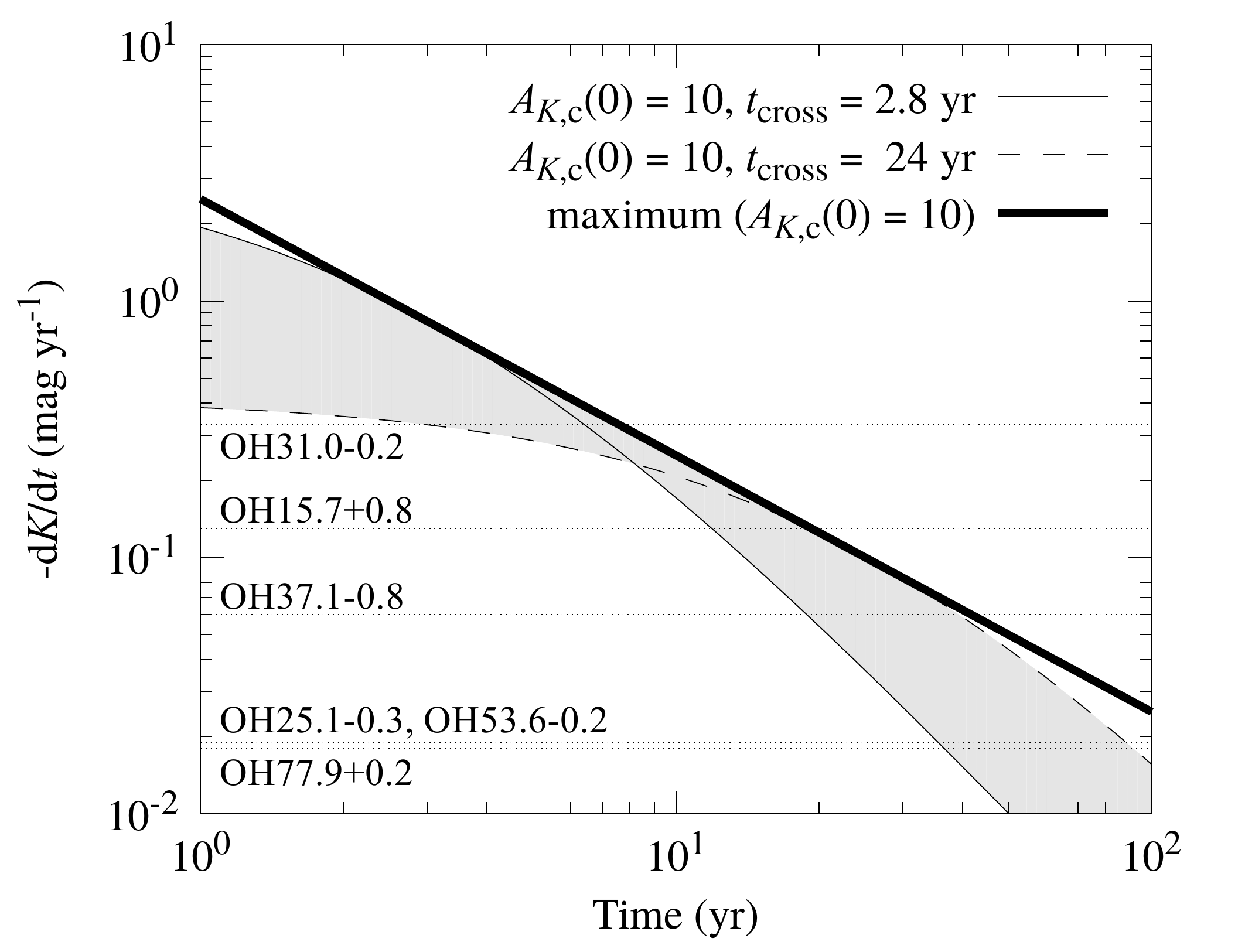}
\vspace{-0.2cm}
\caption{Comparison of {\it K}-band brightening rates between observation and model. Horizontal axis shows the elapsed time from the cessation of dust production. Thin solid and dashed curves show models with initial extinction $A_{K, {\rm c}}(0)$ = 10 mag and crossing time $t_{\rm cross}$ = 2.8 and 24 yr, respectively. Thick solid line shows the maximum rates under the assumption of $A_{K, {\rm c}}(0)$ = 10 mag. Shaded region shows the expected range in the model. {\it K}-band brightening rates observed between 2MASS and UKIDSS observations are shown with horizontal dotted lines.}
\label{fig:compCR}
\end{figure}

If we assume $A_{K,{\rm c}}(0)=$ 10 mag as an example of the optically thickest dust shell of OH/IR stars and $t_{\rm cross}=$ 2.8--24 yr, the time evolution of the {\it K}-band brightening rate is calculated as shown in Figure~\ref{fig:compCR}. In Figure~\ref{fig:compCR}, the thin solid and dashed lines show the calculated results. Based on Equation~(\ref{eq:CR}), the maximum value of the absolute brightening rate at each time $t$ is found to be $A_{K,{\rm c}}(0)/4t$, which occurs in the model with $t_{\rm cross}=t$. This is shown with thick solid line in Figure~\ref{fig:compCR}. The expected range is the region enclosed by these lines and indicated with the shaded region. 

The {\it K}-band brightening rates observed between 2MASS and UKIDSS observations are shown with the horizontal dotted lines in Figure~\ref{fig:compCR}. The {\it K}-band brightening rates are $\lesssim$0.1 mag yr$^{-1}$ except for OH31.0$-$0.2. These values are found in $t=$ 10--90 yr in the model. It means that their brightening rates can be explained as the dispersal effect of the dust shell at such times after the cessation of dust production. However, the large brightening rate of OH31.0$-$0.2, $0.331\pm0.063$ mag yr$^{-1}$, is difficult to explain with this interpretation. Such a value is attained only at $t\lesssim$8 yr as indicated by the thick line in Figure~\ref{fig:compCR}, which appears inconsistent given that this object was recognized as a non-variable OH/IR star by \citet{HH85}, which is more than 10 years earlier than 2MASS observations. 

One possible cause of this large brightening rate is a large $A_{K,{\rm c}}(0)$ (i.e. very optically thick dust shell). As shown in Equation~(\ref{eq:CR}), the absolute value of the brightening rate is proportional to $A_{K,{\rm c}}(0)$. Therefore, a very optically thick dust shell with a large $A_{K,{\rm c}}(0)$ shows a large brightening rate in each elapsed time, and the time range showing a brightening rate can be delayed. To make the elapsed time showing the large brightening rate of OH31.0$-$0.2 (3--8 yr with $A_{K,{\rm c}}(0)$ = 10 mag) to $>$10 yr, we have to assume an extremely high $A_{K,{\rm c}}(0)$ of $\sim$20 mag. Such a high $A_{K,{\rm c}}(0)$ \replaced{means that the dust mass-loss rate of this object is $10^4$ times larger than the usual OH/IR stars,}{corresponds to a very large $\tau_{9.7}$ of $\sim$40,} and this scenario does not seem to be plausible. Therefore, the large brightening rate of OH31.0$-$0.2 implies other mechanisms rather than the dispersal of the dust shell.

\subsection{Rate of color change}
\begin{figure}
\epsscale{1.15}
\plotone{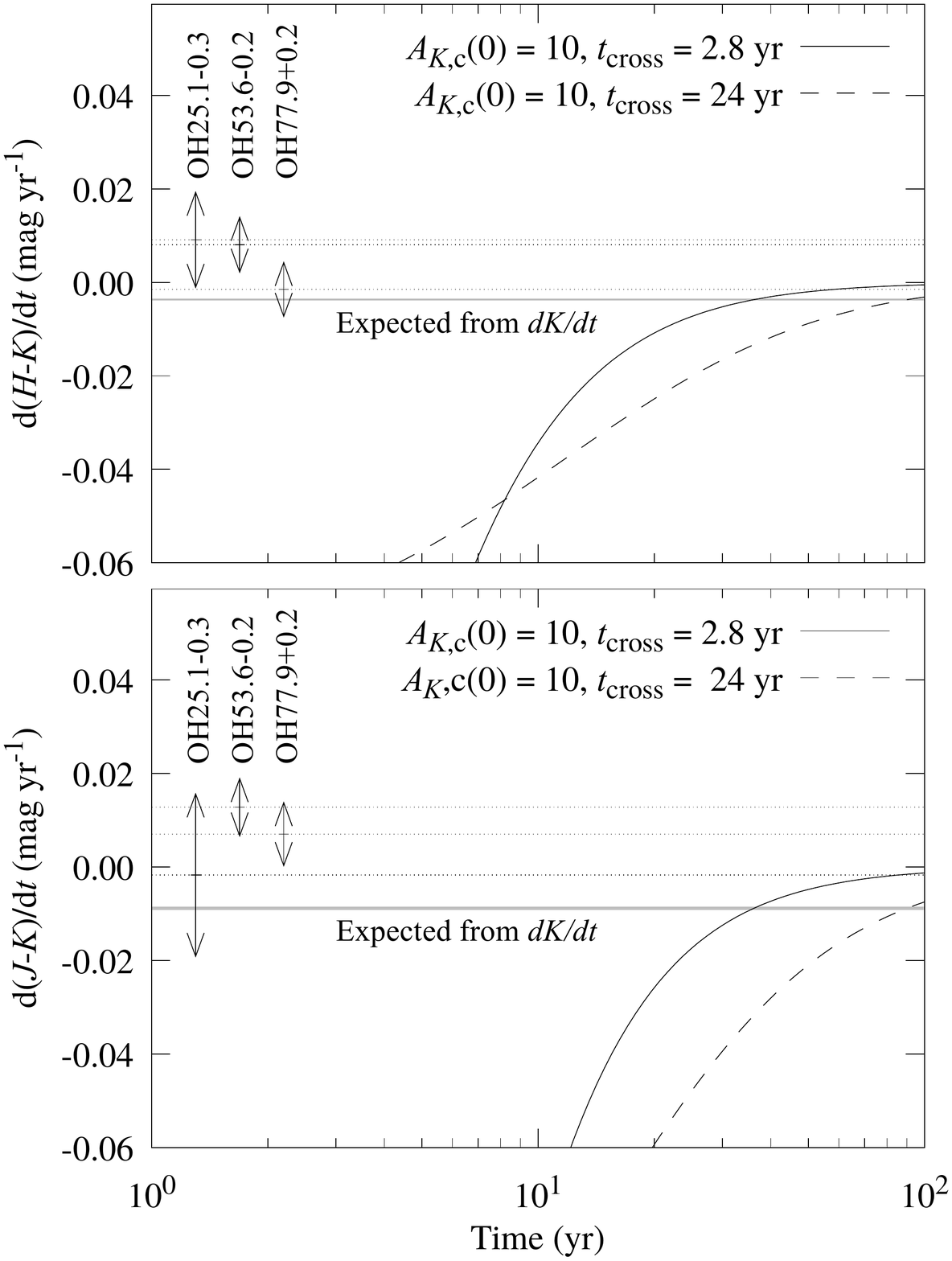}
\caption{Comparison of rates of color change between model and observation. Horizontal axis shows the elapsed time from cessation of dust production. Thin solid and dashed curves show the models with initial extinction $A_{K, {\rm c}}(0)$ = 10 mag and crossing time $t_{\rm cross}$ = 2.8 and 24 yr, respectively. Power-law index of dust opacity, $p$, is assumed to be $-0.7$. Horizontal dotted lines show the observed values. Gray regions show the rates of color change expected from observed {\it K}-band brightening rates of OH25.1$-$0.3, OH53.6$-$0.2, and OH77.9+0.2.}
\label{fig:compCCR}
\end{figure}

Based on Equation~(\ref{eq:CR}) and (\ref{eq:dct2}), we can calculate the rate of color change by additionally assuming the wavelength power-law index $p$. The index $p$ depends on dust model, and various dust models are proposed for oxygen-rich AGB stars \citep{1984ApJ...285...89D, 1990A&A...237..425D, 1992A&A...261..567O, 1999MNRAS.304..389S}. They are well summarized in Figure~1 in the paper presented by \citet{1999MNRAS.304..389S}. Their power-law indices in the NIR are $-0.7$ to $-1.5$, and if we assume $p=-0.7$, the rate of color change is calculated as shown with the solid and dashed lines in Figure~\ref{fig:compCCR}.

Based on Figure~\ref{fig:compCR}, the elapsed time since the cessation of dust production for the three targets whose NIR colors were observed is 40--90 yr depending on \replaced{the assumptions of $A_{K,\rm{c}}(0)$ and $t_{\rm cross}$}{the assumption of $t_{\rm cross}$}. In this range, the NIR color becomes bluer slightly, according to Figure~\ref{fig:compCCR}, as the dust shell continues to disperse. The expected rates of color change can be calculated from the observed {\it K}-band brightening rates based on Equation~(\ref{eq:dct2}) independently from the assumption of $A_{K,{\rm c}}(0)$ and $t_{\rm cross}$. If we assume $p=-0.7$, the rates of change in $(H-K)$ and $(J-K)$ colors for the three objects are calculated to be $-0.0036$ to $-0.0038$ and $-0.0086$ to $-0.0088$ mag yr$^{-1}$, respectively. These values are shown with gray regions in Figure~\ref{fig:compCCR}. We can see that the observed values are greater than expected, indicating that these OH/IR stars did not become bluer as expected.

The difference between the observed and expected values of the rate of change in $(J-K)$ color corresponds to the difference of the observed {\it J}-band brightening rate and the {\it K}-band brightening rate scaled by $(\lambda_J/\lambda_K)^p$, where $\lambda_J$ and $\lambda_K$ are the wavelengths of the {\it J} and {\it K} bands, respectively. Therefore, we evaluate the deviation of the rates of color change by calculating the significance of the difference between the {\it J}-band brightening rate and the {\it scaled} {\it K}-band brightening rate (for $(H-K)$ color, use {\it H} instead of {\it J}).

OH25.1$-$0.3 shows a positive rate of change in $(H-K)$ color which deviates from the expected value with $1.1\sigma$, while in $(J-K)$, it does not show significant deviation. OH77.9+0.2 shows a positive rate of $(J-K)$ color change which deviates from the expectation with $1.9\sigma$ significance, while it shows no significant deviation in the rate of $(H-K)$ color change. The rates of change of OH53.6$-$0.2 are greater than the expected value with significances of 1.9 and 2.9$\sigma$ in $(H-K)$ and $(J-K)$ colors, respectively. This means that OH53.6$-$0.2 significantly became redder than expected.

Based on these results, non-variable OH/IR stars do not seem to have become bluer as expected: some of them seem to have become redder even. At least, OH53.6$-$0.2 clearly shows positive rates, and the observed reddening is difficult to explain with the dispersing dust shell model. These results depend on the assumption of $p$. However, even if we decrease $p$ to $-1.5$, to the end of the $p$ range, the discrepancy between the observations and the model becomes more pronounced. Therefore, we need to consider other mechanisms rather than the dispersing of the circumstellar dust shell.

\subsection{Possibilities beyond the simple model} \label{sec:beyond}
We showed that the observed brightness evolution cannot be explained only with the dispersal of the dust shell and that there should be other mechanisms which were not considered in the simple model. We discuss such possibilities in the following sections.

\subsubsection{Case of warm dust shell}
If the circumstellar dust shell is not yet dispersed, we would expect hot dust grains in the shell. In this case, their thermal emission contributes to the NIR brightness, especially in the {\it K} (and {\it H}) band. However, as the dust shell disperses, the dust temperature decreases, and the dust emission would decrease correspondingly. Therefore, the simple inclusion of dust emission in the simple model does not seem to explain the large brightening rate of OH31.0$-$0.2. The color change of OH53.6$-$0.2 is also difficult to explain. If the dust emission contributes to the NIR brightness, its contribution is larger at longer wavelengths, because the dust emission has a steep positive slope in the NIR. This contribution will decrease as the dust temperature decreases. Then, the NIR color would become bluer. This is contrary to the observed change.

An increase of the stellar temperature and/or luminosity would heat the dust grains and increase the contribution of the dust emission to the NIR brightness. As a result, the {\it K}-band brightness can increase and the NIR color can become redder. These are qualitatively consistent with the observed changes in OH31.0$-$0.2 and OH53.6$-$0.2. If the increase of the stellar temperature and/or luminosity works more effectively than the dispersal effect of the dust shell, such a fast brightening and reddening phenomenon could occur.

\subsubsection{Case of cool dust shell}
If the dust shell is dispersed to an extent where the dust becomes too cool to emit NIR emission, the NIR brightness is dominated by the attenuated stellar emission. In this case, one possibility to explain the fast brightening and reddening is a change of intrinsic stellar properties (i.e. luminosity increase and temperature decrease). Such a change may be caused by a fast stellar expansion as discussed in \S\ref{sec:disc-evol}.

Another possibility is an emergence of dust thermal emission. As discussed above, the thermal emission from hot dust grains contribute more at longer wavelengths, its emergence could make the NIR emission brighter and redder. It can be realized by increasing hot dust grains. One possible way is to heat the dust grains by an increase of the stellar temperature and/or luminosity as discussed above. Another way is a new dust production, which produces hot dust grains around the central star. The emergence of hot dust grains by these effects could result in bright red emission in the NIR.

\subsection{Relation with the stellar evolution}\label{sec:disc-evol}
In the previous section, possible mechanisms for the fast brightening of OH31.0$-$0.2 and the color change of OH53.6$-$0.2 were proposed. We discuss them in the context of the stellar evolution.

\paragraph{The increase of stellar temperature}
The increase of the stellar temperature is in line with the evolutionary scenario after the AGB phase. During the post-AGB phase, it is thought that the stellar temperature monotonically increases with time \citep{1993ApJ...413..641V, 1994ApJS...92..125V, 1995A&A...297..727B, 1995A&A...299..755B, 1998A&A...337..149S, 2016A&A...588A..25M}. Although the definition of the end of the AGB phase remains ambiguous, various timescales are predicted for this transition. \citet{2016A&A...588A..25M} defines the end of the AGB phase as the time when the envelope mass equals to 1\% of the stellar mass. At this time, the stellar temperature is around 4000--5000 K \citep{2016A&A...588A..25M}, and the active pulsation and mass loss are expected to cease \citep{1993ApJ...413..641V, 1994ApJS...92..125V, 1995A&A...297..727B, 1995A&A...299..755B}. \citet{2016A&A...588A..25M} predicts that the stellar temperature increases to $\sim$7000 K on the order of 1--10 kyr after the end of the AGB, while \citet{1994ApJS...92..125V} and \citet{1995A&A...299..755B} predict longer and shorter timescales for the transition, respectively, depending on the particular mass loss treatment adopted during this transition period. The shorter timescales tend to be predicted for stars with larger initial mass \citep{1995A&A...299..755B, 2016A&A...588A..25M}. Therefore, the observed fast brightening and color change may be related to such a fast temperature evolution of intermediate-mass stars. To validate this scenario, we need to reveal the spectral type of these stars.

\paragraph{The increase of stellar luminosity}
The fast increase of the luminosity can occur in the thermal pulse cycles during the AGB phase. In the thermal pulse cycles, stellar properties vary with time. Their temporal variations are theoretically calculated in some cases, and in the early phase of a thermal pulse cycle, fast time variations of the stellar properties are expected to occur \citep{1993ApJ...413..641V, 1995A&A...297..727B, 1998A&A...337..149S, 2019ApJ...879...62M}. A detailed calculation in the early phase of a thermal pulse cycle is given by \citet{2019ApJ...879...62M}. Their model shows that the star rapidly shrinks to half of its radius in a few hundred years and then rapidly expands to greater than the original size in the next few hundred years. Such a fast variation is also seen in the model calculated by \citet{1998A&A...337..149S}. The fast expansion can cause a fast increase of the stellar luminosity by a factor of $>$2 \citep{1993ApJ...413..641V, 1995A&A...297..727B, 1998A&A...337..149S}. This kind of luminosity increase can contribute to the fast brightening and color change by heating dust grains. To quantitatively investigate this possibility, longer-wavelength infrared observations and detailed radiative transfer modeling are needed. 

If this scenario is true, the central star must be in the AGB phase, and the non-variable state must occur in the thermal pulse cycles. \citet{2019MNRAS.482..929T} reports that the dominant pulsation mode can be switched from the fundamental mode (i.e. Mira-like large amplitude pulsation) to the first-overtone mode (i.e. relatively small amplitude pulsation) after such a fast shrink especially in the late AGB phase of low-mass stars. The non-variability may be caused by such a pulsation-mode switching.

\paragraph{The decrease of stellar temperature}
The decrease of stellar temperature is also expected to occur together with the luminosity increase in the fast stellar expansion in the thermal pulse cycles. In the model calculated by \citet{1998A&A...337..149S}, a temperature decrease of a few hundred K in a few hundred years is predicted. If such a temperature decrease occurs when the dust emission is negligible, it will cause the reddening in the NIR. However, the color change of OH53.6$-$0.2 requires a 10 times faster decrease of the stellar temperature. Therefore, this scenario does not seem to be plausible. Nevertheless, it should be investigated with evolutionary calculations with a wide range of parameters.

\paragraph{The new dust production}
The new dust production is not expected after the end of the AGB. However, if the non-variable OH/IR stars are still on the AGB phase, a fast dust production can occur after a temporal reduction of dust production in the thermal pulse cycles.

\citet{1998A&A...337..149S} also calculated the time evolution of the mass-loss rate and SED in the late stage of the AGB phase. In their model, the mass-loss rate decreases with the fast shrinkage of the central star, and then, suddenly increases with the stellar expansion. The enhancement of the stellar mass loss produces hot dust grains and cause the emergence of NIR dust emission. As a result, the {\it K}-band brightness and NIR color suddenly increase. Although the {\it K}-band brightening rate and the rate of $(J-K)$ color change in their model are smaller than the observed values by a factor of 3--4, this behavior is qualitatively consistent with the observed result. If such models with different parameters are investigated, the model which quantitatively matches the observation may be found. 

\subsection{Future investigations}
We have demonstrated that the observed {\it K}-band brightening was generally consistent with the conventional view of non-variable OH/IR stars. However, we have also found it difficult to explain the large brightening rate of OH31.0$-$0.2 and the reddening of OH53.6$-$0.2. To reveal the mechanism of the observed phenomena and understand the currently veiled evolution from the AGB phase to the post-AGB phase, multi-wavelength monitoring, spectroscopy, and distance measurements are needed.

Monitoring observations would segregate the long-term variabilities from possible small-amplitude variability as seen in OH25.1$-$0.2. Multi-wavelength observations are also important. We must observationally establish the NIR color evolution of the non-variable OH/IR stars that exhibit long-term brightening to determine whether the conventional view would still be valid or not.

Stellar properties such as spectral type and luminosity are also important especially to reveal the evolutionary states of non-variable OH/IR stars. However, the spectral type of most non-variable OH/IR stars is largely unknown. We need spectroscopic observations with high sensitivity and high spatial resolution to reveal the spectral type of the objects which are faint in the short-wavelength range and are in the crowded galactic plane. 

The luminosity is also relatively unknown because their distances are uncertain. We do not find reliable distances to these 16 non-variable OH/IR stars even in the Gaia DR2 \citep{2018A&A...616A...1G}. The phase-lag method is a useful way to measure the distance to OH/IR stars. In this method, it is required to measure the phase difference of the variabilities of the two \replaced{OH-maser}{OH maser} emission peaks. However, the non-variable OH/IR stars do not show significant variability of \replaced{OH-maser}{OH maser} emission, and this method is not applicable for these objects \citep{1990A&A...239..193V}. In these circumstances, parallax measurements using maser observations with Very Large Baseline Interferometry (VLBI) is a good way to determine the distance more accurately \citep[e.g.][]{2019IAUS..343..476N}.

Model calculations are also important. The observables derived by combining stellar-evolution models, hydrodynamic stellar-wind models, and radiative-transfer calculations are very useful to interpret the observational results \citep[e.g.][]{1998A&A...337..149S}. Recently, advanced stellar-evolution models \citep[e.g.][]{2016A&A...588A..25M} and hydrodynamic stellar-wind models \citep[e.g.][]{2019A&A...626A.100B} have been developed. The studies on the observables incorporating the recent models will be highly useful to interpret the future observations and possibly resolve the origin of the brightness evolution of non-variable OH/IR stars. At the same time, these numerical models can reproduce the observed behaviors of non-variable OH/IR stars.

\section{Summary} \label{sec:summary}
We investigated the long-term brightness evolution of 16 non-variable OH/IR stars \citep{HH85}. Based on the infrared images taken with 2MASS and {\it SST}/IRAC, their precise positions on the sky were determined. Their recent NIR magnitudes were looked up in the UKIDSS-GPS and OAOWFC data. As a result, multi-epoch data were established for seven objects. However, for one object (OH51.8$-$0.2), we were unable to determine the time evolution of its brightness.

All the remaining six objects appeared to have brightened in a few thousand days. Their {\it K}-band brightening rate ranges from 0.010 to 0.331 mag yr$^{-1}$. The observed {\it K}-band brightening is basically explained with a simple model of a dispersing dust shell with typical parameters of OH/IR stars. However, the largest brightening rate, 0.331 mag yr$^{-1}$, is difficult to explain with this model. This large brightening rate implies the presence of other mechanisms.

For three of the six objects, OH25.1$-$0.3, OH53.6$-$0.2, and OH77.9+0.2, we examined the long-term brightness evolution for more than one band. None of these three objects appeared to have become bluer, and among the three objects, OH53.6$-$0.2 was significantly confirmed to have been reddened. This color change is contrary to the expectation from the dispersal of the dust shell and requires to consider other mechanisms.

An increase of the dust emission could explain both observed phenomena simultaneously. One possible mechanism is an increase of the stellar temperature which is expected to occur in the transition from the AGB phase to the post-AGB phase. Another possibility is an increase of the stellar luminosity or a new dust production which can happen with the fast stellar expansion in the thermal pulse cycles during the AGB phase. To explain the observed reddening, a reduction of the stellar temperature was also proposed.

To understand the cause of the observed variabilities, further investigations in theoretical calculations and observations are needed. Such studies will lead to a better understanding of the evolutionary states of non-variable OH/IR stars and the evolution of AGB stars.


\acknowledgments
This publication makes use of data products from the Two Micron All Sky Survey, which is a joint project of the University of Massachusetts and the Infrared Processing and Analysis Center/California Institute of Technology, funded by the National Aeronautics and Space Administration and the National Science Foundation.
This research has made use of the NASA/IPAC Infrared Science Archive, which is funded by the National Aeronautics and Space Administration and operated by the California Institute of Technology.
This research has made use of the SIMBAD database, operated at CDS, Strasbourg, France.
This research made use of the cross-match service provided by CDS, Strasbourg. 

\vspace{5mm}


\facilities{IRSA, Spitzer, CTIO:2MASS, FLWO:2MASS, UKIRT, OAO:0.91m}

\software{SExtractor \citep{1996A&AS..117..393B}, IRSA, IRSA viewer, SIMBAD \citep{2000A&AS..143....9W}, VizieR, CDS X-match}

\clearpage
\appendix
\section{Systematic Errors} \label{sec:sys}
To further validate the time evolution of the NIR magnitudes of several OH/IR stars found in \S\ref{sec:res}, we examine the systematic errors in the various methods of photometry employed and residual calibration errors.

\subsection{2MASS data} \label{sec:sys-2mass}
\begin{deluxetable*}{ccccccccccc}
\tablecaption{
Detailed list of 2MASS data used for OH25.1$-$0.3, OH53.6$-$0.2, and OH77.9+0.2.
\label{tab:2mass-detail}}
\tablehead{
\colhead{Object} & \colhead{Band} & \colhead{[jhk]\_m$^a$} & \colhead{[jhk]\_msigcom} & \colhead{ph\_qual} & \colhead{rd\_flg} & \colhead{bl\_flg} & \colhead{cc\_flg} & \colhead{[jhk]\_psfchi} & \colhead{[jhk]\_m\_stdap$^a$} & \colhead{[jhk]\_msig\_stdap}
\\
\colhead{} & \colhead{} & \colhead{(mag)} & 
\colhead{(mag)} & \colhead{} & \colhead{} & 
\colhead{} & \colhead{} & \colhead{} &
\colhead{(mag)} & \colhead{(mag)}
}
\startdata
OH25.1$-$0.3 & {\it J} & 15.852 & 0.107 & B & 2 & 2 & p & 1.31 & 15.133 & 0.092 \\
 & {\it H} & 13.121 & 0.044 & A & 2 & 1 & 0 & 1.91 & 13.061 & 0.026 \\
 & {\it K}$_{\rm S}$ & 11.528 & 0.035 & A & 2 & 1 & 0 & 2.38 & 11.511 & 0.017 \\
OH53.6$-$0.2 & {\it J} & 14.332 & 0.029 & A & 2 & 1 & 0 & 1.28 & 14.313 & 0.025 \\
 & {\it H} & 12.912 & 0.024 & A & 2 & 1 & 0 & 0.97 & 12.893 & 0.030 \\
 & {\it K}$_{\rm S}$ & 11.965 & 0.021 & A & 2 & 1 & 0 & 1.10 & 11.960 & 0.027 \\
OH77.9+0.2 & {\it J} & 15.196 & 0.042 & A & 2 & 1 & s & 1.04 & 15.183 & 0.051 \\
 & {\it H} & 13.463 & 0.027 & A & 2 & 1 & s & 1.58 & 13.479 & 0.049 \\
 & {\it K}$_{\rm S}$ & 11.877 & 0.017 & A & 2 & 1 & 0 & 1.43 & 11.895 & 0.022 \\
\enddata
\tablecomments{$^a$ Magnitudes in the 2MASS system.}
\end{deluxetable*}

2MASS magnitudes are available for OH25.1$-$0.3, OH53.6$-$0.2, and OH77.9+0.2. They are summarized in Table \ref{tab:2mass-detail} with relevant flags. The {\it [jhk]\_m} is the default magnitude used in the 2MASS catalog. The {\it rd\_flg} shows the measurement method used in deriving the {\it [jhk]\_m} magnitude. The {\it rd\_flg} of these data is always 2, which means that the {\it [jhk]\_m} magnitudes are derived by profile-fitting measurements \citep{2MASS.Suppl., 2006AJ....131.1163S}. The profile fitting is performed to fit the pixel values in a circle with a radius of 2 pixels (2$\arcsec$) centered on the detected source with multiple point-spread-functions (PSFs). The 2MASS PSC also gives magnitudes measured with aperture photometry, {\it [jhk]\_m\_stdap}, which are curve-of-growth-corrected magnitudes measured with a 4$\arcsec$ radius aperture. We compare these two magnitudes to assess the systematic errors and missing flux.

Both of the {\it [jhk]\_m} and {\it [jhk]\_m\_stdap} magnitudes of OH53.6$-$0.2 and OH77.9+0.2 in all bands are consistent with each other within the corresponding uncertainty. The {\it K}$_{\rm S}$-band magnitudes of OH25.1$-$0.3 are also consistent with each other. This means that these {\it [jhk]\_m} are good values of the object brightness without any significant amount of missing fluxes. 

The {\it [jhk]\_m\_stdap} magnitudes of OH25.1$-$0.3 in the {\it J} and {\it H} bands are found to be brighter than the {\it [jhk]\_m} magnitudes by 0.719 and 0.060 mag with significances of $\sim$5.1 and $\sim$1.2$\sigma$, respectively. As shown in Figure~\ref{fig:id-b}, since the region around OH25.1$-$0.3 is crowded, emission from surrounding stars can contaminates the source magnitude measured with the 4$\arcsec$ radius aperture. Therefore, the difference between the {\it [jhk]\_m} and {\it [jhk]\_m\_stdap} magnitudes is likely caused by the contamination, and the {\it [jhk]\_m} magnitudes are thought to be appropriate for the brightness of the object. 

The reduced chi-squared value and the number of required sources of the profile fitting are referred to {\it [jhk]\_psfchi} and {\it bl\_flg}, respectively. Except for the {\it J}-band data of OH25.1$-$0.3, {\it [jhk]\_psfchi} is $\lesssim$2 and {\it bl\_flg} is 1. This means that the sources are well fitted with one PSF, and it is consistent with the result that the {\it [jhk]\_m} magnitudes do not have any significant amount of missing flux. As for the {\it J}-band data of OH25.1$-$0.3, {\it bl\_flg} is 2. This suggests that some extended component can contribute to the source brightness, and this may be related to the large difference between the {\it [jhk]\_m} and {\it [jhk]\_m\_stdap} magnitudes. 

Table \ref{tab:2mass-detail} also shows that some data have non-zero {\it cc\_flg} indicating possible contamination or confusion. OH77.9+0.2 shows an {\it s} flag in the {\it J} and {\it H} bands. This indicates possible contamination caused by stripe patterns around bright stars. In the case of OH77.9+0.2, a bright star is found at 2$\arcmin$ north from the target. We confirm that stripe patterns caused by the bright star is not apparent in the 2MASS images at the target position and concluded that the {\it J}- and {\it H}-band magnitudes are reliable enough. OH25.1$-$0.3 has a {\it p} flag in the {\it J}-band data. It indicates possible contamination caused by a persistence pattern caused by bright stars. We cannot find such a pattern clearly in the 2MASS image, but it is difficult to find it in the crowded region around OH25.1$-$0.3. Therefore, the {\it J}-band magnitude may not be as reliable. 

Based on these assessments, we conclude that all 2MASS data except for the {\it J}-band data of OH25.1$-$0.3 are reliable. If we rely only on the {\it H}- and {\it K}-band magnitudes, OH25.1$-$0.3 is thought to become red as observed for OH53.6$-$0.2.

\subsection{UKIDSS data} \label{sec:sys-ukidss}
\begin{figure*}
\gridline{\fig{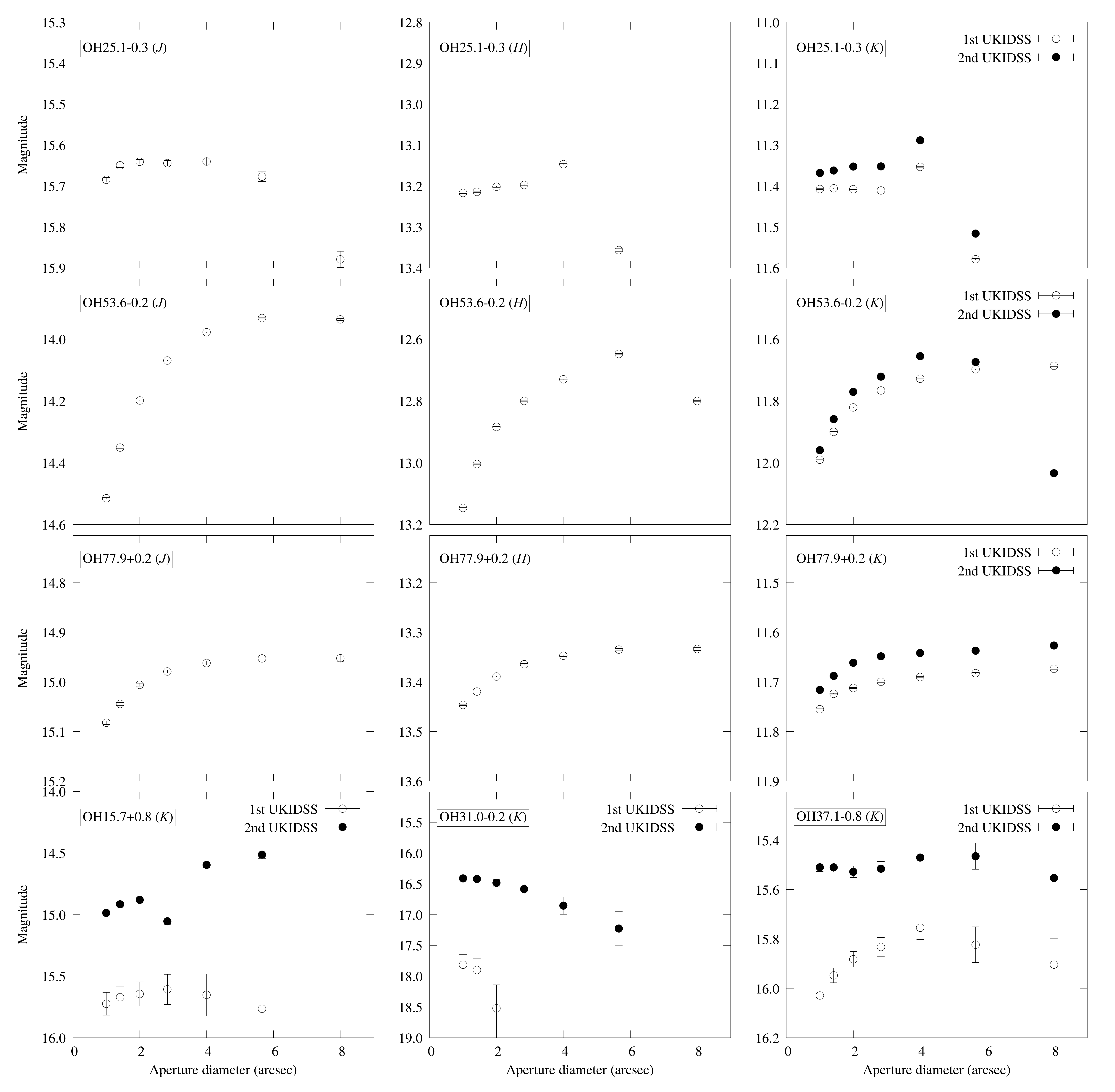}{0.92\textwidth}{}}
\vspace{-1cm}
\caption{
Aperture-size dependence of the aperture-photometry-based magnitudes obtained by UKIDSS. Upper three rows show those for OH25.1$-$0.3, OH53.6$-$0.2, and OH77.9+0.2 in the {\it J}, {\it H}, and {\it K} bands. The bottom panels show those for OH15.7+0.8, OH31.0$-$0.2, and OH37.1$-$0.8 in the {\it K} band. Object name and observation band are shown in the upper left corner in each panel. In the {\it K}-band plots, the first and second UKIDSS observations are plotted with open and filled circles, respectively. The default magnitude, {\it AperMag3}, is the third point from left in each panel.
\label{fig:UKIDSS-detail}}
\end{figure*}

\begin{figure*}
\gridline{\fig{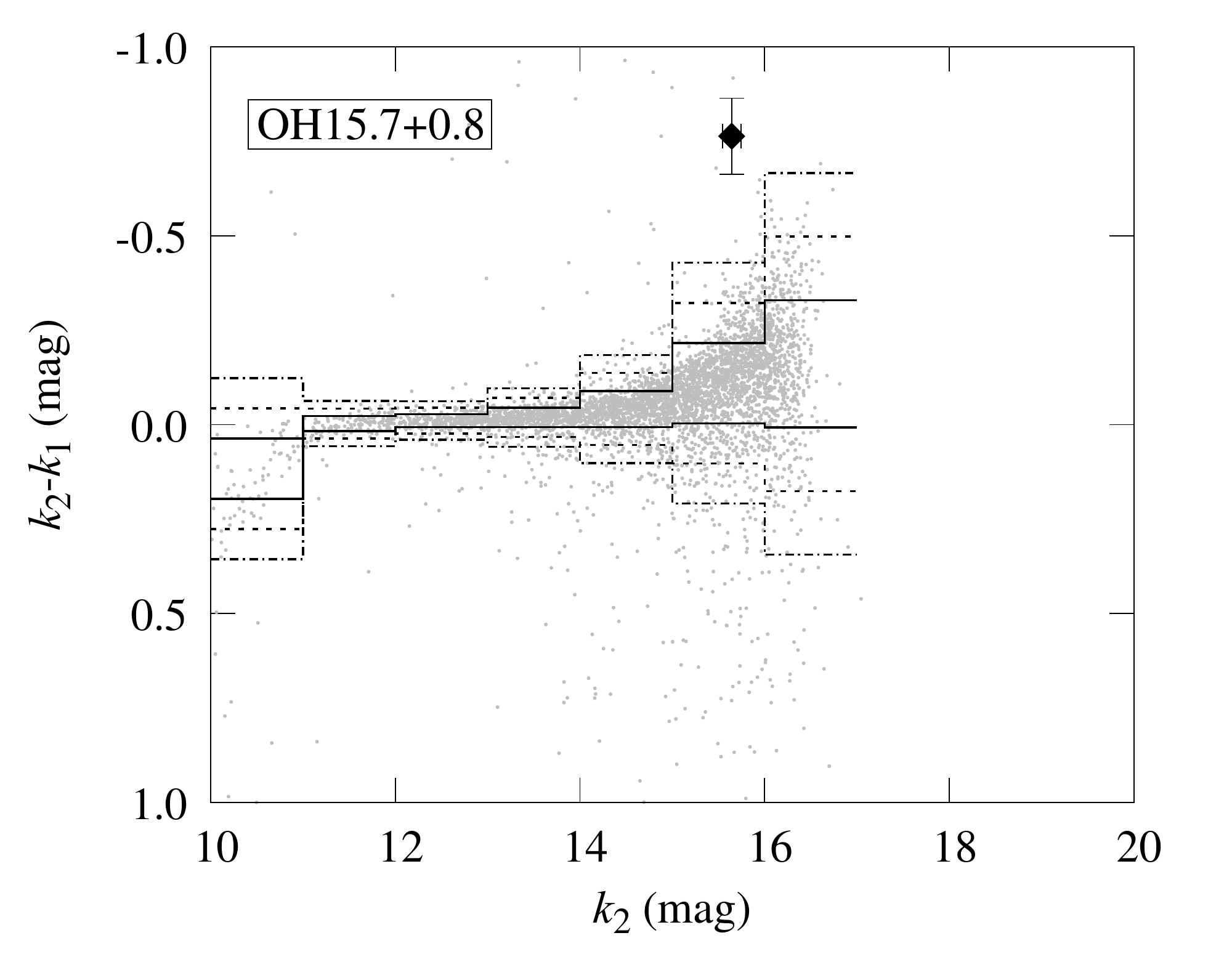}{0.33\textwidth}{}
          \fig{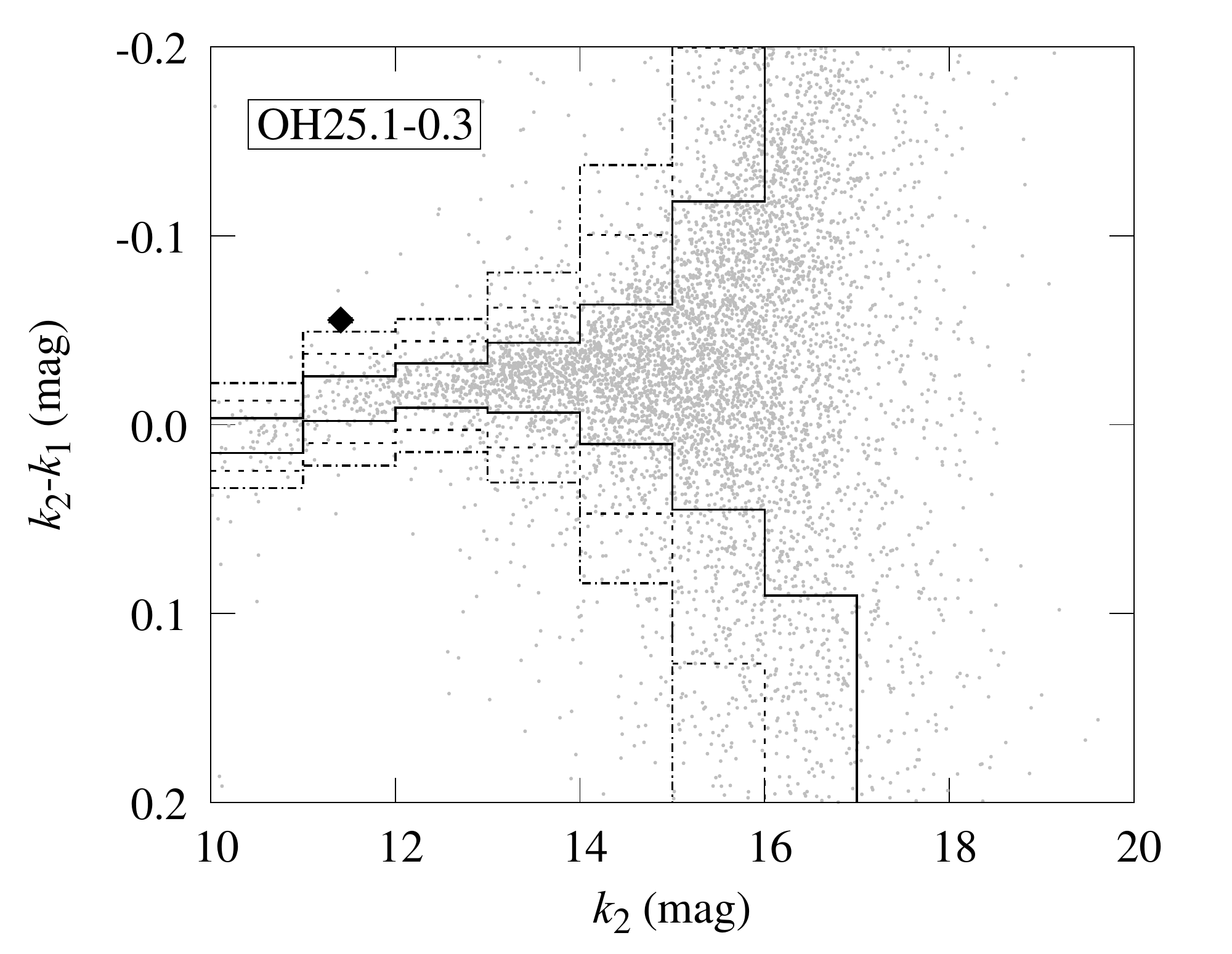}{0.33\textwidth}{}
          \fig{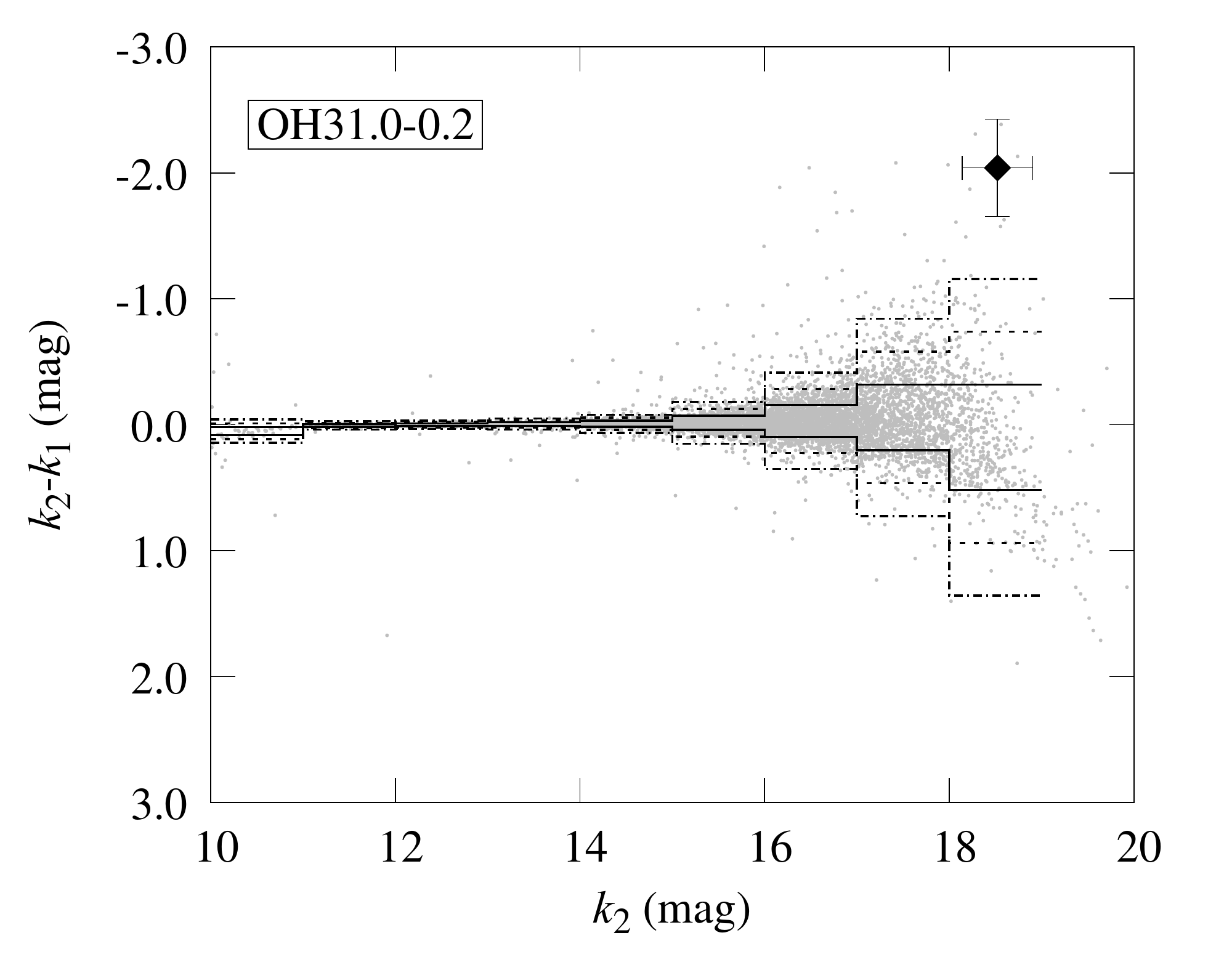}{0.33\textwidth}{}}
\vspace{-1cm}
\hspace{-0.5cm}
\gridline{\fig{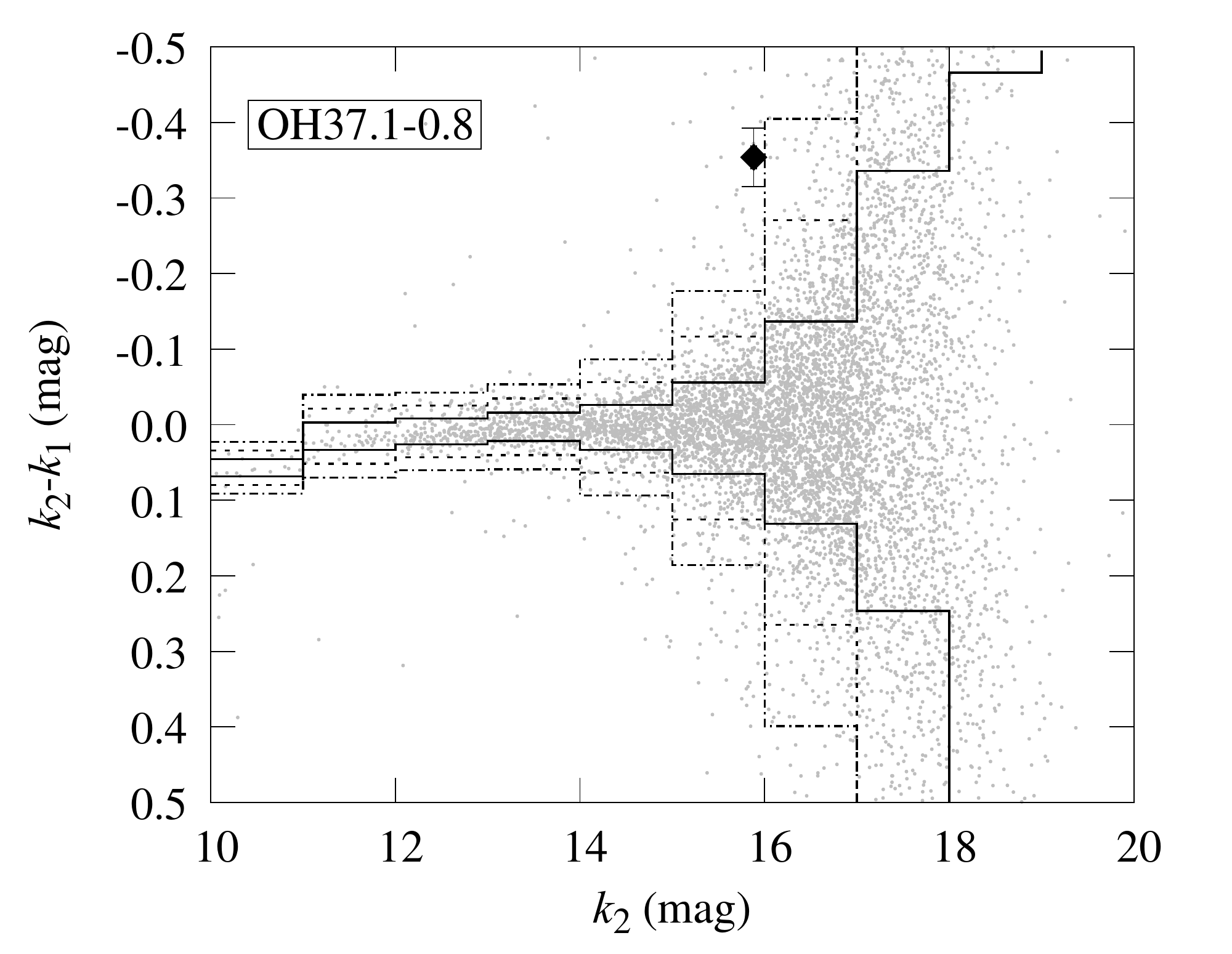}{0.33\textwidth}{}
          \fig{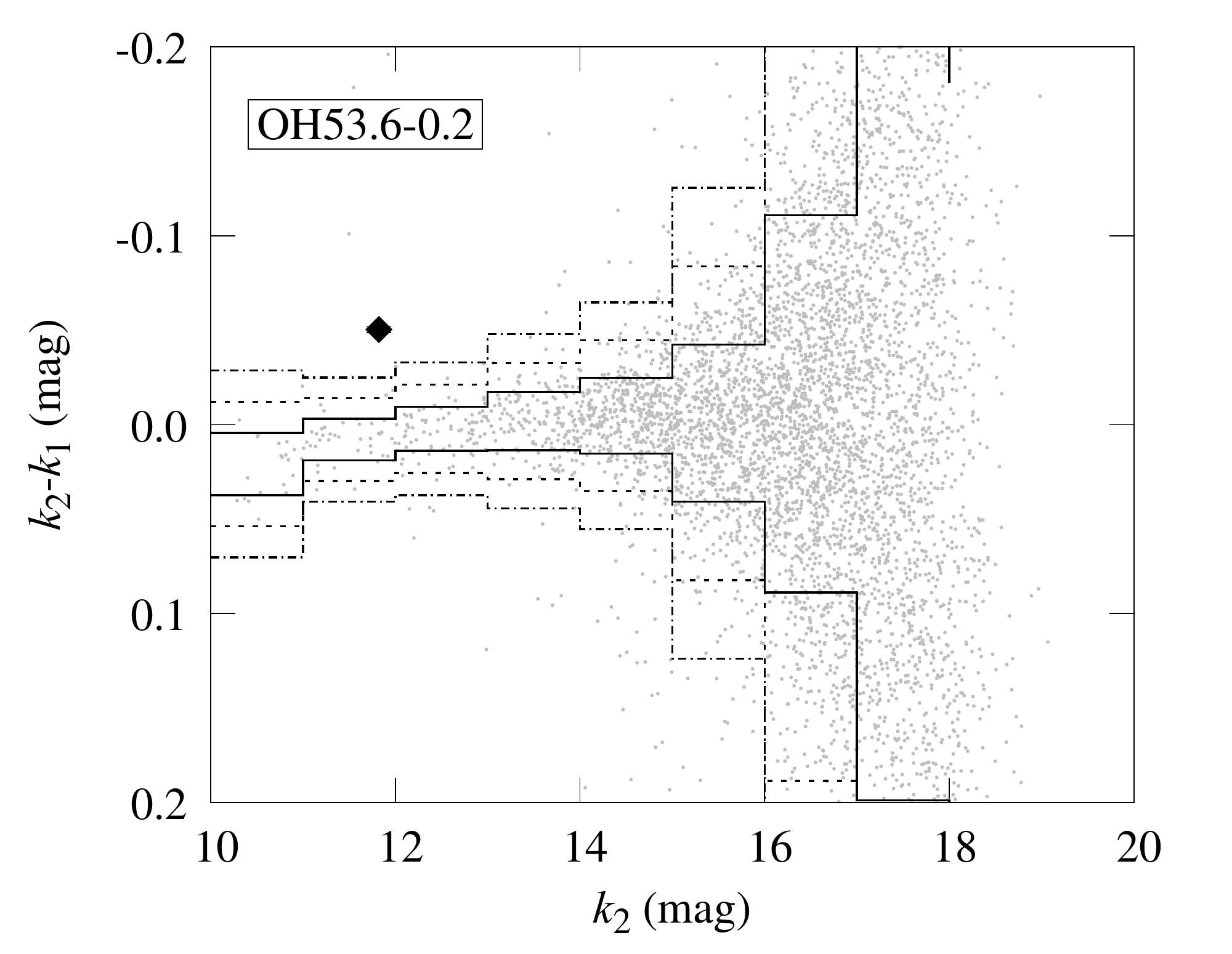}{0.33\textwidth}{}
          \fig{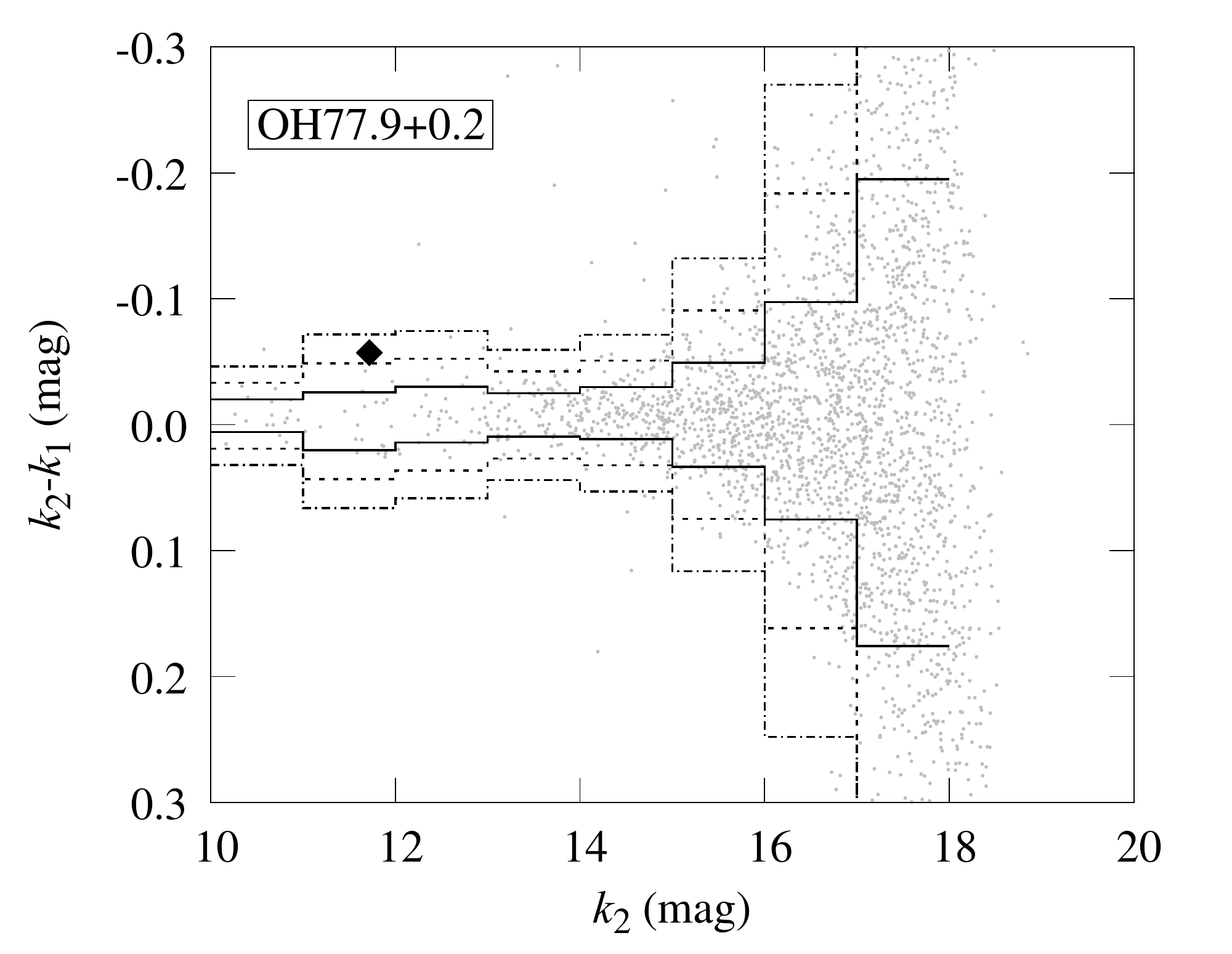}{0.33\textwidth}{}}
\vspace{-0.6cm}
\caption{Magnitude differences between two UKIDSS observations plotted against observed magnitude in the {\it K} band. $k_1$ and $k_2$ show the {\it K}-band magnitude observed in the first and second UKIDSS observations, respectively. Gray dots show the objects around each target within 3$\arcmin$. Solid, dotted, and dash-dotted lines show the 1, 2, and 3$\sigma$ regions from average calculated in 1-mag bins. The data of the target is shown with black diamond in each panel. The error bars show the photometric uncertainties without systematic errors. 
\label{fig:sys-ukidss}}
\end{figure*}

The UKIDSS magnitudes are measured based on the aperture photometry with various aperture radii. The aperture-photometry-based magnitudes are given as {\it AperMag[1-13]} measured with aperture diameters of 1, $\sqrt{2}$, 2, $2\sqrt{2}$, 4, $4\sqrt{2}$, 8$\dots$ arcsec \citep{2004SPIE.5493..411I, 2008MNRAS.391..136L}. The default magnitude, {\it AperMag3}, is measured with a 2$\arcsec$ diameter aperture. To check the validity of {\it AperMag3}, we examine the aperture-diameter dependence of the measured magnitudes. 

Figure~\ref{fig:UKIDSS-detail} shows the aperture-size dependence of the measured magnitudes. A smaller aperture setting can give more reliable magnitudes of sources in crowded regions. However, if the source has a radial profile thicker than a PSF because of some causes such as seeing and intrinsic diffuse emissions, a small aperture setting can \replaced{results}{result} in underestimating the source brightness. In such a case, the aperture-size dependence will show an increasing trend. On the other hand, a large aperture setting can lead to unstable measurement of the source brightness, because it is easily affected by contamination from surrounding stars and the poorly estimated sky background. Therefore, reliable magnitudes are thought to be in between these two cases, and such region will appear as a plateau region in Figure~\ref{fig:UKIDSS-detail}.

In the case of OH25.1$-$0.3, the default magnitude is on the plateau in all bands and seems to be reliable as the source brightness. However, for OH53.6$-$0.2 and OH77.9+0.2, the default magnitude is not on the plateau, and the magnitudes measured with larger apertures may be better for the total source brightness. If we employ the magnitudes measured with larger apertures, the source brightness increases by 0.1--0.3 mag, and the brightening since the 2MASS era will be more significant. In the case of OH15.7+0.8, the default magnitude for the first UKIDSS observation is on the plateau. As for the second observation, the magnitudes measured with apertures larger than the default are unstable. Therefore, the default magnitude seems to be good as the source magnitude. OH31.0$-$0.2 shows decreasing trends with increasing aperture diameter in both data, and magnitudes measured with smaller apertures seem to be reliable. If we use the magnitudes measured with aperture diameters of 1 and $\sqrt{2}$ arcsec, the source will get brighter by 0.6--0.7 and 0.06--0.07 mag in the first and second data, respectively. In the case of OH37.1$-$0.8, the default magnitude is at the plateau in the second data, but in the first data, it is at the edge of the plateau. The magnitudes measured with aperture diameters larger than 4$\arcsec$ are unstable. Therefore, the magnitude measured with an aperture diameter of $2\sqrt{2}$ arcsec seems to be most reliable, but it is consistent with the default magnitude within the error. 

The magnitude differences between two UKIDSS {\it K}-band observations are almost independent from the aperture settings. However, as for the case of OH31.0$-$0.2, the magnitude difference between the two UKIDSS observations may be overestimated when using the default magnitudes. If we employ the magnitudes measured with smaller apertures, the magnitude difference between the two UKIDSS observations decreases by $\sim$0.6 mag, but the difference (brightening) is still significant. This correction also decreases the brightening rate of OH31.0$-$0.2 to 0.23 mag yr$^{-1}$. This value still requires a short elapsed time of $\lesssim$11 yr since the cessation of dust production, which is difficult to explain with the dispersal of the dust shell as discussed in \S\ref{sec:disc-brighteningrate}. Therefore, this correction does not change the conclusion that OH31.0$-$0.2 is showing a very fast brightening which cannot be explained with the dispersal of the dust shell.

\begin{figure*}
\gridline{\fig{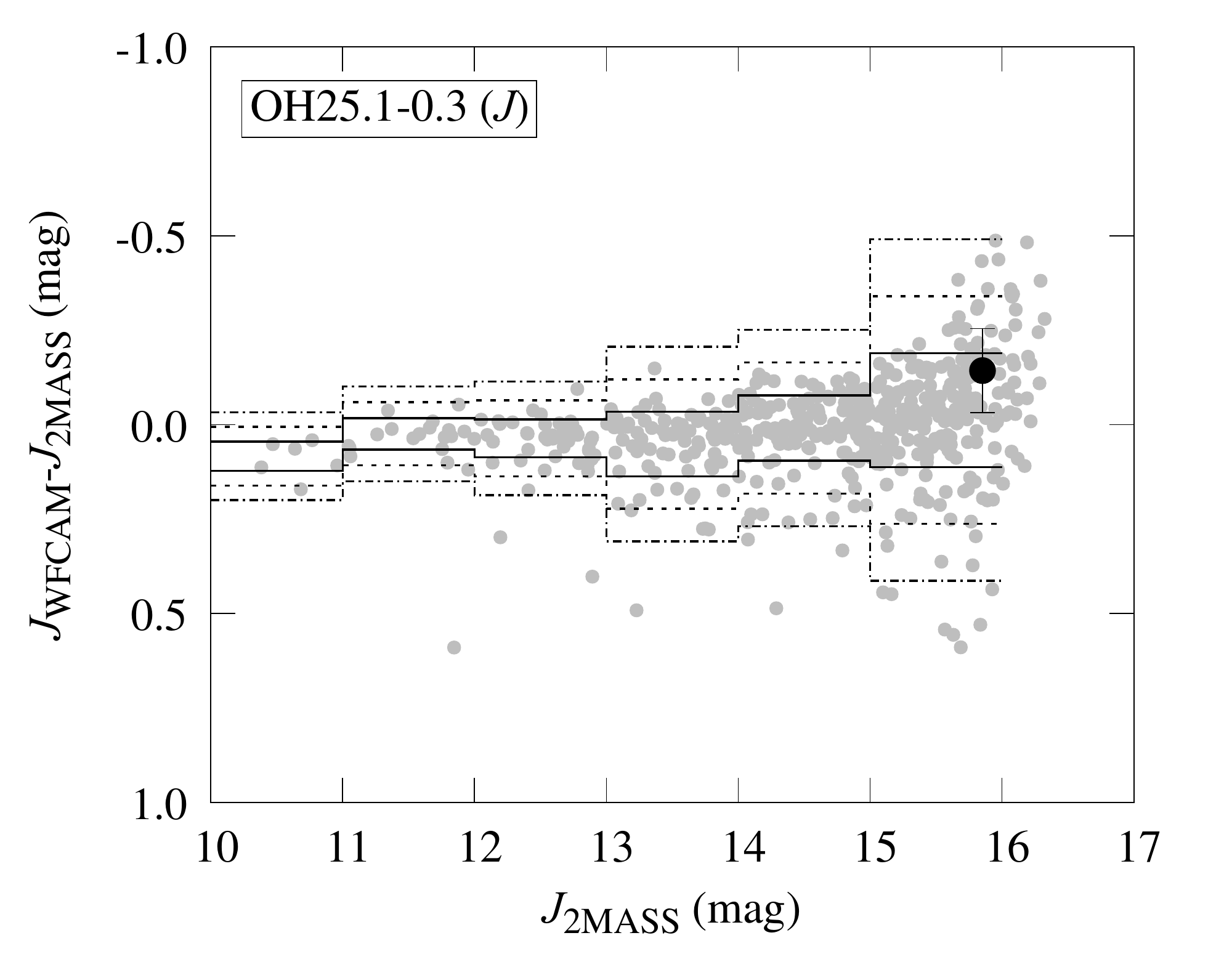}{0.33\textwidth}{}
\hspace{-1.cm}\fig{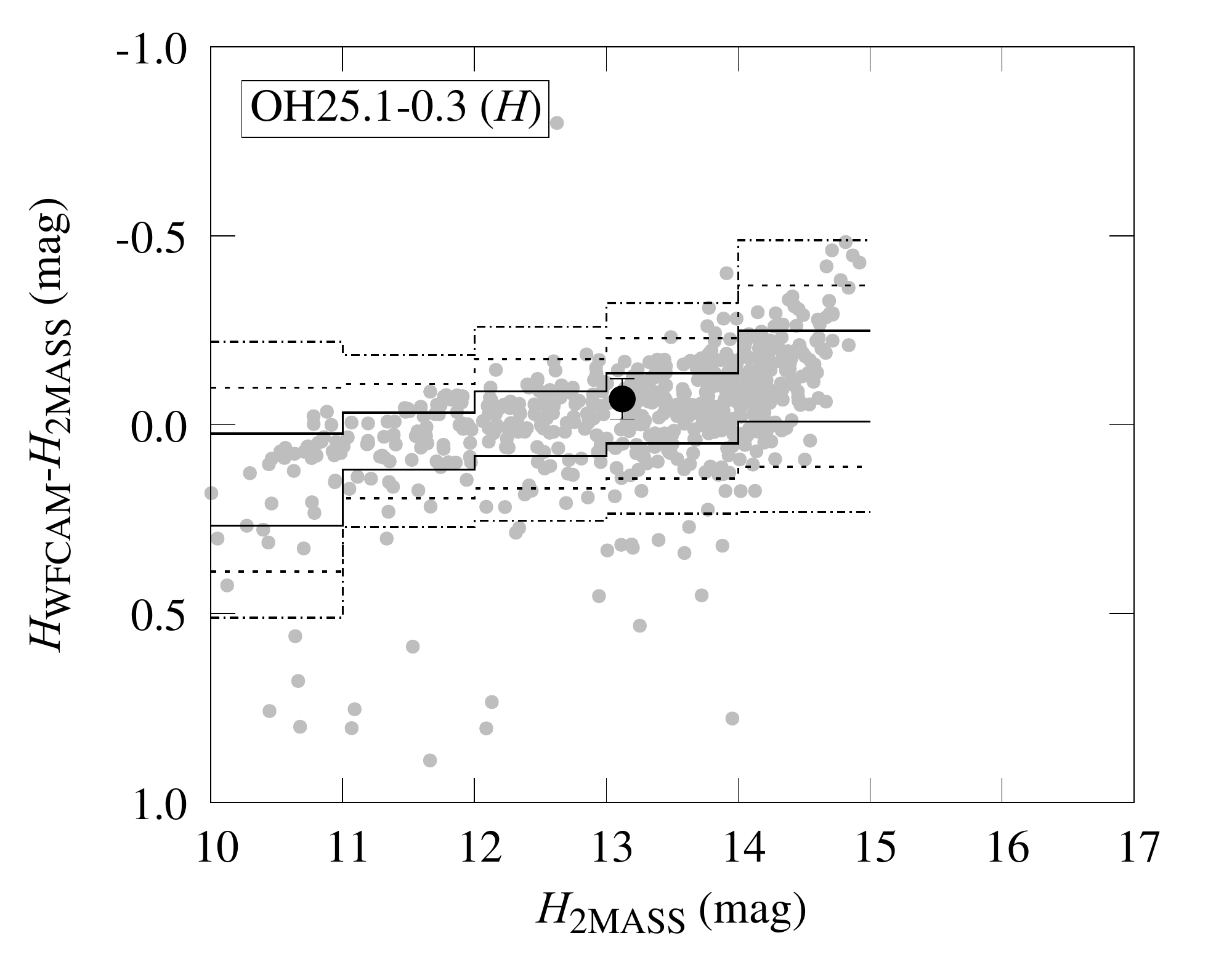}{0.33\textwidth}{}
\hspace{-1.cm}\fig{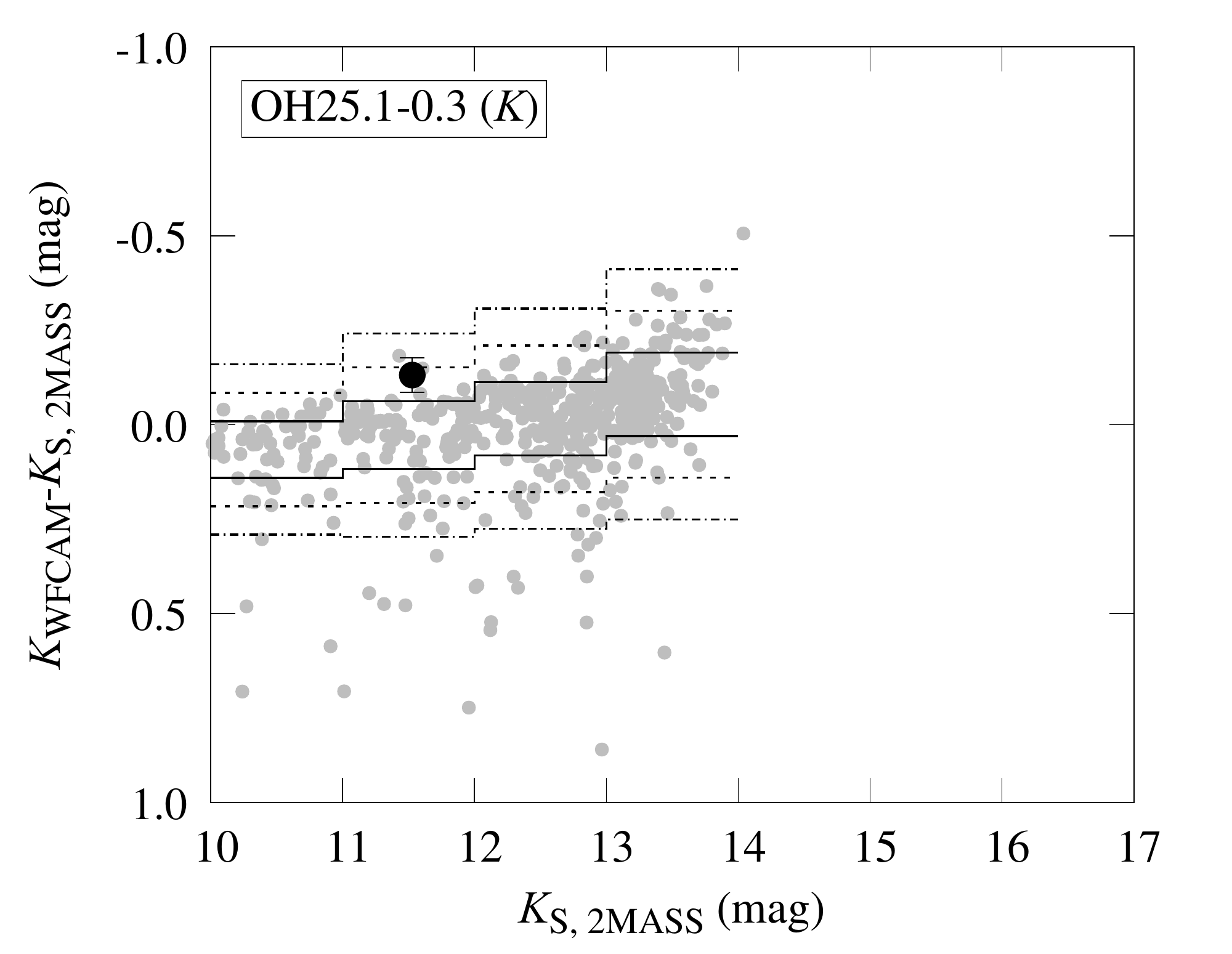}{0.33\textwidth}{}}
\vspace{-1.3cm}
\hspace{-0.5cm}
\gridline{\fig{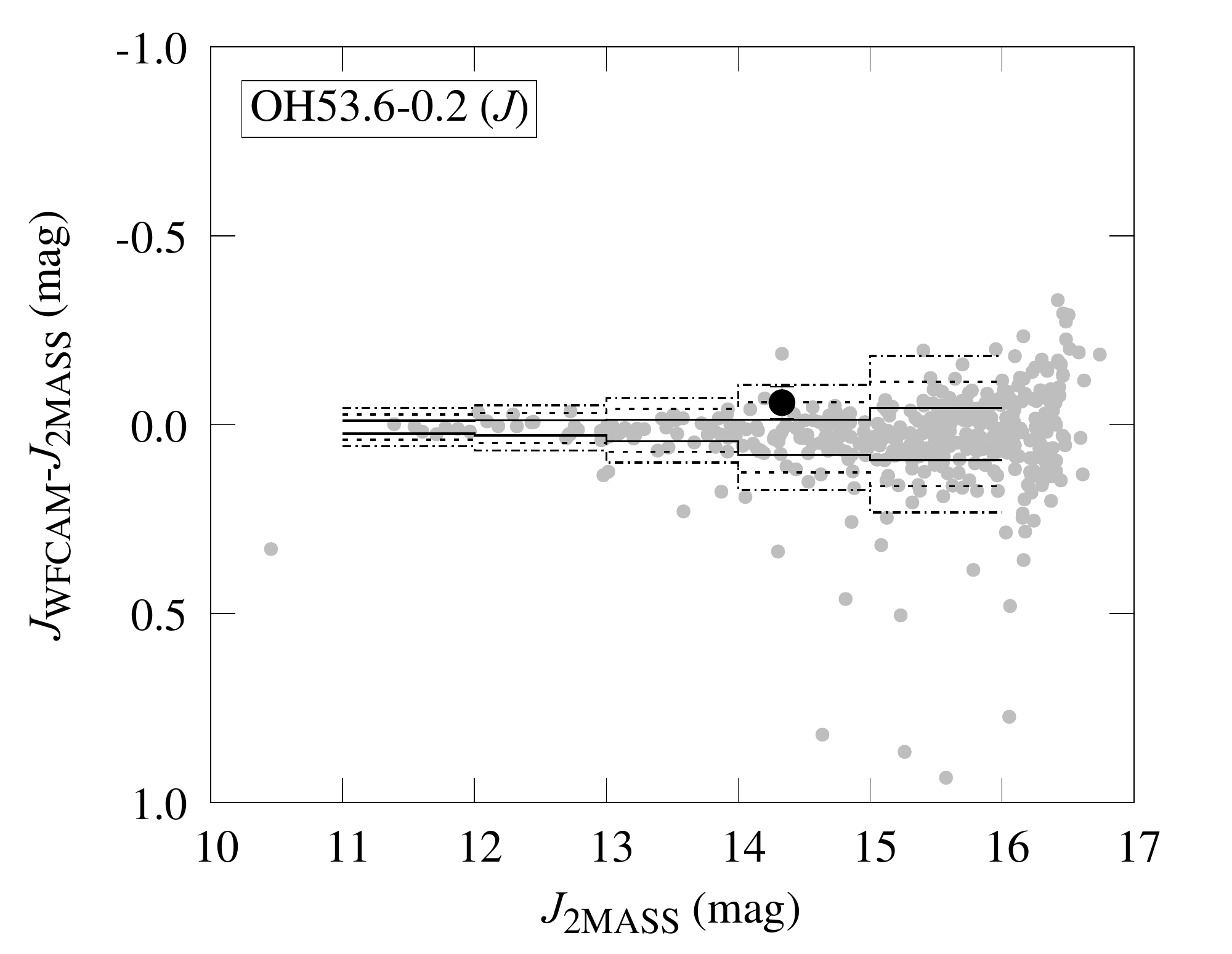}{0.33\textwidth}{}
\hspace{-1.cm}\fig{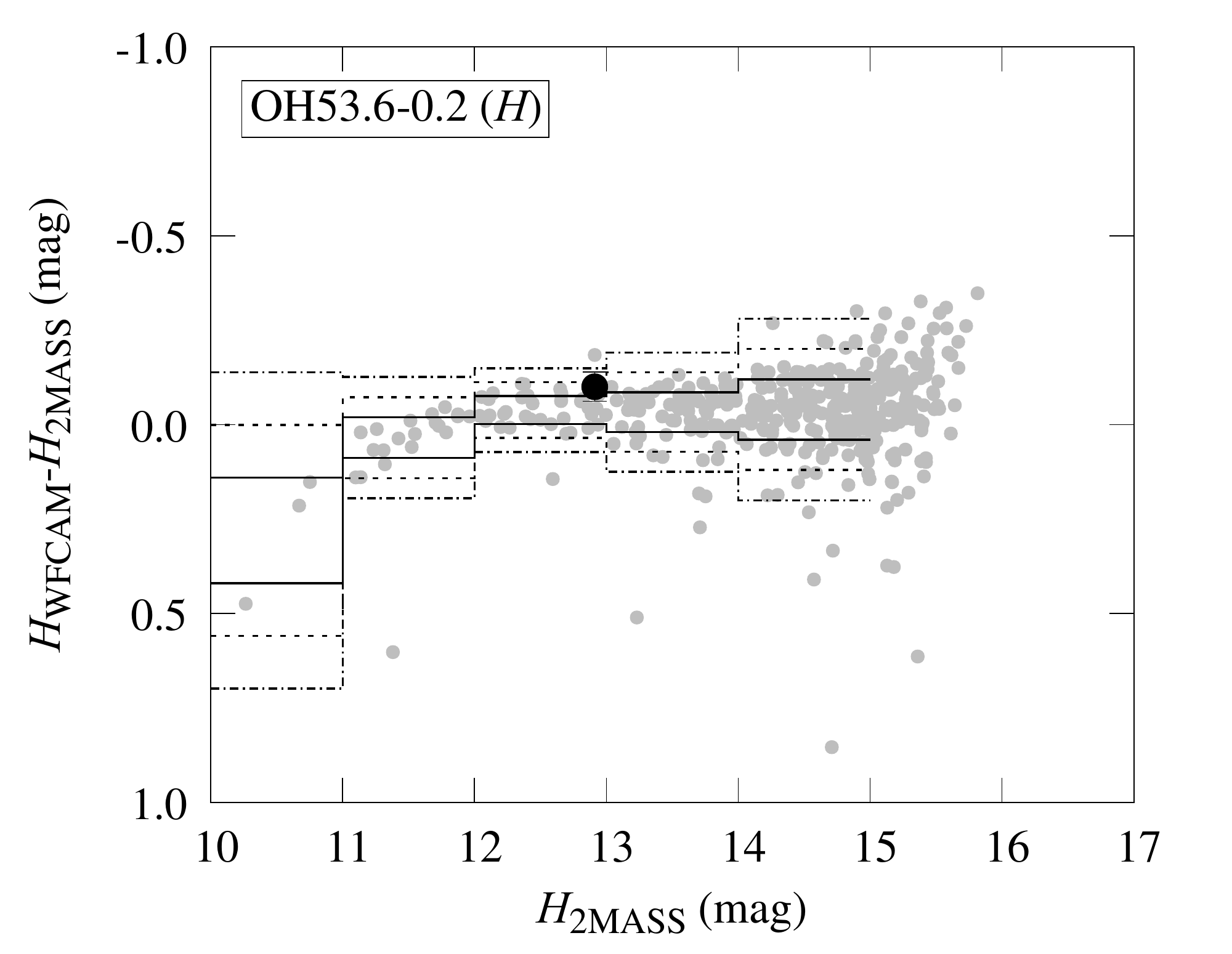}{0.33\textwidth}{}
\hspace{-1.cm}\fig{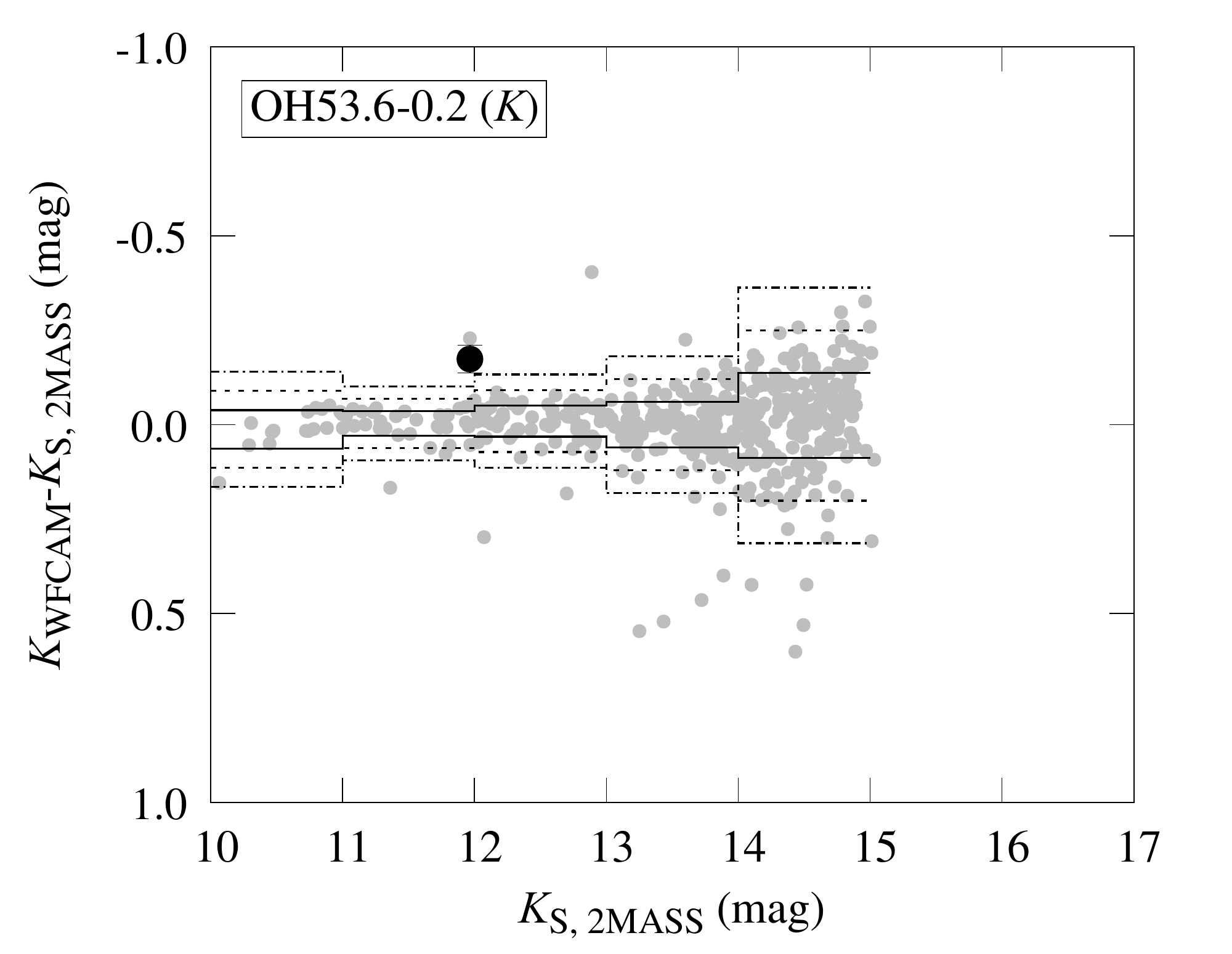}{0.33\textwidth}{}}
\vspace{-1.3cm}
\hspace{-0.5cm}
\gridline{\fig{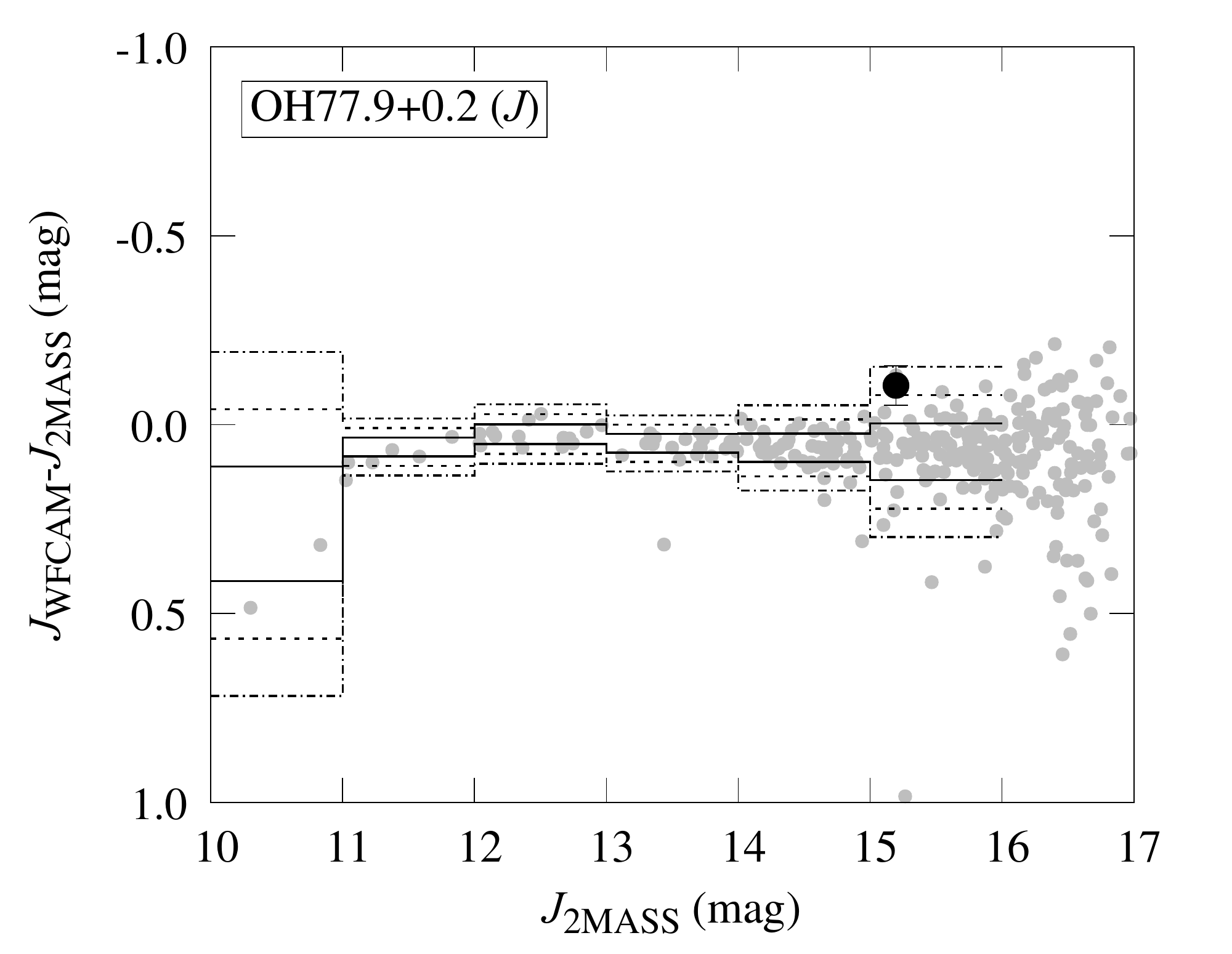}{0.33\textwidth}{}
\hspace{-1.cm}\fig{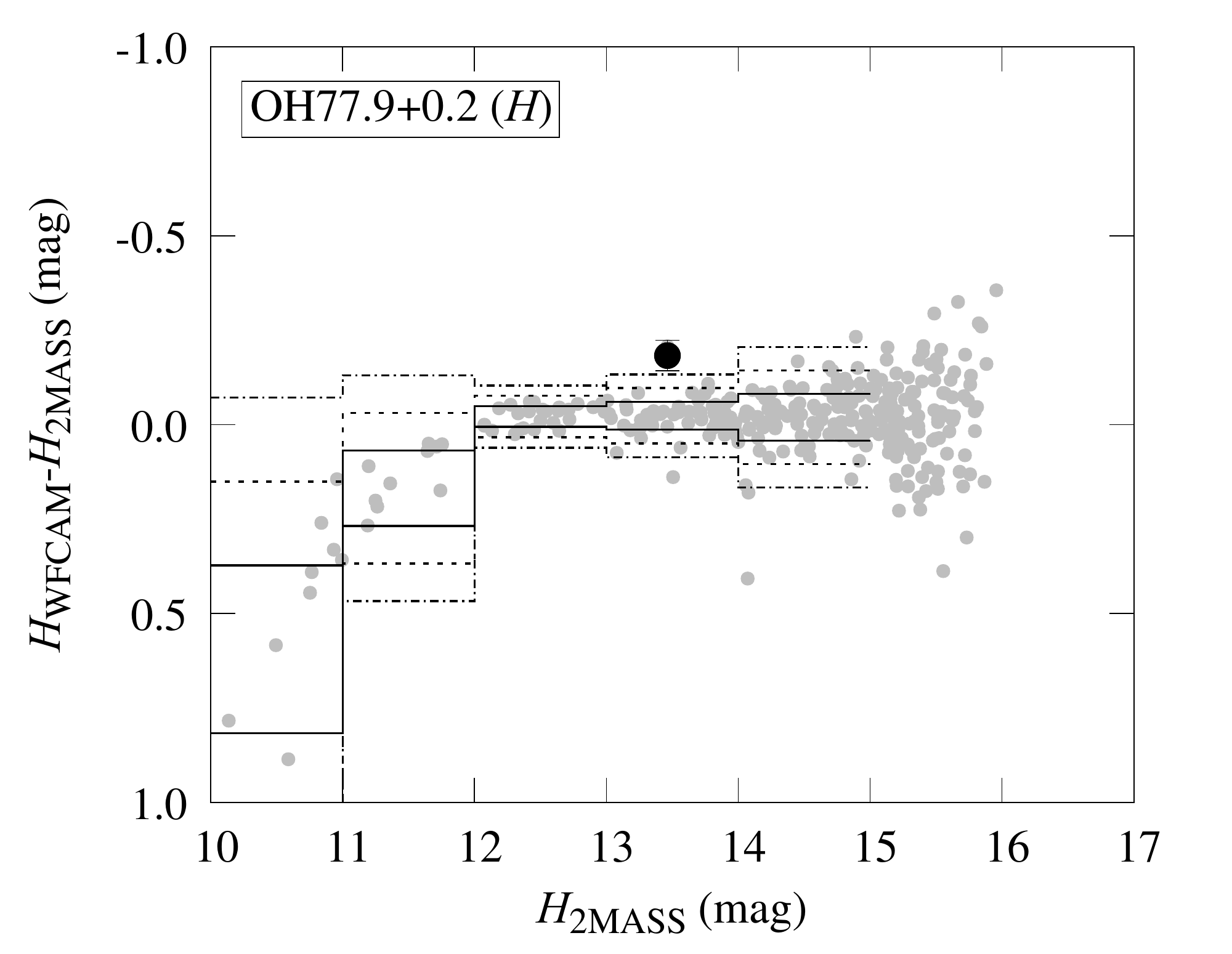}{0.33\textwidth}{}
\hspace{-1.cm}\fig{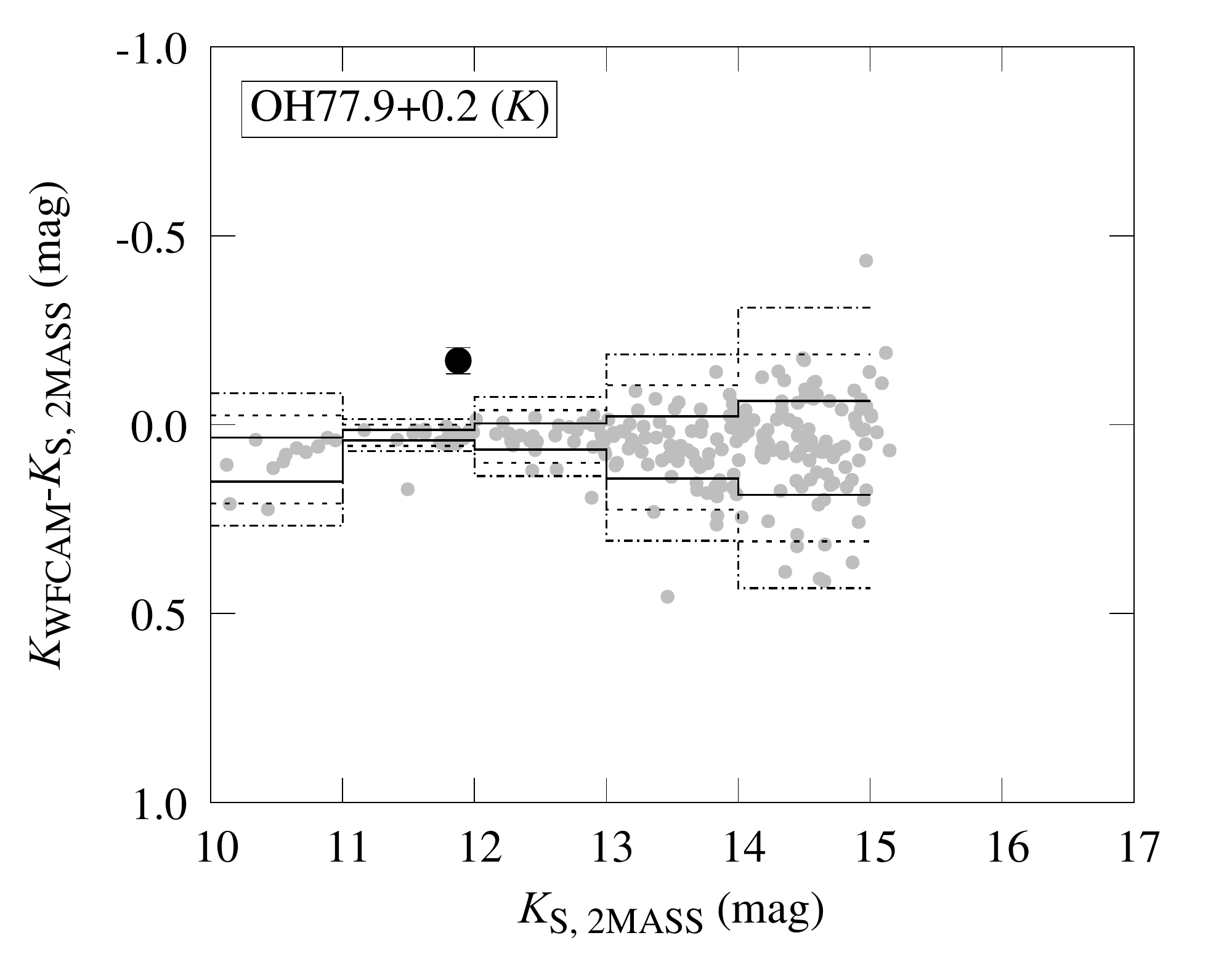}{0.33\textwidth}{}}
\vspace{-0.6cm}
\caption{Residual systematic magnitude difference between 2MASS and UKIDSS observations plotted against 2MASS magnitudes. Object name and observation band are shown in the left upper corner of each panel. In the derivation of the magnitude difference, 2MASS magnitudes are converted to the WFCAM photometric system. Average and standard deviation of the magnitude difference are calculated in each 1-mag bin in the horizontal direction. Solid, dotted, and dash-dotted lines show the 1, 2, and 3$\sigma$ regions from the average values. The targets are shown with black circles, and the error bars show only photometric uncertainties.
\label{fig:sys-2mass-ukidss}}
\end{figure*}

The magnitude differences between two UKIDSS {\it K}-band observations of the surrounding objects are shown in Figure~\ref{fig:sys-ukidss} to examine the systematic errors between the magnitude difference and object brightness. To examine the local systematic errors, we select objects within $3\arcmin$ from the target OH/IR stars and compute the average and standard deviation of the magnitude difference for each 1-mag bin (Figure~\ref{fig:sys-ukidss}). The data for the non-variable OH/IR stars are also shown in black diamonds, and we can find that the observed magnitude differences of our targets significantly deviate from the average. Therefore, we conclude that the brightening observed between two UKIDSS observations are real even if the systematic errors are considered.

\subsection{2MASS-UKIDSS difference} \label{sec:sys-2u}
\begin{figure*}
\gridline{\fig{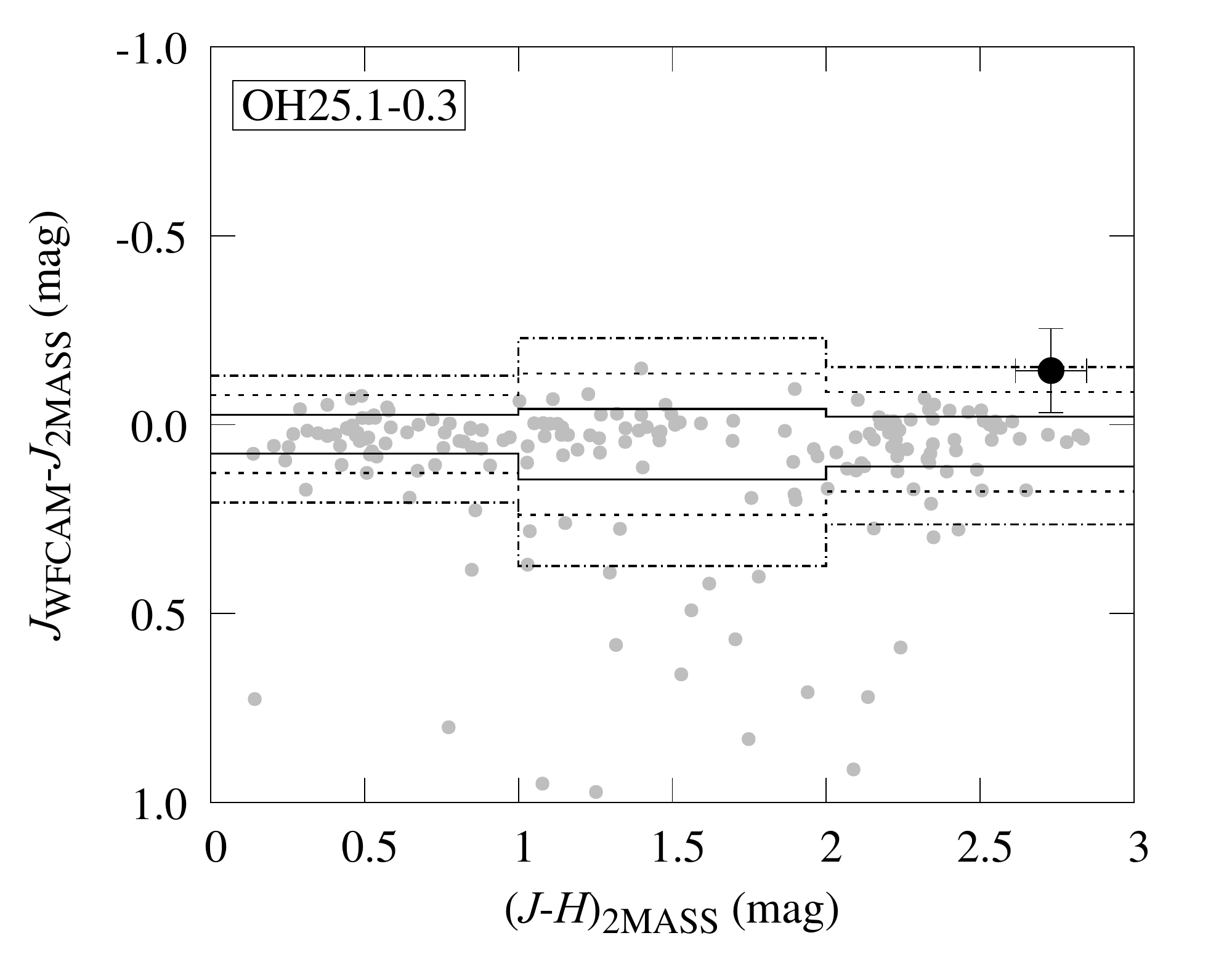}{0.33\textwidth}{}
\hspace{-1.cm}\fig{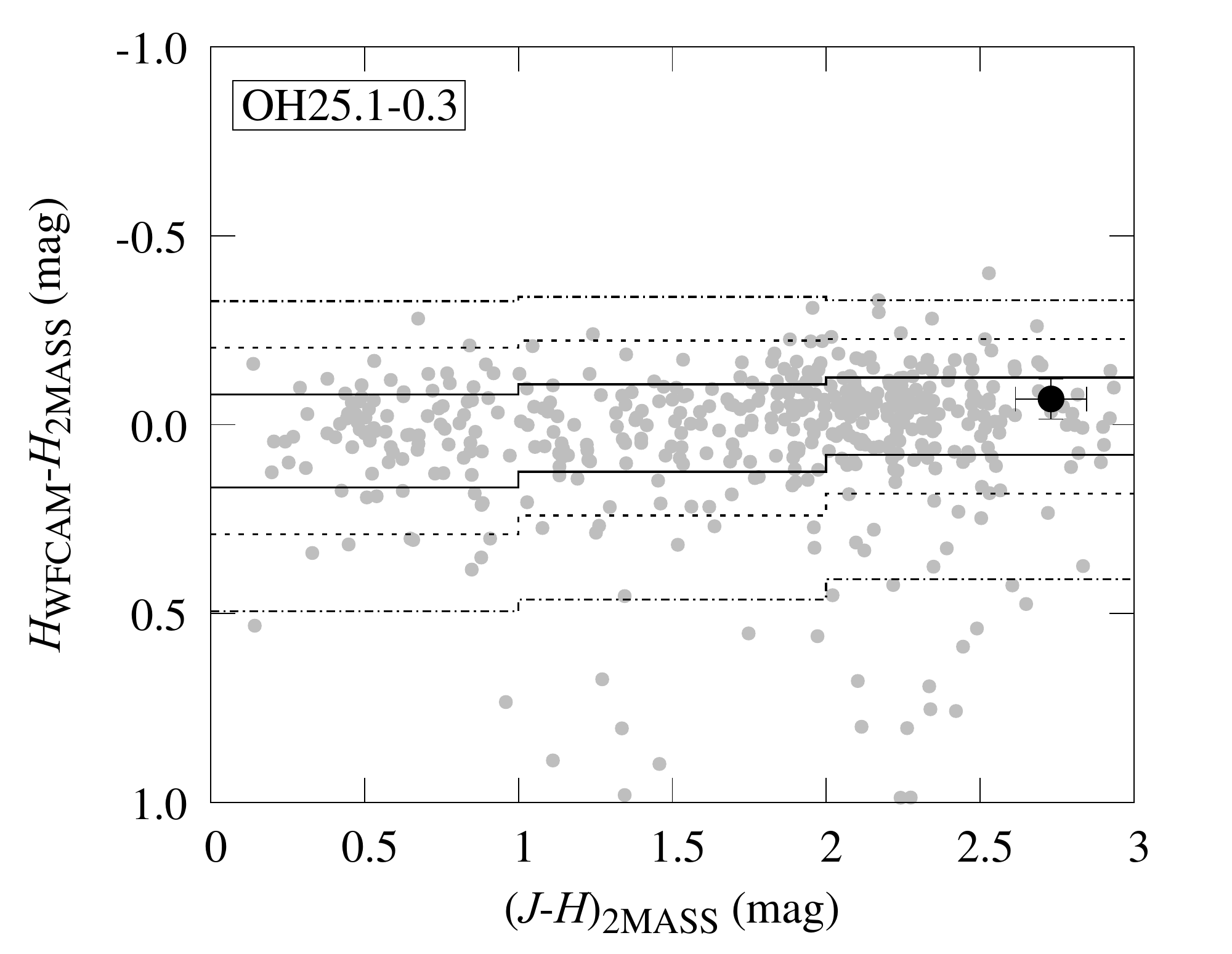}{0.33\textwidth}{}
\hspace{-1.cm}\fig{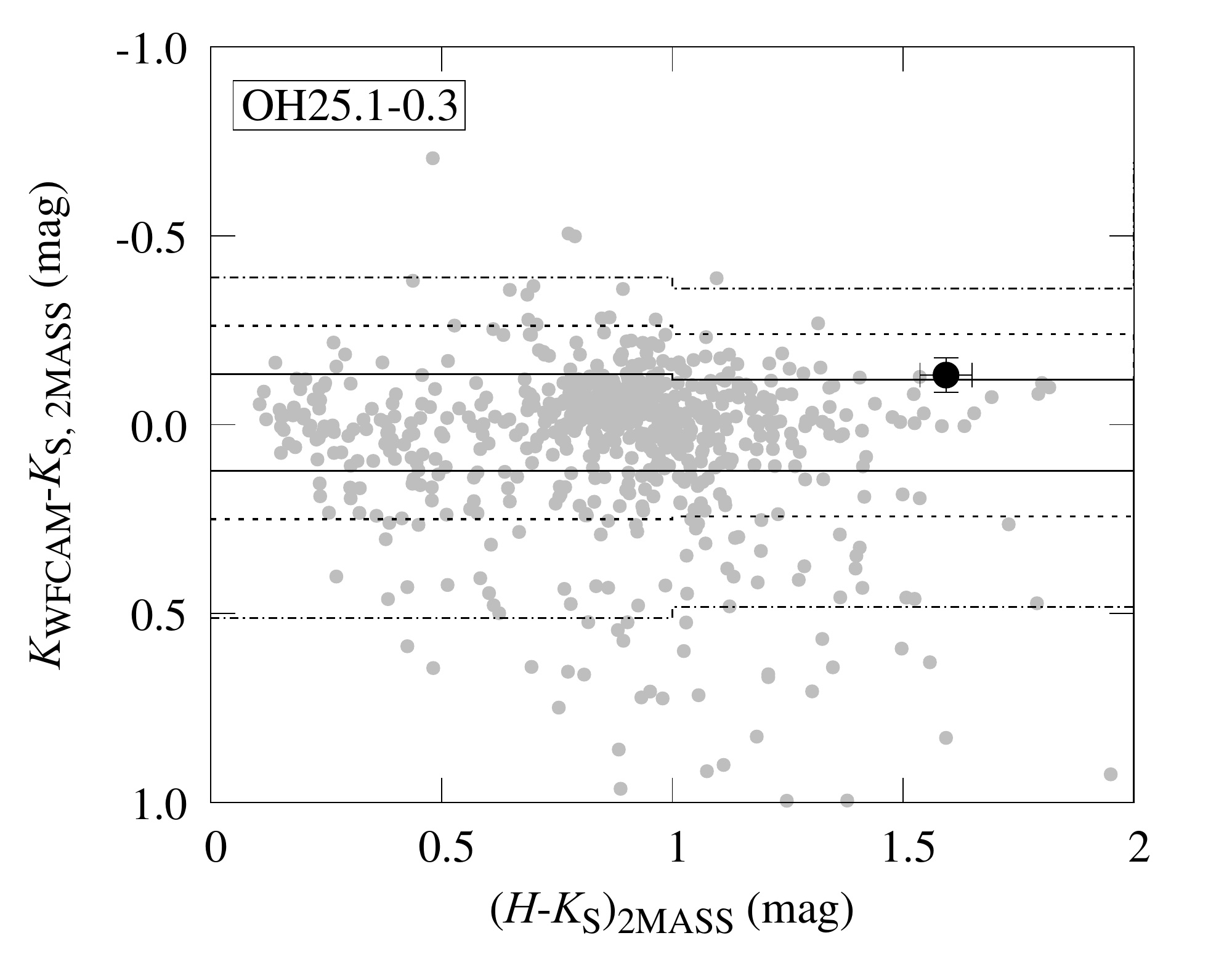}{0.33\textwidth}{}}
\vspace{-1.3cm}
\hspace{-0.5cm}
\gridline{\fig{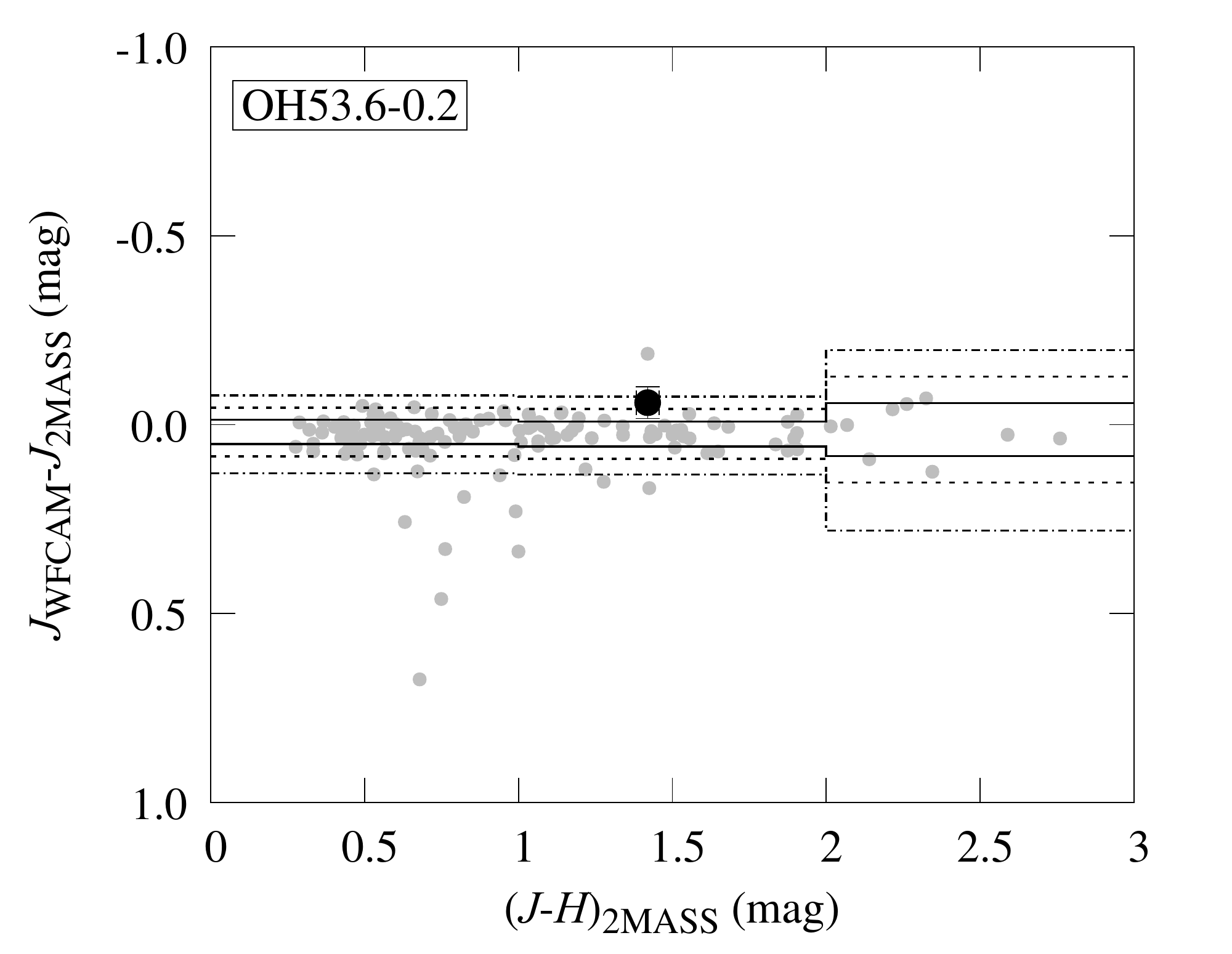}{0.33\textwidth}{}
\hspace{-1.cm}\fig{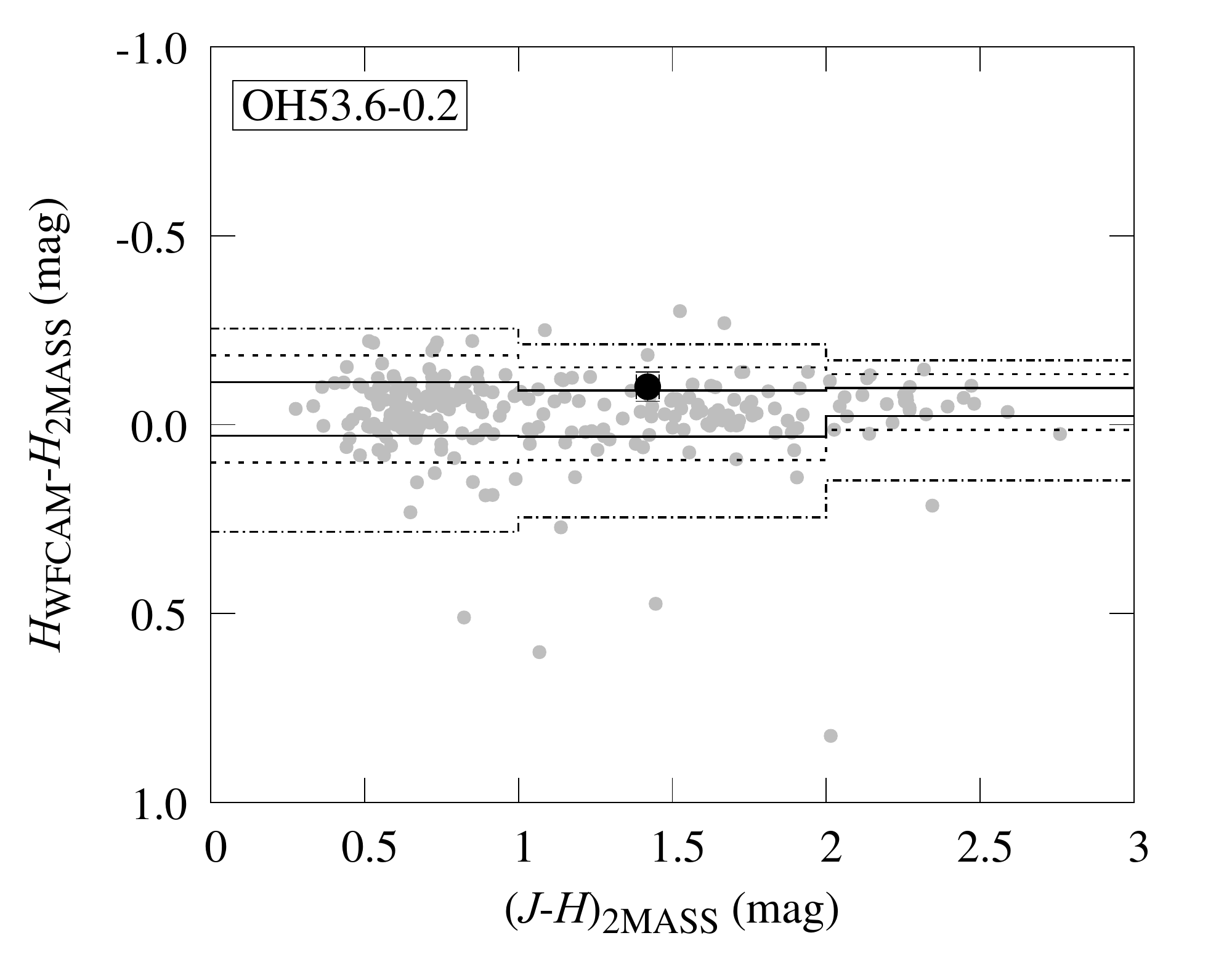}{0.33\textwidth}{}
\hspace{-1.cm}\fig{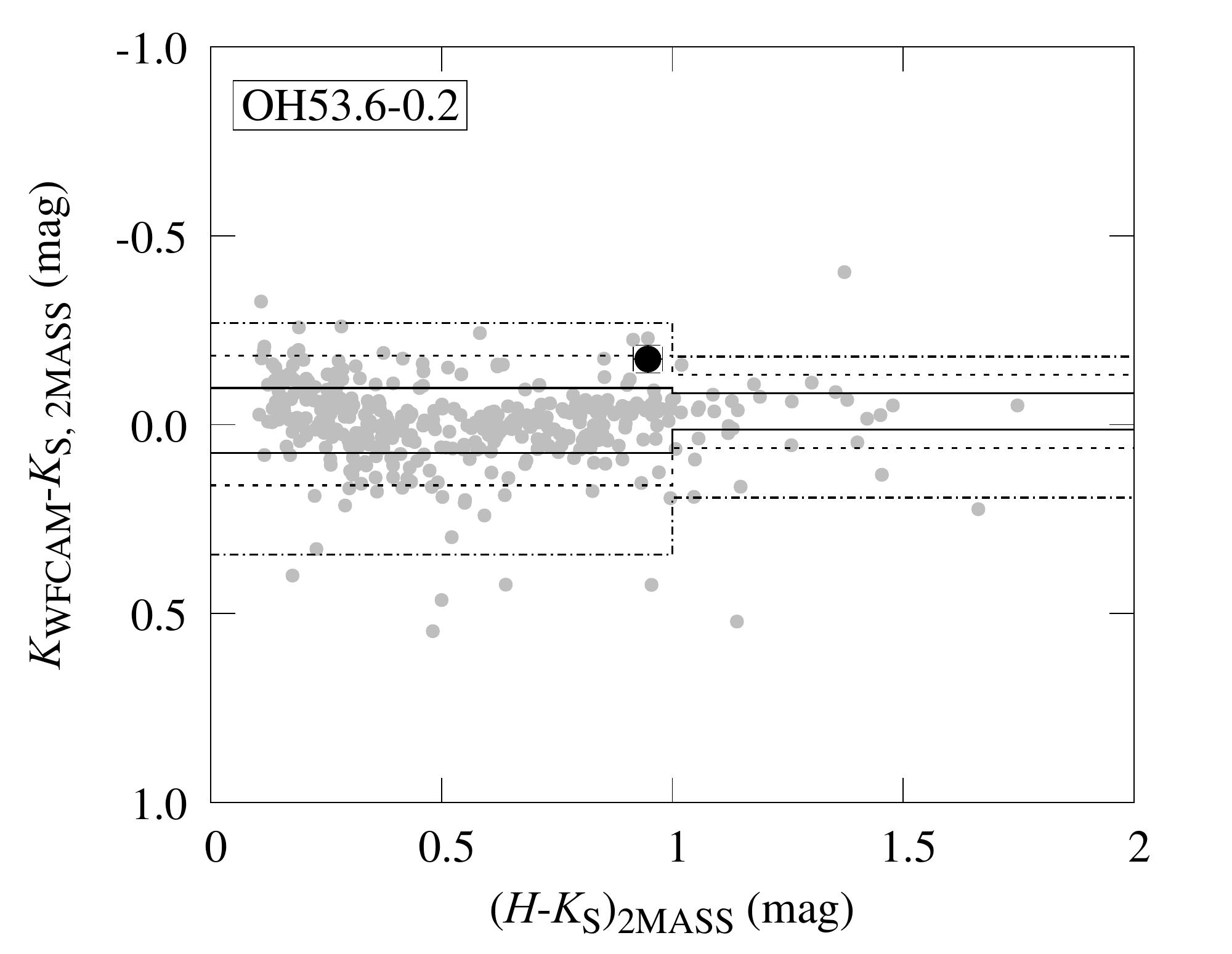}{0.33\textwidth}{}}
\vspace{-1.3cm}
\hspace{-0.5cm}
\gridline{\fig{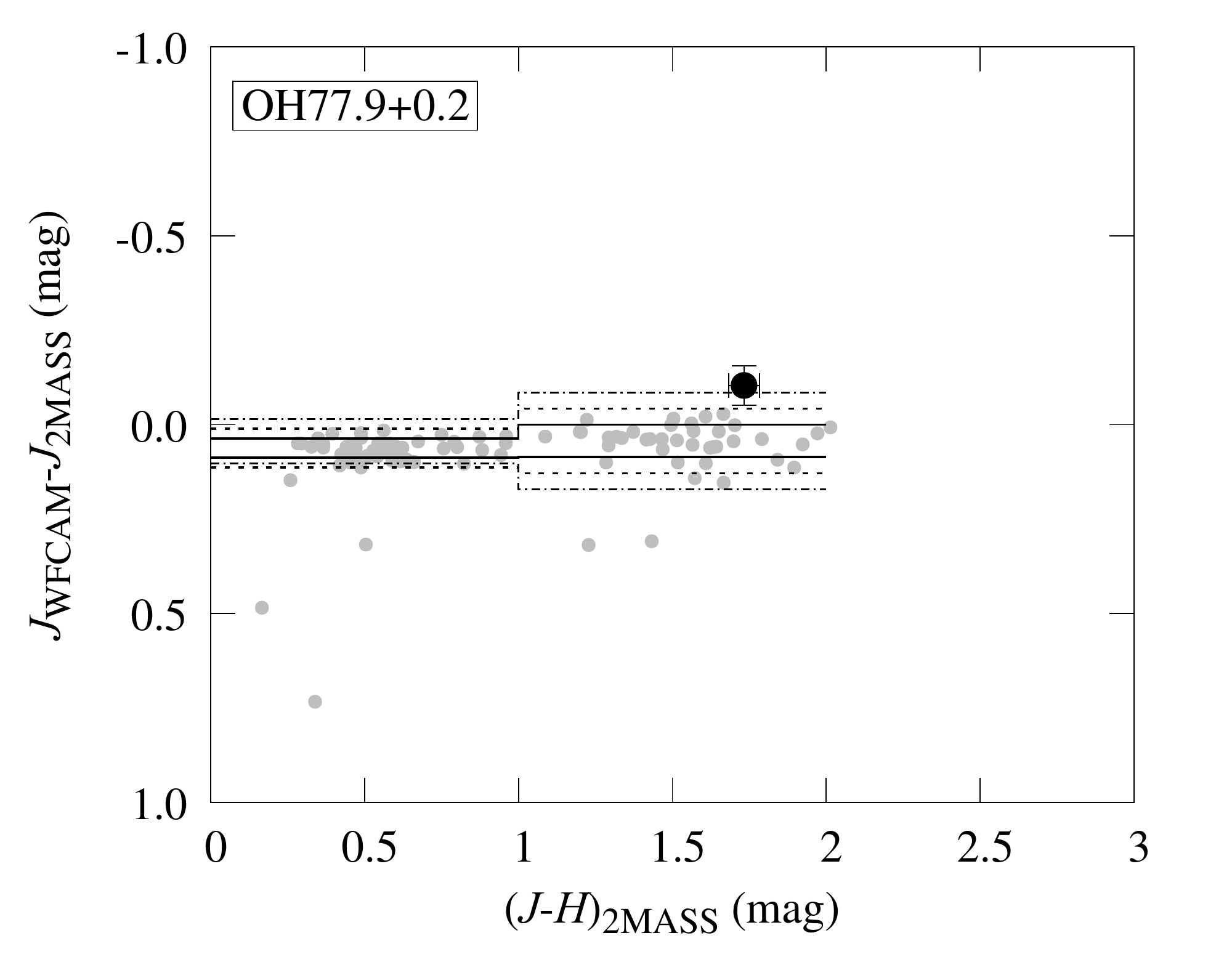}{0.33\textwidth}{}
\hspace{-1.cm}\fig{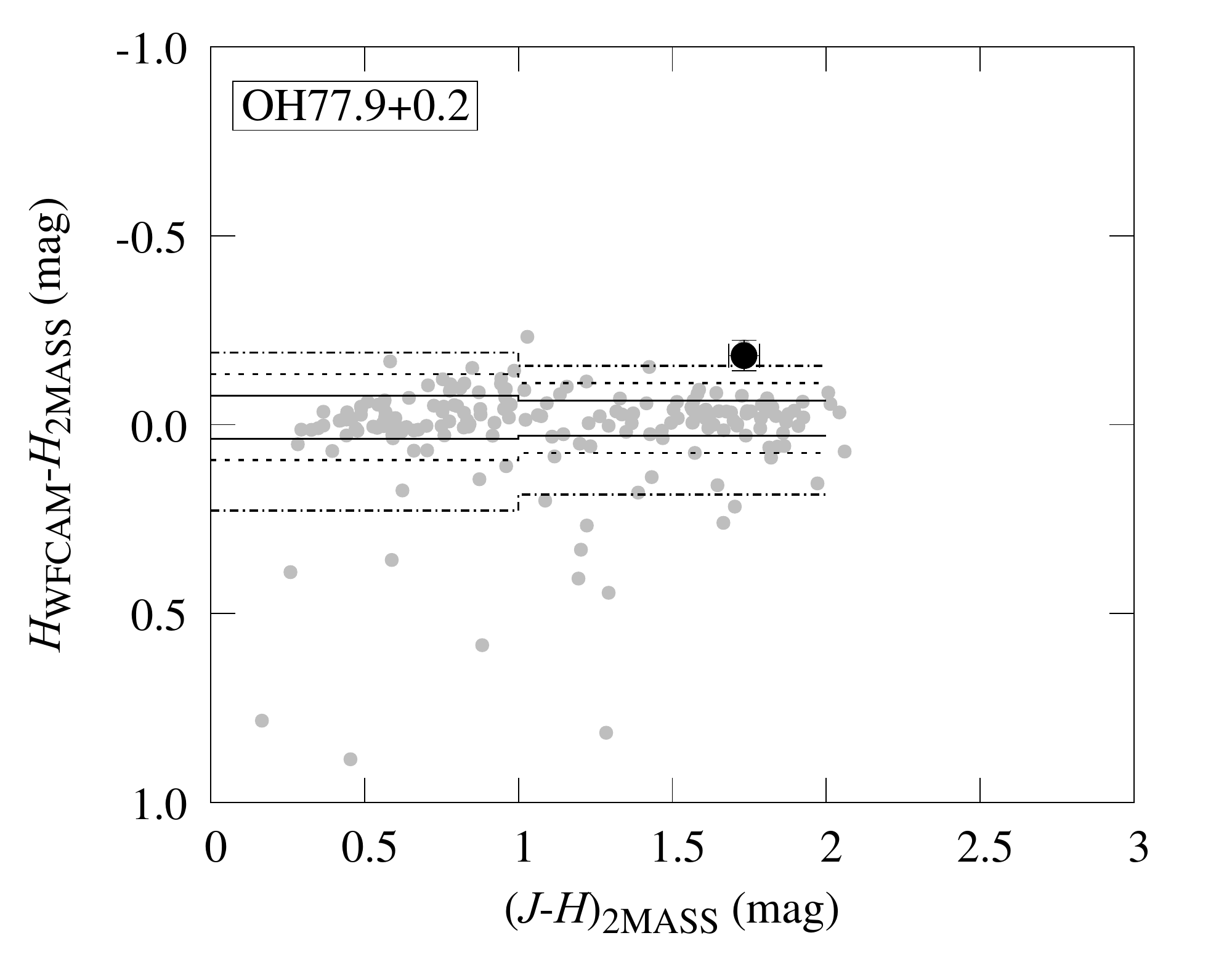}{0.33\textwidth}{}
\hspace{-1.cm}\fig{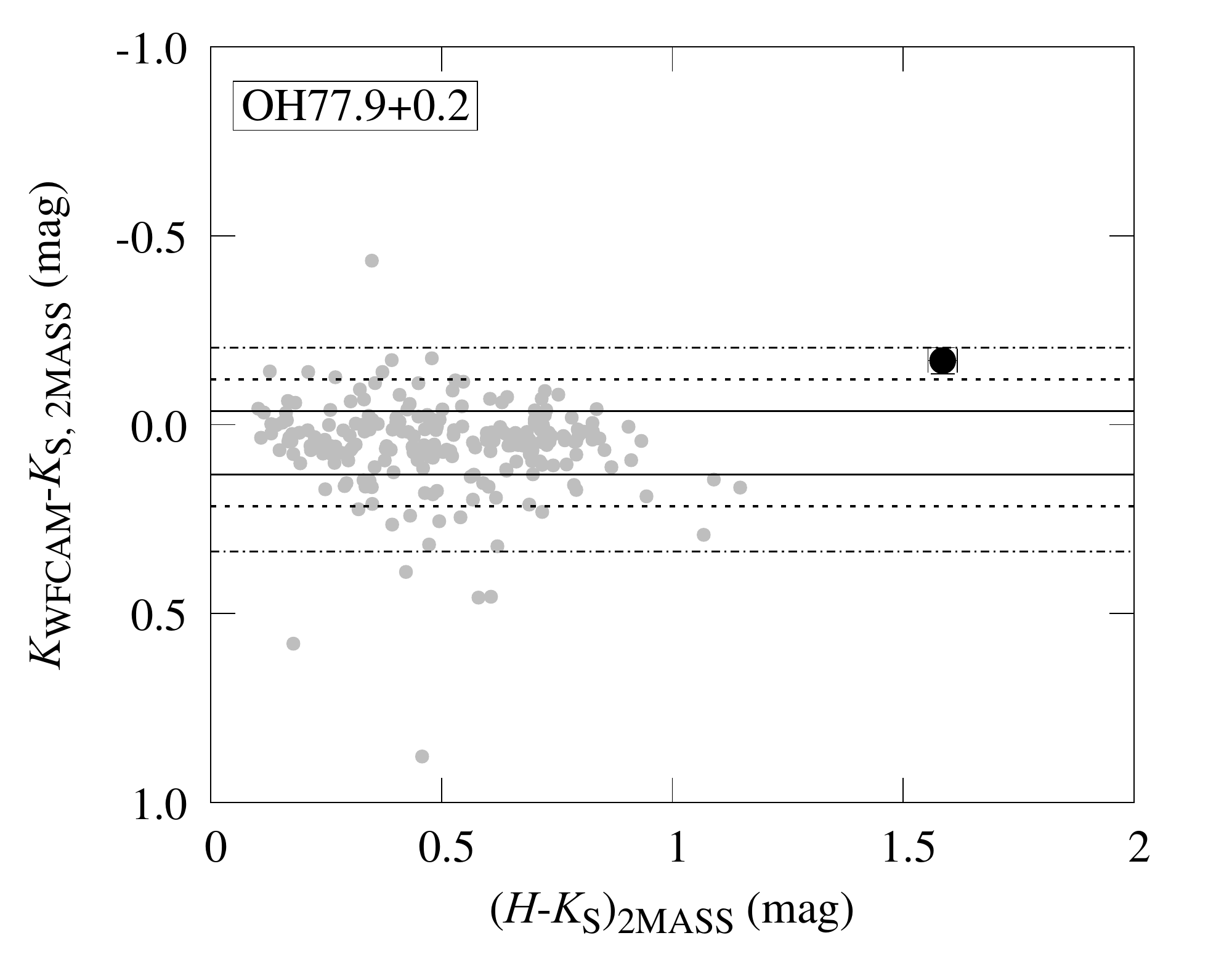}{0.33\textwidth}{}}
\vspace{-0.6cm}
\caption{Same as Figure~\ref{fig:sys-2mass-ukidss}, but plotted against 2MASS colors. Average and standard deviation in the {\it K}-band plot for OH77.9+0.2 (bottom right panel) are calculated in 2-mag color bin.
\label{fig:sys-2mass-ukidss-color}}
\end{figure*}

The systematic difference between the 2MASS and WFCAM systems are corrected by the system conversion as described in \S\ref{sec:variability}. However, there can be residual systematic difference which cannot be corrected with the system conversion. To assess such a residual systematic error, we examine the magnitude differences between the 2MASS and UKIDSS observations.

Figure~\ref{fig:sys-2mass-ukidss} and \ref{fig:sys-2mass-ukidss-color} show the magnitude difference plotted against 2MASS magnitudes and 2MASS colors, respectively. Gray dots represent the data corresponding to objects within 3$\arcmin$ from the target. To reduce the degree of data scatter, data with 2MASS photometric errors of $>$0.2 mag are removed. In Figure~\ref{fig:sys-2mass-ukidss-color}, only bright objects are plotted: the magnitude threshold is set to 14, 15, and 15 mag for OH25.1$-$0.3, OH53.6$-$0.2, and OH77.9+0.2, respectively. The difference is calculated after converting the 2MASS magnitudes to the WFCAM photometric system. The average and standard deviation are calculated in each 1-mag bin in the horizontal direction, and the 1, 2, and 3$\sigma$ regions are indicated with solid, dotted, and dash-dotted lines in each panel. In the {\it K}-band plot for OH77.9+0.2 in Figure~\ref{fig:sys-2mass-ukidss-color}, the average and standard deviation of the magnitude difference is calculated in a 2-mag color bin because of the small number of samples.

For OH25.1$-$0.3, systematic trends with a positive slope are seen in all bands in Figure~\ref{fig:sys-2mass-ukidss}. The magnitude difference in the {\it K} band slightly deviates from the systematic trend, while those in the {\it J} and {\it H} bands are not significant. The average magnitude differences in the bins containing the target are $-0.039$, $-0.043$, and 0.028 mag in the {\it J}, {\it H}, and {\it K} bands, respectively. These are within the systematic calibration error expected from the systematic uncertainties of 2MASS and UKIDSS observations. The color dependence of the magnitude difference is not significant. If we correct the measured systematic errors, the brightening in the {\it J} and {\it H} bands become insignificant, while that in the {\it K} band increases by $\sim$20\%. The color changes also get more significant.

In the plots for OH53.6$-$0.2, the data shows flat patterns in the range around the target position in both Figure~\ref{fig:sys-2mass-ukidss} and \ref{fig:sys-2mass-ukidss-color}. OH53.6$-$0.2 shows deviation from the trend in all bands, especially in the {\it K} band. The absolute values of the mean magnitude differences in the bins where the target is contained are 0.004--0.039 and are in the range expected from calibration errors. Since this kind of error is already taken into account, we do not need extra correction of these offsets.

\begin{figure*}
\gridline{\fig{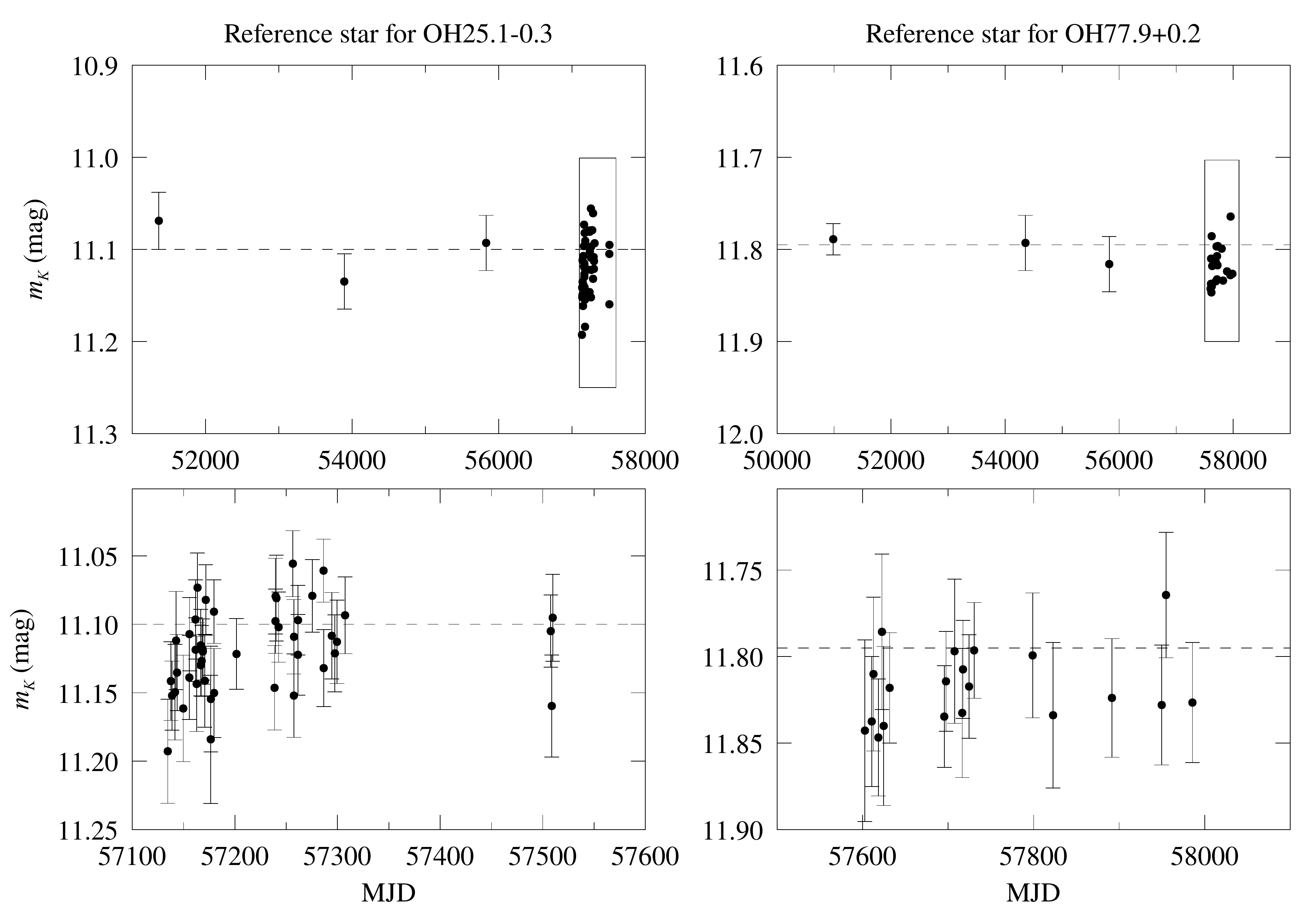}{0.8\textwidth}{}}
\vspace{-1cm}
\caption{{\it K}-band light curves of reference stars. Left and right panels show the light curves of reference stars for OH25.1$-$0.3 and OH77.9+0.2, respectively. Bottom panels are magnified plots of the OAOWFC data. The magnified regions are shown with squared regions in the top panels. Dashed lines show the mean magnitude derived from 2MASS and UKIDSS observations. All of the magnitudes are converted to the WFCAM system.
\label{fig:K-bandLC-refs}}
\end{figure*}

The plots for OH77.9+0.2 also shows flat patterns around the target position in both Figure~\ref{fig:sys-2mass-ukidss} and \ref{fig:sys-2mass-ukidss-color}, and the data of OH77.9+0.2 significantly deviates from the trends seen in the surrounding stars. However, in Figure~\ref{fig:sys-2mass-ukidss}, faint stars show slightly large offsets in the {\it J} band. In the bin containing the target, the average magnitude difference is 0.072 in the {\it J} band. This is a large value to explain with the systematic uncertainties of 2MASS and UKIDSS observations, while other bands show only small differences. By correcting the large offset in the {\it J} band, the brightening of OH77.9+0.2 in the {\it J} band increases by a factor of 1.7, and the rate of change in $(J-K)$ color changes to $-0.001\pm0.007$ mag yr$^{-1}$. Even if this correction is applied, the result that OH77.9+0.2 does not show significant color change is not affected.

Based on the discussion above, the findings of this work, the {\it K}-band brightening and the absence of the expected blueing, are not affected by the possible residual systematic errors.

\subsection{OAOWFC data} \label{sec:oaowfcunc}

To assess the systematic errors of the OAOWFC data, we examine the light curve of the reference stars in detail. Figure~\ref{fig:K-bandLC-refs} shows the closeup plots of the light curves. The dashed lines are the average magnitudes calculated from the 2MASS and UKIDSS observations. 

The standard deviations of the OAOWFC data are 2--3\% and consistent with the measurement errors shown with the error bars in the bottom panels of Figure~\ref{fig:K-bandLC-refs}. It means that the photometric measurements and error evaluation of the OAOWFC data are properly performed. 

The 2MASS and UKIDSS data distribute within $\pm$3--4\% around the average magnitudes. This is reasonable based on the systematic uncertainties of 2MASS and UKIDSS data of 2--3\%. The typical deviations of the OAOWFC data from the average magnitudes are $\sim$3\% as seen in Figure~\ref{fig:K-bandLC-refs}. In addition, the OAOWFC data seem to show slightly fainter magnitudes than the average magnitudes. This tendency can be the reason of the magnitude difference between the OAOWFC observations and the 2MASS-UKIDSS trend seen in Figure~\ref{fig:K-bandLC}. We do not make further investigations of the cause of this systematic error because it is out of the scope of this study, but if we correct this systematic error, the brightening of the target objects becomes more significant.

Finally, the light curve of the reference star for OH25.1$-$0.3 does not show periodic variabilities unlike OH25.1$-$0.3. This means that the periodic variability seen in the light curve of OH25.1$-$0.3 can be a real variability. Since the number of observations is small, continuous and dense monitoring observations are required to validate the variability.


\listofchanges

\end{document}